\documentclass{article}
\usepackage[utf8]{inputenc}
\pdfoutput=1  
\usepackage{jcappub}
\usepackage{graphicx}
\usepackage{amsmath,amssymb}
\graphicspath{{./figures/}}
\usepackage{appendix}
\usepackage{placeins}
\usepackage[normalem]{ulem}
\usepackage{multirow}
\usepackage{colortbl}

\usepackage{mathtools}
\usepackage{amsfonts}
\usepackage{upgreek}
\usepackage{latexsym}
\usepackage{stfloats}
\usepackage{afterpage}
\usepackage{caption}
\usepackage{subcaption}
\usepackage{enumitem}
\usepackage[table]{xcolor}

\newcommand{\be}{\begin{equation}}
\newcommand{\ee}{\end{equation}}
\newcommand{\beq}{\begin{equation}}
\newcommand{\eeq}{\end{equation}}
\newcommand{\bea}{\begin{eqnarray}}
\newcommand{\eea}{\end{eqnarray}}

\newcommand{\HH}{{\cal H}}

\newcommand{\de}{\delta}

\newcommand{\class}{{\sc class}}

 \definecolor{magenta}{rgb}{0.1,0.98,0.6}

\definecolor{dgreen}{rgb}{0,0.6,0.0}

\definecolor{cyan}{rgb}{0,0.6,0.6}

\definecolor{mypink}{rgb}{0.9,0.4,0.7}

\title{The halo bias for number counts on the light cone from relativistic N-body simulations}
\author[a]{Francesca Lepori,}
\author[a]{Sebastian Schulz,}
\author[a]{Julian Adamek,}
\affiliation[a]{Institute for Computational Science, Universit\"at Z\"urich, Winterthurerstr.\ 190, 8057 Z\"urich, Switzerland}
\author[b]{and Ruth~Durrer}
\affiliation[b]{D\'epartement de Physique Th\'eorique and Centre for Astroparticle Physics, Universit\'e de Gen\`eve,
24 quai Ernest-Ansermet, 1211 Gen\`eve 4, Switzerland}

\emailAdd{francesca.lepori2@uzh.ch}
\emailAdd{sebastian.schulz@uzh.ch}
\emailAdd{julian.adamek@uzh.ch}
\emailAdd{ruth.durrer@unige.ch}

\abstract{We present the halo number counts and its two-point statistics, the observable angular power spectrum, extracted for the first time from relativistic N-body simulations. The halo catalogues used in this work are built from the relativistic N-body code
\textit{gevolution}, and the observed redshift and angular positions of the sources are computed using a non-perturbative ray-tracing method, which includes all relativistic scalar contributions to the number counts. 
We investigate the validity and limitations of the linear bias prescription to describe our simulated power spectra. In particular, we assess the consistency of different bias measurements on large scales, and we estimate up to which scales a linear bias is accurate in modelling the data, within the statistical errors.
We then test a second-order perturbative bias expansion for the angular statistics, on a range of redshifts and scales previously unexplored in this context, that is $ 0.4 \le \bar{z} \le 2 $ up to scales $\ell_\mathrm{max} \sim 1000$. 
We find that the angular power spectra at equal redshift can be modelled with high accuracy with a minimal extension of the number of bias parameters, that is using a two-parameter model comprising linear bias and tidal bias.
We show that this model performs significantly better than a model without tidal bias but with quadratic bias as extra degree of freedom, and that the latter is inaccurate
at $\bar{z} \ge 0.7$. 
Finally, we extract from our simulations the cross-correlation of halo number counts and lensing convergence. 
We show that the estimate of the linear bias from this
cross-correlation is consistent with the measurements based on the clustering statistics alone, and that it is crucial to take into account the effect of magnification in the halo number counts to avoid systematic shifts in the computed bias. 
}

\date{\today}

\begin{document}

\maketitle

\section{Introduction}
At present and in the near future, many very large galaxy surveys are being and will be performed. The Dark Energy Survey (DES) recently released the Year 3 data products~\cite{DES:2021wwk}; the Dark Energy Spectroscopic 
Instrument (DESI)~\cite{DESI:2016fyo} completed its Survey Validation observations, in preparation for the 5-year survey that just started. Future experiments such as the Vera C. Rubin Observatory's Legacy Survey of Space and Time (LSST)~\cite{LSSTScience:2009jmu,LSSTDarkEnergyScience:2012kar}, the Euclid satellite~\cite{EUCLID:2011zbd, Amendola:2016saw}, 
the Spectro-Photometer for the History of the Universe, Epoch of Reionization, and Ices Explorer (SPHEREx)~\cite{Dore:2014cca}, and the Square Kilometre Array (SKA)~\cite{Maartens:2015mra} are designed to map the galaxy distribution in the Universe on the largest scales, from early until present time.

One of the main challenges in cosmology is to develop a robust theoretical model for our observables that will allow us to extract information from  ultra-large scales down to  intermediate and small scales that we will observe with stunning precision, thanks to the vast volume of observational data available in the near future. 
In the concordance model for cosmology, the galaxy distribution that we observe originates from primordial perturbations generated in the early Universe, see e.g.~\cite{RuthBook}. Therefore, in order to correctly 
 interpret observations from a galaxy survey, we need three key ingredients: i) an accurate description for the evolution of the matter density field and the geometry of the Universe, ii) an accurate description of the relation between the matter distribution and the distribution of galaxies (or other observable tracers), and iii)
 a way to account for all observational effects, including light-cone projection, selection effects and other systematics possibly present in our data. 
The first ingredient can be obtained either with analytical methods, within the framework of perturbation theory~\cite{Bernardeau:2009bm} and the Effective Field Theory of Large Scale Structure~\cite{Baumann:2010tm, Carrasco:2012cv, Porto:2013qua}, or with numerical methods by running cosmological simulations, where the continuous dark matter distribution is represented via discrete N-body particles~\cite{Nbody:1988,Angulo:2021kes}. 
Concerning the second point, the relation between matter and galaxies is encoded in the so-called `galaxy bias'~\cite{Desjacques:2016bnm}. 
The galaxy bias depends on the complex process of galaxy formation~\cite{mo_van_den_bosch_white_2010}. However, it is well established that galaxies reside inside dark matter halos --- virialized objects that form through gravitational instability. Dark matter halos also constitute biased tracers of the large-scale dark matter distribution. Galaxy and halo bias can be studied and modelled using similar methods, with the simplification that halos are composed of dark matter and therefore baryonic effects do not affect halos to the same extent as galaxies. 
The third point encodes different effects. One important observational effect is that we perform our observations on our past light cone, therefore relativistic corrections such as redshift-space distortions~\cite{Kaiser:1987qv} and lensing magnification~\cite{Matsubara:2004fr} need to be modelled properly. 

Fully relativistic number counts have been derived first in linear perturbation theory in Refs.~\cite{Yoo:2009au, Yoo:2010ni, Bonvin:2011bg, Challinor:2011bk, Jeong:2011as}, then later on extended to higher orders in Refs.~\cite{Yoo:2014sfa, Umeh:2014ana, Bertacca:2014hwa, DiDio:2014lka, Nielsen:2016ldx, Magi:2022nfy}, generalized to include vector modes~\cite{Durrer:2016jzq}, and to describe spatially curved geometries~\cite{DiDio:2016ykq}. Furthermore, the same formalism has been applied to other tracers of the Large-Scale Structure (LSS)~\cite{Bonvin:2008ni, Bernardeau:2009bm, Hall:2012wd, Irsic:2015nla, Alonso:2021obj}. 
The linear two-point statistics of the relativistic number counts has also been implemented in the publicly available Boltzmann codes Cosmic Linear Anisotropy Solving System ({\sc class})~\cite{DiDio:2013bqa, DiDio:2013sea, Lesgourgues:2011re, Blas:2011rf} and Code for Anisotropies in the Microwave Background ({\sc camb})~\cite{Challinor:2011bk, Challinor:2011, Lewis:1999bs}. 
However, the linear treatment of cosmological 
perturbations has a limited range of validity. For this reason, it is essential to compare our perturbative framework against fully nonlinear and non-perturbative results that can be obtained with cosmological N-body simulations. 
In this direction, Newtonian simulations have long been a cornerstone in the study of lensing and clustering statistics in the nonlinear regime, see for example Refs.~\cite{Fosalba:2007mf, Hilbert:2008kb, Fosalba:2013mra, Fosalba:2013wxa}. 
More recently, new numerical codes have been developed to include the effects of General Relativity (GR) in cosmological simulations~\cite{Bentivegna:2015flc, Giblin:2015vwq, Adamek:2015eda, Adamek:2016zes, Macpherson:2016ict, Barrera-Hinojosa:2019mzo, Adamek:2020jmr}, and several studies have been carried out to explore the nonlinear regime of cosmological observables including GR effects~\cite{Giblin:2017ezj, Borzyszkowski:2017ayl, Adamek:2018rru, Lepori_2020b,Lepori:2021lck, Barrera-Hinojosa:2021msx, Macpherson:2021gbh, Rasera:2021mvk, Tian:2021qgg}. 

In this context, some of us have studied the angular statistics of the weak-lensing observables (convergence, ellipticity and rotation) in Ref.~\cite{Lepori_2020b},
and the number counts of N-body particles in Ref.~\cite{Lepori:2021lck}, extracted directly from high-resolution N-body simulations performed with the relativistic code \textit{gevolution}~\cite{Adamek:2015eda,Adamek:2016zes}.
In this work we extend the analysis of Ref.~\cite{Lepori:2021lck} to study the relativistic number counts of dark matter halos. Dark matter halos are biased
tracers of the LSS. Therefore, in this work, we tackle one
of the main difficulties in interpreting galaxy surveys: the biasing problem. 


In this paper, we present a detailed analysis of the halo bias, extracted directly from the observable angular power spectrum. Halos are collapsed density peaks of matter and therefore much easier to model than actual galaxies. We work with a complete population of halos and do not consider additional observational biases that would come with the selection function of a realistic galaxy survey, see e.g.~\cite{Hui:2007cu, Schmidt:2009ri, Schmidt:2009b}.
For the first time, we investigate the nonlinear bias in the observable statistics in directly observable harmonic space, using relativistic simulations to ray-trace our past light cone. We test the validity and limitations of the linear bias model, and we compare our simulation data to a second-order perturbative bias expansion. Furthermore, we study the interplay of nonlinear biasing and redshift-space distortions (RSD) and the halo bias in the cross-correlation of number counts and lensing convergence. 
The angular harmonic space has the advantage that is constructed from the truly observable coordinates (redshift and angular positions), while
in a 3D analysis in Fourier or configuration space one needs to assume a cosmological model to convert the observed redshifts and angular positions into distances. On the other hand, it is not convenient to use if the redshifts of the objects are known very precisely: 
in order to fully exploit the redshift information, one would need to split the sample in many thin redshift bins and run a full tomographic analysis (i.e.\ including also cross-correlations between unequal-redshift bins).  This would make the analysis computationally prohibitive and would lead to large shot noise per redshift bin. For this reason, the specific application we have in mind for
this analysis are photometric galaxy surveys, such as 
DES~\cite{DES:2021wwk}, LSST~\cite{LSSTScience:2009jmu,LSSTDarkEnergyScience:2012kar} and Euclid~\cite{EUCLID:2011zbd, Amendola:2016saw}, that will 
extract cosmological information from a tomographic analysis of the LSS, cross-correlating galaxy clustering and cosmic shear ($3\times2$pt analysis). Presently, the linear bias approximation is widely adopted both in the data analysis (e.g.\ in the DES fiducial analysis~\cite{DES:2021wwk}) and in cosmological forecasts~\cite{Euclid:2019clj, Euclid:2021qvm, Euclid:2021rez}. 
Our work is a first step towards a better understanding of biasing in the observable angular power spectrum, with the aim of improving the currently available models and helping the scientific community to extract information from a wider range of scales. For a photometric survey the radial resolution is $\gtrsim 50\,h^{-1}$Mpc so that RSD which are purely radial can be treated with linear theory and velocity bias is not relevant. However, for density fluctuations which are studied also on small transversal scales a treatment beyond the linear bias is necessary as we shall demonstrate.

The rest of this paper is structured as follows. In Sec.~\ref{s:sim}
we describe the N-body code, the simulations, the halo catalogues, and the ray-tracing method used to obtain the observed positions of halos and particles on the light cone. 
In Sec.~\ref{sec:maps} we discuss the method we apply to construct the sky maps of number counts, and in Sec.~\ref{sec:method-spectra-cov} we describe our estimator to extract the angular power spectrum and its covariance. 
In Sec.~\ref{s:lin-bias} we test the linear bias model against our simulation data. 
In particular, in Sec.~\ref{sec:3.1} we study the consistency of different methods to estimate the linear bias, comparing measurements on the light cone from the angular power spectrum and measurements on the snapshots, coming from the power spectrum in Fourier space. In Sec.~\ref{sec:3.2} we 
study up to which scales we can robustly model the simulated spectra with the linear bias prescriptions. 
In Sec.~\ref{s:nl-bias} we explore minimal extensions of the linear bias model for the angular power spectrum, and we show how they improve the modelling of our
simulation data. 
In Sec.~\ref{s:bias-cross} we discuss the cross-correlations of halo number counts and 
the lensing convergence. We show how to extract the linear bias from them, and 
we assess the importance of lensing magnification.
In Sec.~\ref{s:con} we summarize our results and sketch future developments of this work. 
Additional details about our analysis are provided in the appendices.

\section{The simulations}\label{s:sim}

The simulation used in this work is the \texttt{unity2} simulation described in Ref.~\cite{Lepori:2021lck} which has the following properties. It contains $5760^3$ N-body particles in a volume of $(4032\,\mathrm{Mpc}/h)^3$, which yields a mass resolution of $3\times 10^{10}\,M_\odot/h$.
The grid resolution is $700\,\mathrm{kpc}/h$ which is equal to the mean particle separation, i.e.\ the simulation has the same number of N-body particles as grid points. 
The baseline cosmology is a $\Lambda$CDM model
with two massive and one massless neutrino species. 
The massive states $m_2, m_3$ are in normal hierarchy, with masses $m_2 = 0.008689$\,eV,
and $m_3 = 0.05$\,eV.
The fiducial cosmological parameters are based on the 
\textit{Planck} 2013 analysis~\cite{Planck:2013pxb}. They are fixed to the values
$\Omega_\mathrm{cdm} = 0.26858$, $\Omega_\mathrm{b} = 0.049$, $h = 0.67$, $T_\mathrm{CMB} = 2.7255$ K, $A_s = 2.215 \times 10^{-9}$ and $n_s = 0.9619$. 
The initial conditions are generated from the linear transfer functions produced with
the code {\sc class}.

The simulation uses the public version 1.2\footnote{\url{https://github.com/gevolution-code/gevolution-1.2}} of the relativistic N-body code \textit{gevolution}. This code solves Einstein's equations together with the geodesic equation for the particle ensemble, and evolves all six degrees of freedom of the metric in Poisson gauge under the weak-field assumptions. All scalar, vector and tensor perturbations are included at first order, while the first and second spatial derivatives are included to all orders for the scalar potentials. Since we are in $\Lambda$CDM cosmology, relativistic corrections are expected to be extremely small and our results should be in good agreement with a Newtonian calculation.
Details on the code can be found in Refs.~\cite{Adamek:2015eda,Adamek:2016zes}. 
Our observer is chosen to occupy one
of the corners of the periodic domain, and the data on the light cone are stored on the fly.
Comoving positions and peculiar velocities of the particles are recorded when their trajectories intersect the background past light cone, while the metric perturbations are stored on spherical pixelised maps, that are concentric around the observer and spaced by a radial separation equal to the grid resolution. The pixelisation is handled by the \texttt{HEALPix} library \cite{Gorski:2004by}.
These data are used by a post-processing routine that integrates the null geodesic equations
backwards in time from the observer to each target object. 
The null geodesic equations are solved using a scheme which is non-perturbative for the scalar part of the gravitational potential $\phi$. The gravitomagnetic vector potential and the transverse and traceless tensor modes are neglected in this post-processing step for simplicity, but they could easily be included. 
The method is identical to the one described in detail in Ref.~\cite{Lepori_2020b},
with the only difference that the target objects are not N-body particles but halos. In the context of ray tracing, the fact that we use a relativistic simulation has the benefit that the calculation is automatically self-consistent, and we do not need to consider subtleties of interpreting the data as in Newtonian simulations \cite{Adamek:2019aad}.

\begin{figure}
\begin{center}
  \includegraphics[width=0.6\textwidth]{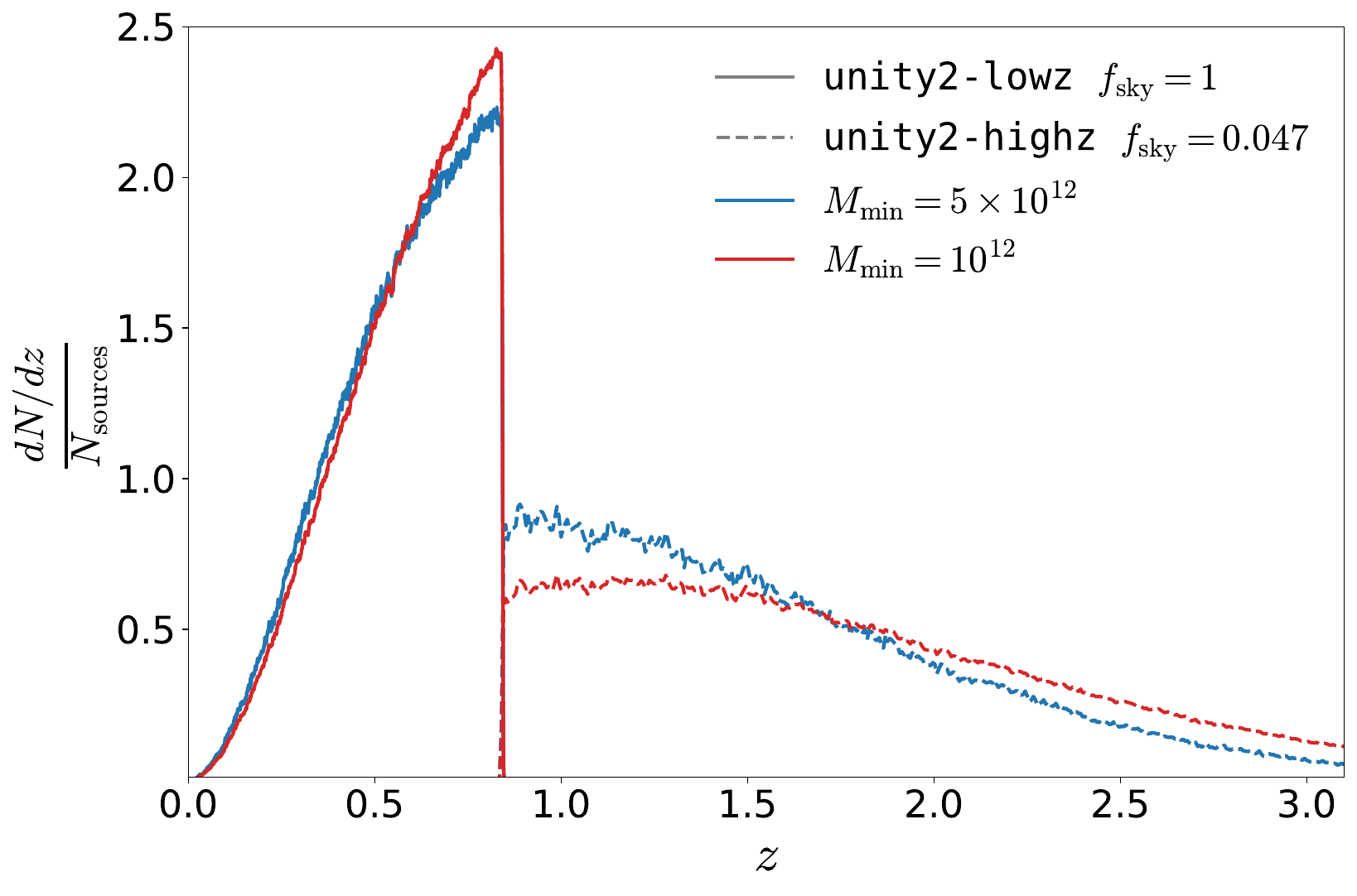}
  \end{center}
  \caption{Normalised redshift distributions of the halo catalogues used in this work, \texttt{unity2-lowz} and \texttt{unity2-highz}, for two values of the minimum halo mass, that is $M_\text{min} = 10^{12}\,M_\odot/h$ (red lines) and  $M_\text{min} = 5\times10^{12}\,M_\odot/h$ (blue lines). 
  The total number of halos (sources) for these catalogues are $N_\mathrm{sources} \approx 5.9\times10^7$ (\texttt{unity2-lowz}) and $N_\mathrm{sources} \approx 1.1\times10^7$ (\texttt{unity2-highz}) for the halo population with masses above $M_\text{min} = 10^{12}\,M_\odot/h$, whereas $N_\mathrm{sources} \approx 1.3\times10^7$ (\texttt{unity2-lowz}) and $N_\mathrm{sources} \approx 1.6\times10^6$ (\texttt{unity2-highz}) for the halo population with masses above $M_\text{min} = 5\times10^{12}\,M_\odot/h$.
  The sky fractions of the catalogues are denoted by $f_{\rm sky}$.}
  \label{fig:halo-dNdz}
\end{figure}

The halo population that we analyse in this work is produced from 
the particle data on the light cone using the halo finder {\sc rockstar}~\cite{Behroozi_2012}. This algorithm first identifies friends-of-friends groups in six-dimensional phase space and then computes halo properties based on spherical overdensity criteria in three-dimensional comoving position space. Perturbations to the metric (i.e. relativistic effects) are ignored in this step, but these would only cause corrections of order $\phi \sim 10^{-5}$ to the matter density, similar to the kinetic energy $\sim v^2$ which is also commonly ignored. However, other errors to the mass measurement are expected to be much larger.
Two halo catalogues are produced for the same observer. For $z \lesssim 0.85$ we have a catalogue with full-sky coverage , labelled \texttt{unity2-lowz}. At higher redshift, that is $0.85 \lesssim z \lesssim 3.5$, a second catalogue, labelled \texttt{unity2-highz}, covers $1932$ square degrees which correspond roughly to
$5\%$ of the sky. 

From the full catalogues produced by \textsc{rockstar} we first remove all subhalos and then apply a selection based on halo mass. For the latter, we use the property $M_{200\mathrm{b}}$ as a proxy, which is the mass enclosed within a spherical overdensity of $200$ times the background density. We note that the mass estimates from our simulation are biased low because of the limited resolution, and the numerical value of $M_{200\mathrm{b}}$ should therefore not be taken literally. Its value is not important, but rather the fact that we are selecting the most massive objects according to some threshold for some mass proxy.  

In Fig.~\ref{fig:halo-dNdz} we show the normalized redshift distribution 
for \texttt{unity2-lowz} and \texttt{unity2-highz} for two mass thresholds, $M_\text{min} = 10^{12}\,M_\odot/h$ (red) and  $M_\text{min} = 5\times10^{12}\,M_\odot/h$ (blue).
For the more restrictive threshold $M_\text{min} = 5\times10^{12}\,M_\odot/h$, 
a larger percentage of sources are found at lower redshift. 
Most of the results presented in this manuscript are based on the catalogue 
with that threshold.

\subsection{From the halo catalogue to sky maps of number counts}
\label{sec:maps}

\begin{figure*}
\centering
\begin{subfigure}[b]{0.7\textwidth}
\includegraphics[width=\textwidth]{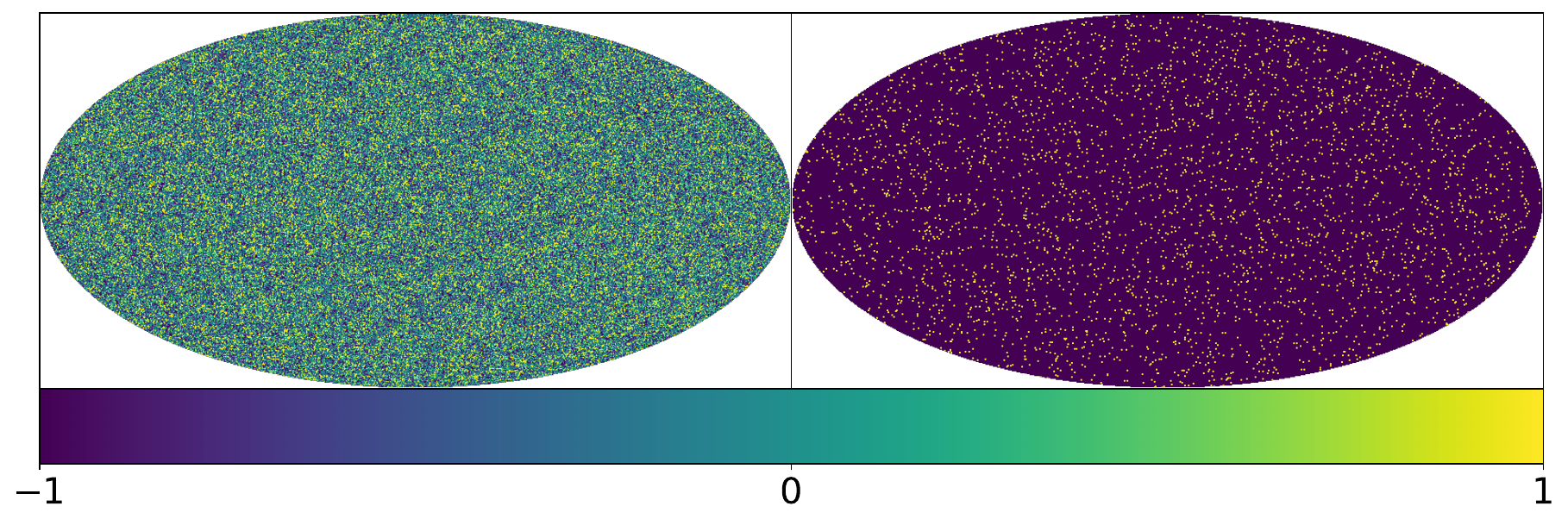}
\caption[]%
{{\small Maps from \texttt{unity2-lowz} ($f_\mathrm{sky} = 1$) at $\bar{z} = 0.5$. }}    
\label{fig:map1}
\end{subfigure}
\\
\begin{subfigure}[b]{0.7\textwidth}  
\centering 
\includegraphics[width=\textwidth]{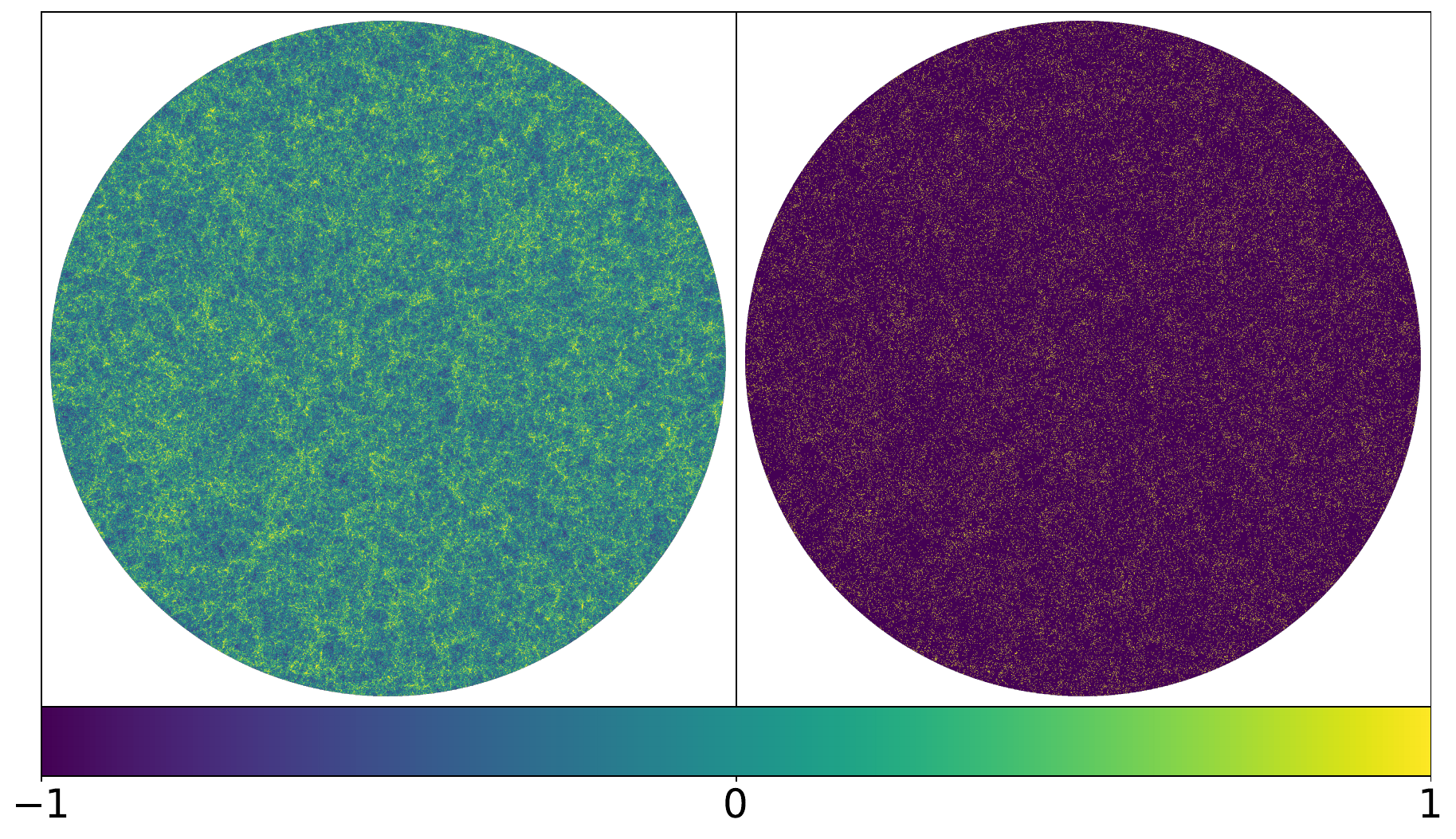}
\caption[]%
{{\small  Maps from \texttt{unity2-highz} ($f_\mathrm{sky} = 0.047$) at $\bar{z} = 1$.  }}    
\label{fig:map2}
\end{subfigure}
\caption[  ]
{\small Maps of number count fluctuations extracted from the particle (left panels) and halo (right panels) distributions. 
The top panels (a) represents a Mollweide projection of full-sky maps. The maps in the bottom panels (b) each subtend a solid angle of $1932$ square degrees, and they are shown in a gnomonic projection.  
The mass selection applied to generate these maps corresponds to $M_\text{min} = 5\times10^{12}\,M_\odot/h$. } 
\label{fig:maps}
\end{figure*}
From the halo catalogues described above, we extract sky maps of number counts
in tomographic redshift bins. We follow a similar procedure as in Ref.~\cite{Lepori:2021lck}, where it has been applied to extract the number counts of N-body particles, that is:
\begin{enumerate}
\item We select all halos with mass above a chosen threshold $M_\text{min}$,
in a redshift bin centred at $\bar{z}$ and with half-bin width 
$\sigma_z$. We apply a sharp cut-off in the redshift selection, which means we use \lq tophat\rq~bins.
\item We project the number of objects $n$
onto a pixelised map. This projection uses the \texttt{HEALPix} framework \cite{Gorski:2004by}.
\item We compute the average number of objects
per pixel $\bar{n}$, and from that we construct the map of the counts fluctuation in each pixel $i$  as
$\Delta^i \equiv (n^i - \bar{n})/\bar{n}$. 
\item When \texttt{unity2-highz} is analysed, we apply a Boolean mask on the map according to the fraction of sky $f_\mathrm{sky}$ covered by the pencil beam. 
\end{enumerate}

In our analysis, the chosen half-bin width is $\sigma_z = 0.05\,(1 + \bar{z})$, which is similar to the typical redshift resolution of a photometric survey. Fig.~\ref{fig:maps} shows sky maps of number counts extracted from the halo catalogues and the corresponding particle catalogues, at two reference redshifts
for \texttt{unity2-lowz} and \texttt{unity2-highz}.

\subsection{Estimation of the angular power spectrum and its covariance}
\label{sec:method-spectra-cov}

The angular power spectra are computed from the  spherical harmonics coefficients of the map. For a full-sky map we use the \texttt{HEALPix} estimator \texttt{anafast}, while for maps that cover a smaller fraction of the sky we
run the code \texttt{PolSpice} \cite{Szapudi:2000xj, Chon:2003gx} which 
corrects for the effect of the mask.

Assuming that perturbations are Gaussian, the covariance of the angular power spectrum is
\begin{equation}
    \mathrm{Cov}_{[\ell, \ell'][(ij)(pq)]} = \delta_{\ell, \ell'}\frac{C^{\mathrm{est}, ip}_\ell C^{\mathrm{est}, jq}_{\ell'} + C^{\mathrm{est}, iq}_\ell  C^{\mathrm{est}, jp}_{\ell'}}{f_\mathrm{sky}\Delta \ell (2\ell + 1)}, \label{eq:cov}
\end{equation}
where for $i, j, p, q$ run over the 
list of cross-correlated maps, $C^{\mathrm{est}}_\ell$ is the angular power spectrum estimated from the number counts maps, and $\Delta \ell$ is the multipole bandwidth, and $f_\mathrm{sky}$ is the fraction of sky. 
The autocorrelations are strongly affected by shot-noise, i.e.\ $C^{\mathrm{est}, ij}_\ell = C^{ij}_\ell + \delta_{ij}/N_i$, for a redshift bin with $N_i$ objects per steradian.

For the autocorrelation of a map, this translates into the error
\begin{equation}
    \sigma^2_{\ell, i} = \frac{2}{f_\mathrm{sky}\Delta \ell (2\ell + 1)} \left(C^i_\ell + \frac{1}{N_i}\right)^2. 
    \label{eq:cov-auto}
\end{equation}

In order to remove the shot noise, we estimate the angular power spectrum at
a fixed redshift bin using the jackknife resampling technique. We randomly split the objects in our redshift bin into two sub-samples with roughly equal number of sources, we build maps of the number counts for the two sub-samples separately, and we estimate the autocorrelation as the cross-correlation of the two maps, see e.g.\ \cite{Inman:2015pfa}.
Using this method, we largely reduce the impact of the shot noise because, even if the sub-samples have half as many galaxies, their cross-correlation is not affected by shot noise coming from self-pairs.

We estimate the error on the angular power spectrum from Eq.~\eqref{eq:cov}, that is
\begin{equation}
    \sigma^2_{\ell, i\,\mathrm{cross}} = \frac{2}{f_\mathrm{sky} \Delta \ell (2\ell + 1)} \left[\left(C^i_\ell + \frac{1}{N_i}\right)^2 + \frac{1}{N^2_i}\right].
    \label{eq:cov-cross}
\end{equation}

Comparing Eq.~\eqref{eq:cov-auto} and Eq.~\eqref{eq:cov-cross}, we see that
the jackknife estimator removes the shot noise from the mean, but it 
slightly increases the statistical fluctuations around the mean. 
In Appendix~\ref{a:jk}, we show the effectiveness of the jackknife method to remove the noise from our measurements. 

\section{The linear bias model in harmonic space} 
\label{s:lin-bias} 

In full generality, the relation between a biased field, such as the halo overdensity, and the underlying matter overdensity is an intricate function of the nonlinear density and tidal fields and their derivatives, see for example Ref.~\cite{Desjacques:2016bnm}. 
On large scales, where linear perturbation theory still remains valid, this relation simplifies to the linear bias model, i.e.
\begin{equation}
\delta_\mathrm{h}(z, \mathrm{x}) = b_1 (z) \delta (\tau(z), \mathrm{x}),
\end{equation}
where $\delta_\mathrm{h}$ is the local halo overdensity, $\delta$
is the density contrast and $b_1$ is the linear bias, which simply depends on redshift. Note that $\delta_\mathrm{h}$ and $\delta$ are defined in a matter gauge and that a gauge correction appears when one uses the density contrasts defined in a different gauge \cite{Fidler:2018geb}.

The linear bias model has been widely used in the literature, both in data analysis and forecasts for future cosmological surveys~\cite{DES:2021wwk, Euclid:2019clj, Euclid:2021qvm, Euclid:2021rez}.
Nevertheless, it is well known that on small scales this model is not accurate and, for this reason, a stringent scale cut must be applied in order to avoid systematic biases in the analysis.  
While the validity of the linear bias model has been tested in Fourier space for the galaxy power spectrum and in configuration space for the 2-point correlation function, an analysis based on the angular power spectrum is lacking, and our work aims to fill this gap.
With this purpose in mind, we extract the angular power spectra in tomographic redshift bins from \texttt{unity2-lowz} and \texttt{unity2-highz}. 
The angular power spectra are binned in band powers with $\Delta \ell = 5$ for
the full sky catalogue \texttt{unity2-lowz} and $\Delta \ell = 20 \approx f_\text{sky}^{-1}$ for \texttt{unity2-highz}. 
The angular power spectra are compared to their theoretical prediction, computed with the {\sc class} code which implements the number counts in linear perturbation theory including relativistic effects. 
In our theoretical model, we include nonlinearities in the matter density using the \texttt{HMCODE} recipe implemented in {\sc class} \cite{Mead:2016zqy}, with the model parameters fitted to the Cosmic Emulator dark matter only simulation~\cite{Heitmann:2013bra}.

In Sec.~\ref{sec:3.1} we estimate the bias by comparing the halo and the particle number counts, while in Sec.~\ref{sec:3.2} we extract the linear bias by fitting the halo angular power spectrum to our theoretical model, and we study at which scales the linear bias approximation breaks down.

\subsection{Linear bias: Fourier space versus harmonic space}
\label{sec:3.1}
    
As a first consistency check, we compare the linear halo bias
estimated from two different statistics: the power spectrum in Fourier space, and the angular power spectrum in harmonic space.
For both measurements, we apply the same mass selection, excluding from the halo catalogue all sources with masses below $M_\mathrm{min} = 5\times10^{12}\,M_\odot/h$. 

We estimate the linear bias in Fourier space from
the halo and the particle snapshots. 
The $k$-dependent bias in Fourier space is computed by taking the ratio of the cross-spectrum of halos and matter $P^\text{hm}$ with the matter power spectrum $P^\text{m}$, at a fixed redshift, that is
\begin{equation}
b(k, \bar{z}) = \frac{P^\text{hm}(k, \bar{z})}{P^\text{m}(k, \bar{z})}.
\label{eq:bias_fourier}
\end{equation}
The linear bias is extracted by fitting this function to
a constant in the range of scales between $k_\mathrm{min} = 0.02\,h/\rm{Mpc}$ and
$k_\mathrm{max} = 0.08\,h/\rm{Mpc}$. Note that $k_\mathrm{min}$ is some two orders of magnitude larger than the horizon scale, so that the gauge dependence of the power spectrum, which is relevant mainly on scales of the order of the horizon and beyond, has a negligible effect. More details on the estimation of the linear bias from the snapshots are given in Appendix~\ref{ap:Fourier-bias}.

        \begin{figure*}
        \centering
        \begin{subfigure}[b]{0.48\textwidth}
            \centering
            \includegraphics[width=\textwidth]{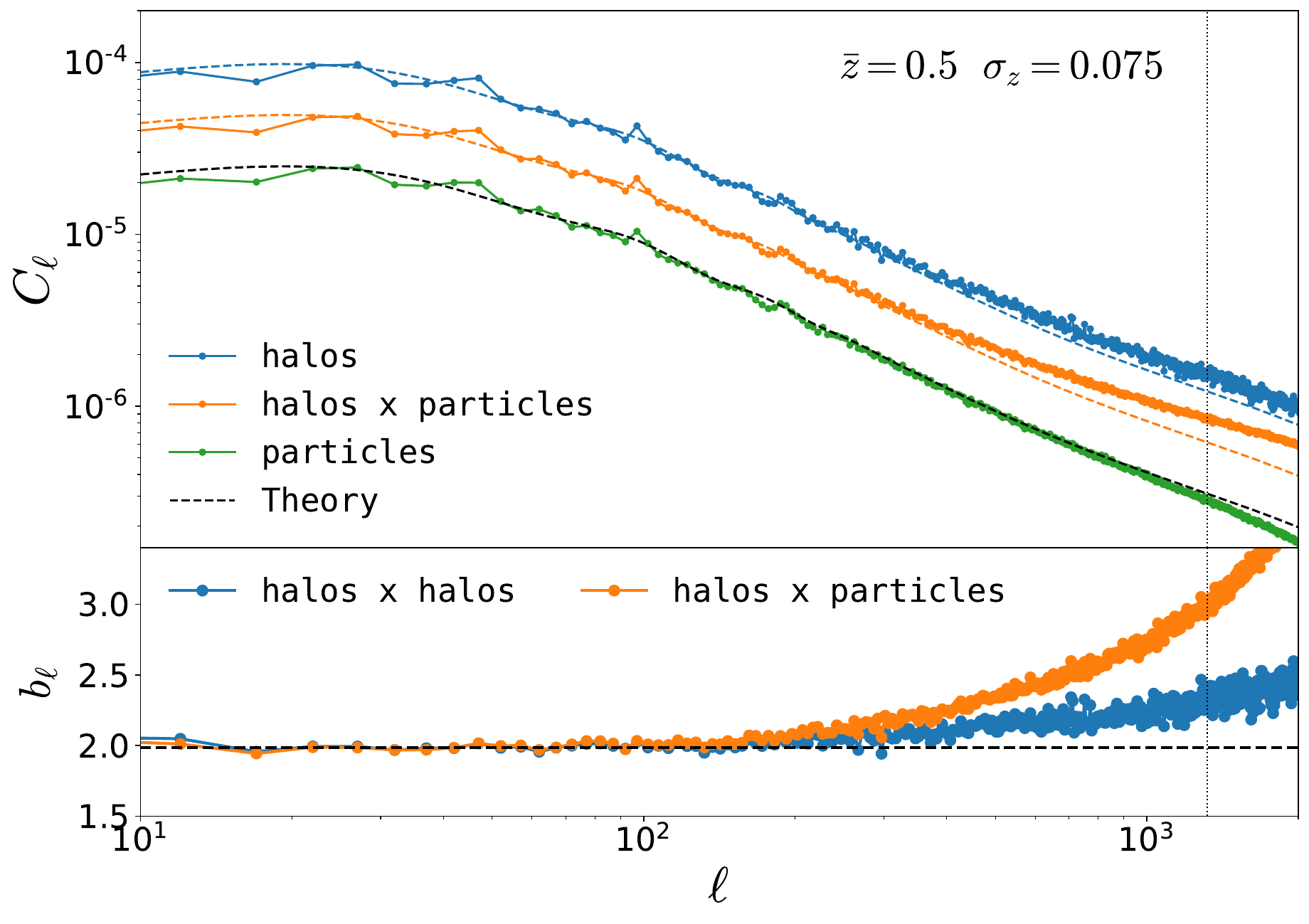}
            \caption[]%
            {{\small  \texttt{unity2-lowz}}}    
            \label{fig:lin-bias-lowz}
        \end{subfigure}
        \hfill
        \begin{subfigure}[b]{0.48\textwidth}  
            \centering 
            \includegraphics[width=\textwidth]{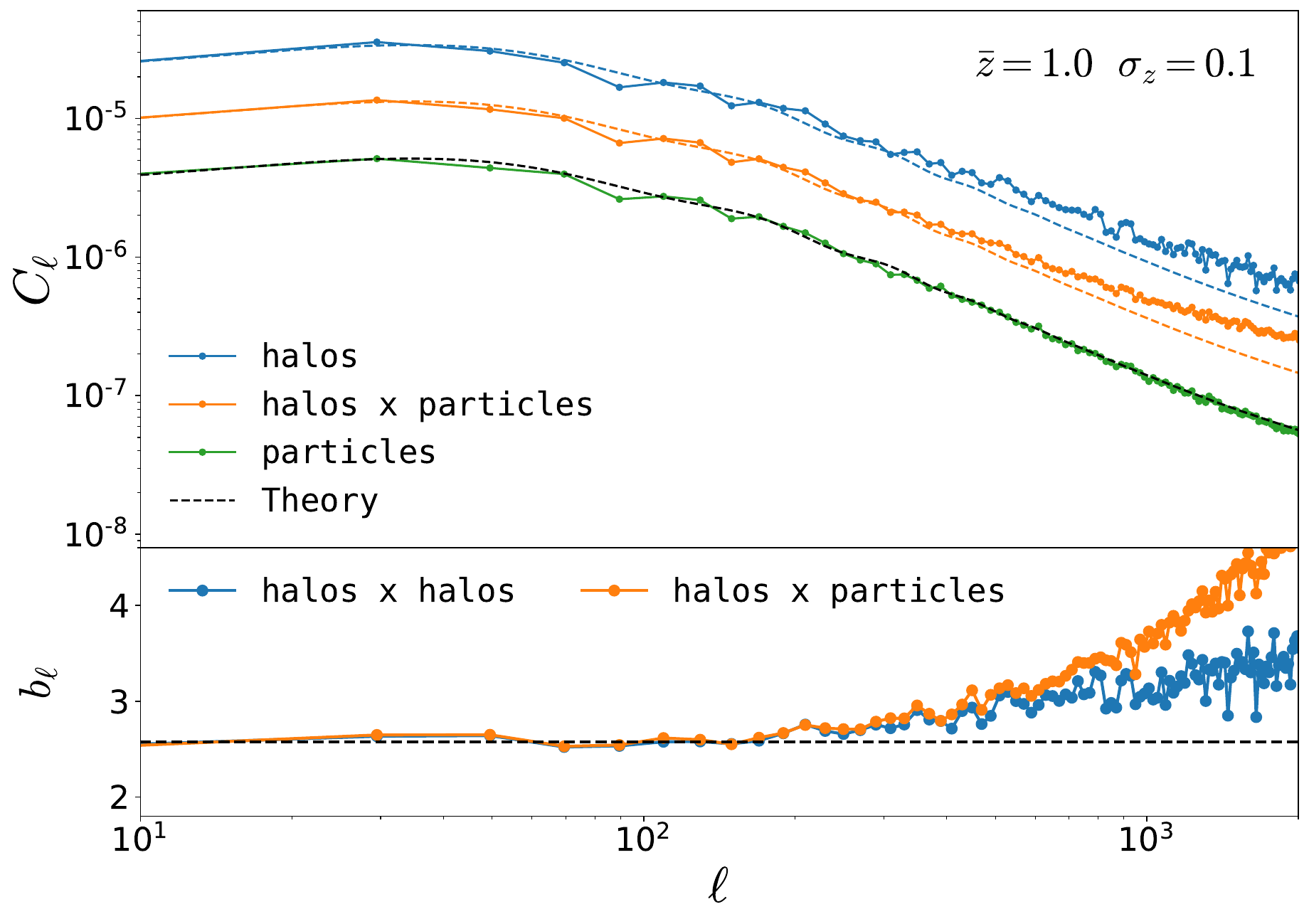}
            \caption[]%
            {{\small  \texttt{unity2-highz}}}    
            \label{fig:lin-bias-highz}
        \end{subfigure}
        \caption[ ]
        {\small Examples of angular power spectra estimated from (a) \texttt{unity2-lowz} and (b) \texttt{unity2-highz}. \underline{Top panels}: green, orange and blue data points are the measurements from the autocorrelation of the particle maps, the cross-correlation of halo and particle maps, and the autocorrelation of halo maps, respectively. The different amplitudes of three curves are due to the halo bias. Dashed lines represent the prediction from {\class}  using \texttt{HMCODE} for the angular power spectrum of matter (in black) and the biased angular power spectra, where the bias has been estimated through a fit as described in the text. \underline{Bottom panels}: Bias functions constructed from the autocorrelation of halo maps (blue) and the cross-correlation of halos and particles (orange). A horizontal dashed black line shows the average $ (b^\text{auto}_1 +  b^\text{cross}_1)/2$, where 
        $b^\text{auto}_1$ and $b^\text{cross}_1$ are the best-fit linear bias from auto and cross-correlation. The black vertical dotted line represents the scales where finite-grid resolution effects suppress the spectra at the level of $5\%$. As in Ref.~\cite{Lepori:2021lck}, this is estimated to be $\ell_5(\bar{z})\simeq \ell_{\text{Nyq}}(\bar{z})/\sqrt{20}$, where $\ell_{\text{Nyq}}$ is the multipole corresponding to the Nyquist frequency of the simulation.
        In the right panel, this line is not visible because at $\bar{z} = 1$ it falls beyond the range of multipoles shown in the plot. }
        \label{fig:lin-bias-1}
    \end{figure*}

In order to extract the linear bias from the statistics in harmonic space, we neglect here light-cone effects. We construct maps of halo number counts and particle number counts using the comoving positions of the sources and their background redshift. In a given redshift bin, we estimate the autocorrelation of the map of halo number counts, the autocorrelation of the map of particle number counts, and their cross-correlation. We construct the bias functions
\begin{equation}
b^\text{auto}_\ell(z) =  \sqrt{\frac{C^\text{hh}_\ell(z)}{C^\text{mm}_\ell(z)}}, \qquad b^\text{cross}_\ell(z) =  \frac{C^\text{hm}_\ell(z)}{C^\text{mm}_\ell(z)},
\end{equation}
where we remove the shot noise from the angular power spectra with the jackknife method described in Sec.~\ref{sec:method-spectra-cov} and Appendix~\ref{a:jk}. 

On large scales these $\ell$-dependent functions 
are well described  by a constant bias
that we estimate with a fit in the 
range of multipoles $\ell \in [20, \ell_\mathrm{max}]$, where $\ell_\mathrm{max} = 80, 130$ for  
\texttt{unity2-lowz} and \texttt{unity2-highz}, respectively. We apply a different cut at small scales for the two catalogues to ensure that enough data points are included in the fit for \texttt{unity2-highz}. By excluding low multipoles $\ell < 20$, which are also most affected by sample variance, any residual gauge dependence of the calculation should be very small.
We have verified that in the selected range of multipoles the linear bias model provides a good fit. 

\begin{figure}
\begin{center}
  \includegraphics[width=0.5\textwidth]{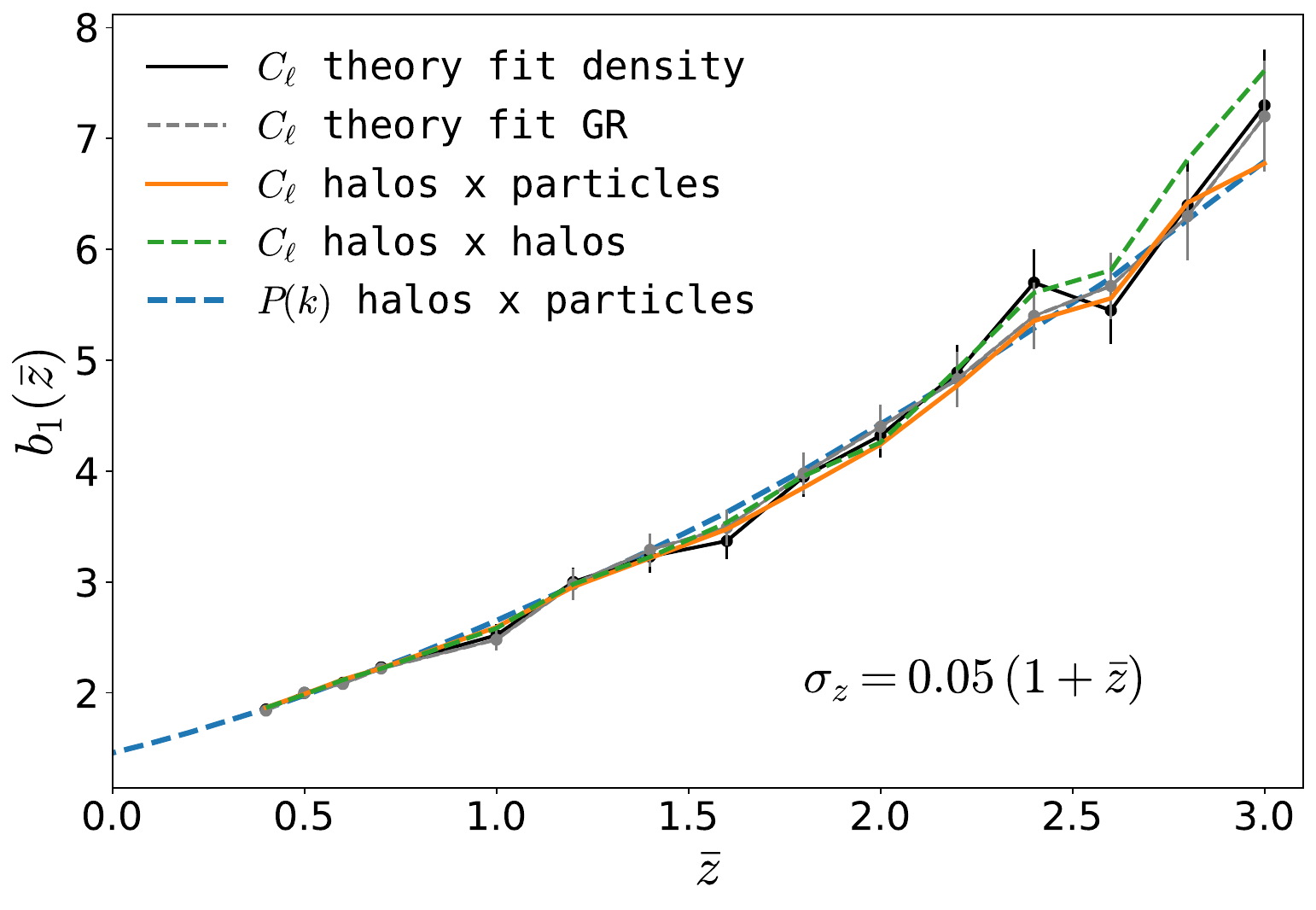}
  \end{center}
  \caption{Comparison of different measurements of the redshift-dependent linear bias,
  for halos with masses above $M_\mathrm{min} = 5\times10^{12}\,M_\odot/h$. $P(k)$ denotes measurements from power spectra on snapshots of a simulation in Fourier space (see Appendix \ref{ap:Fourier-bias}), whereas $C_\ell$ denotes measurements from angular power spectra. In both cases, we show a measurement using the cross-correlation ``halos $\times$ particles'' (dashed blue and solid orange, respectively), and for $C_\ell$ we additionally show a measurement using the auto-correlation ``halos $\times$ halos'' (green dashed). Instead of using the simulated number counts of particles as reference, we also constrain the bias using only the halo counts and a theoretical model that either accounts for local density fluctuations only, ``theory fit density'' (solid black), or includes the effects of redshift-space distortions and lensing on the light cone, ``theory fit GR'' (dashed grey), as explained in detail in Sec.~\ref{sec:3.2}. 
  }
  \label{fig:lin-bias-comp}
\end{figure}

While the Fourier space analysis measures 
the linear bias at a fixed conformal time (or background redshift), the angular statistics measures the correlation of the
overdensity fields projected onto
a radial redshift bin:
\begin{equation}
\tilde{\delta}_\text{h} (\bar{z}, \mathbf{\hat{n}}) = \int \frac{dN}{dz} W(z, \bar{z}) \delta_\text{h}(z,  \mathbf{\hat{n}}) dz,
\end{equation}
where $dN/dz$ is the redshift distribution of the sources, while $W$ encodes the redshift selection we apply to our catalogue. In our case, we apply a simple sharp cut in redshift (tophat), selecting all sources with 
background redshift in the range $z \in [\bar{z} - \sigma_z,\bar{z} + \sigma_z]$. 
The angular power spectra strongly depend on the bin size $\sigma_z$, since radial modes are washed out for wide
redshift bins. However, both the halo and the matter density fields are affected in the same way by the radial projection and, at least on linear scales, we have verified that the bias function depends only weakly on the bin size. 
Therefore, we expect the bias estimation from the harmonic and Fourier space spectra to be consistent with each others. 

In Fig.~\ref{fig:lin-bias-1} we show the angular power spectra for the halo number counts, the particle number counts, and their cross-correlation, at $\bar{z} = 0.5$ (Fig.~\ref{fig:lin-bias-lowz}) and $\bar{z} = 1$ (Fig.~\ref{fig:lin-bias-highz}).
The bottom panels show the functions $b^\text{auto}_\ell$ (blue) and $b^\text{cross}_\ell$ (orange), from which we extract the linear bias in harmonic space. For the scales below $\ell \sim 150$ the bias functions are roughly constant, and $b^\text{auto}_\ell$ overlaps with $b^\text{cross}_\ell$ as expected. 
Therefore, we estimate the best-fit linear bias $b^\text{auto/cross}_1$ from auto and cross-correlation by fitting  $b^\text{auto/cross}_\ell$ to a constant in the large-scale regime. 

On smaller scales, nonlinear biasing manifests itself with an excess of power in the simulation data, compared to the  prediction from the linear bias model. Furthermore, we see that the bias functions $b^\text{auto}_\ell$  and $b^\text{cross}_\ell$ increase due to higher-order corrections in the bias expansion that affect auto and cross-correlations in different ways.

In Fig.~\ref{fig:lin-bias-comp} we compare the linear bias estimated from the snapshots using the Fourier space statistics, $P(k)$, to the measurements in angular harmonic $C_\ell$, using either the autocorrelation ``halos $\times$ halos'' of halo maps or the cross-correlation ``halos $\times$ particles.''

The measurements in harmonic space from the halo autocorrelation and the cross-correlation ``halos $\times$ particles'' give consistent results up to $\bar{z}\sim 2$.
At higher redshift, the small discrepancy between the measurements 
increases to reach about $\sim 10\%$ at $\bar{z} = 3$. 

\begin{figure*}
        \centering
        \begin{subfigure}[b]{0.5\textwidth}
            \centering
            \includegraphics[width=\textwidth]{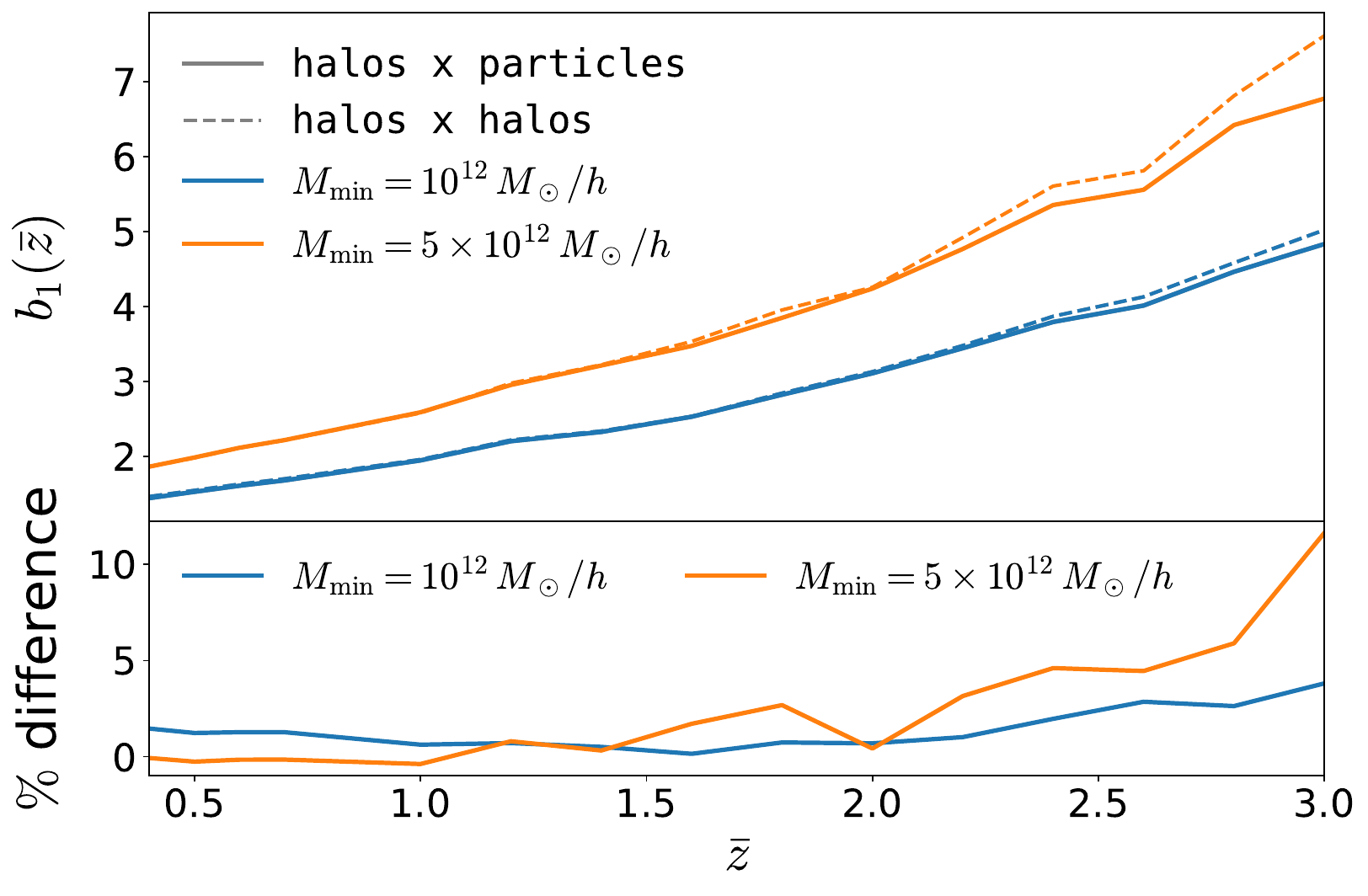}
        \end{subfigure}
        \hfill
        \begin{subfigure}[b]{0.48\textwidth}  
            \centering 
            \includegraphics[width=\textwidth]{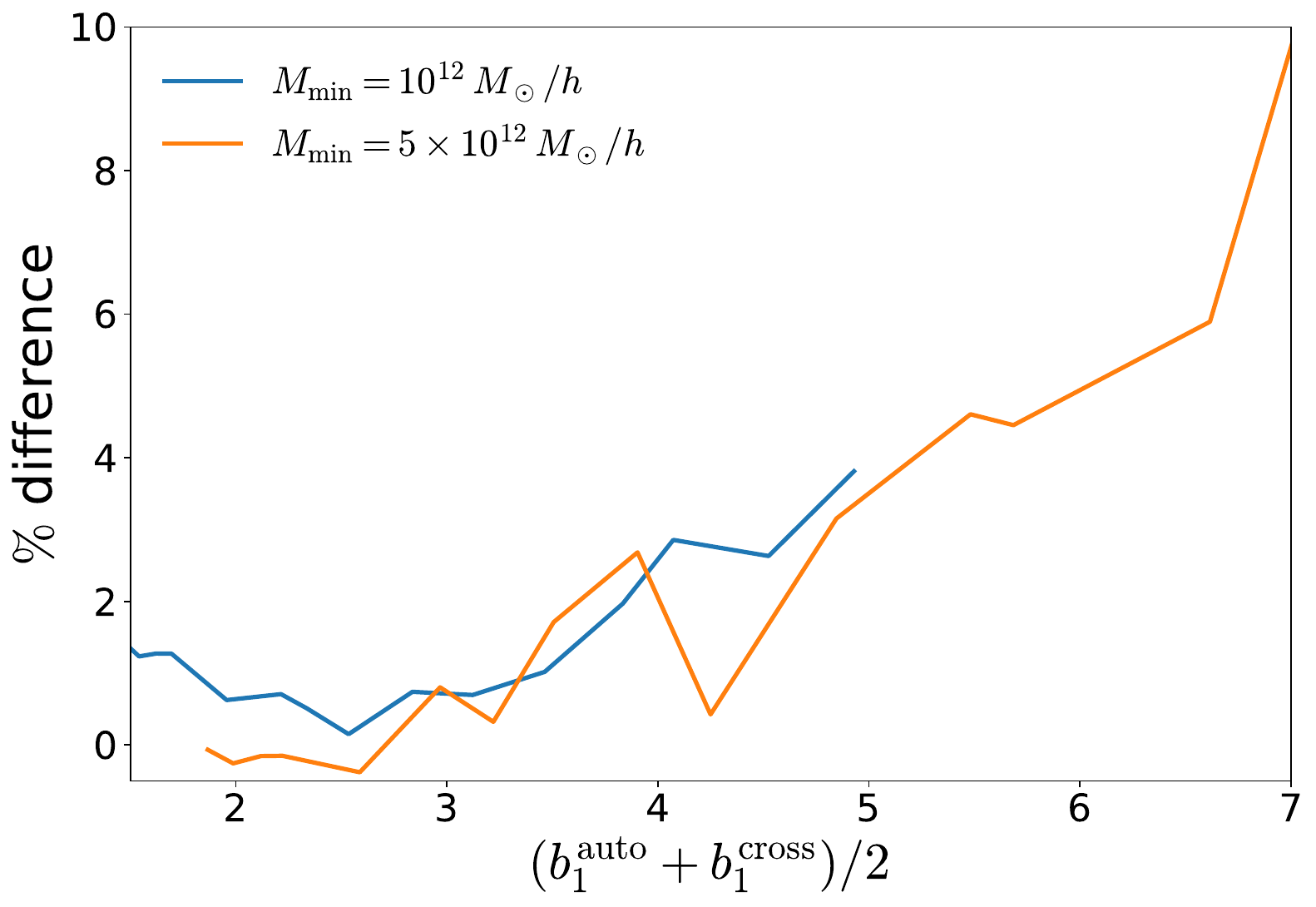}
        \end{subfigure}
        \caption[]
        {\small Comparison of the linear bias estimation for two halo populations, with $M_\mathrm{min} = 10^{12}\,M_\odot/h$ (blue) and $M_\mathrm{min} = 5\times10^{12}\,M_\odot/h$ (orange). \underline{Left}: In the top panel we show the redshift-dependent linear bias estimated from the autocorrelation of halo maps and cross-correlation of halos and particles for the two halo populations. The bottom panel displays the percentage difference in the best-fit estimation from auto and cross-correlation as a function of the mean redshift, where ``$\%$ difference'' $= 100 \times (b^\text{auto}_1 - b^\text{cross}_1)/[(b^\mathrm{auto}_1 + b^\mathrm{cross}_1)/2]$. \underline{Right}: We show the same percentage discrepancy as a function of the averaged linear bias $(b^\mathrm{auto}_1 + b^\mathrm{cross}_1)/2$. }
        \label{fig:lin-bias-2}
    \end{figure*}

This analysis refers to a halo population with masses larger than $5\times10^{12}\,M_\odot/h$. For comparison, we repeat the analysis for a population of smaller objects ($M_\mathrm{min} = 10^{12}\,M_\odot/h$).
The redshift-dependent linear bias from the angular statistics for the two halo populations is shown on the left plot in Fig.~\ref{fig:lin-bias-2}. 
As expected, halos with the smaller mass threshold have a smaller linear bias. For the population of smaller objects
the discrepancy between the best-fit bias from auto and cross-correlations, shown in the bottom panel of Fig.~\ref{fig:lin-bias-2}, is smaller compared to the more massive halos. On the right plot in Fig.~\ref{fig:lin-bias-2} we show this discrepancy (in percent) as a function of the average best-fit bias from auto and cross-correlation. We see that for both populations the difference increases with the values of the bias in a similar way. 
As theoretically, measuring the linear bias via the halo-particle correlation or as the square root of the halo-halo divided by the particle-particle correlation should be equivalent, the difference we find for highly biased halos highlights that
higher-order bias parameters are important even on large scales for these objects and thus contaminate the measurement of the linear bias. Our results suggest that the impact of these higher-order corrections grows roughly proportionally to the bias.
It is also not surprising that linear bias fares worse at higher redshift. While the fluctuations themselves are smaller and therefore more linear at higher redshifts, massive halos are rarer and therefore more strongly biased. This does not only increase the linear bias but also the contribution from higher-order bias terms,  for which the simple relation $P^\text{hm}/P^\text{m}=\sqrt{P^\text{hh}/P^\text{m}}$ no longer holds. 

\subsection{Testing the validity of the linear bias model for angular statistics}
\label{sec:3.2}
In the previous section, we estimate the linear bias by simply comparing the halo and the matter
statistics. While this procedure is straightforward when the data come from a numerical simulation and particles information is available, the unbiased matter density cannot be observed. Therefore, in real observations the statistics of the biased number counts is fitted to a theoretical prediction, where the bias can be measured as one of the free parameters of the theoretical model. 

In this section, we estimate the linear bias of our sample by fitting the simulation data to a one-parameter
model. We want to assess whether we can accurately 
recover the linear bias estimated from the particle distribution using our theoretical  prediction, whether the bias estimation is affected by light-cone effects such as redshift-space distortions (RSD), and up to which multipoles the linear bias prescription can be adopted without spoiling the quality of the fit. 

The theoretical model used in this section is based on the linear number counts implemented in the {\sc class} code~\cite{DiDio:2013bqa}, and the notation for the different contributions to the fluctuation of the counts follows Ref.~\cite{Lepori:2021lck}. We write the full halo number counts as
\begin{equation}
\Delta_\mathrm{h}= \delta_\text{h} + \delta_\text{rsd} + \delta_{\kappa} + \delta_\text{gr},
\label{eq:nc-full}
\end{equation}
where $\delta_\text{h}$ is the local overdensity of halos,
\begin{equation}
  \delta_\text{h} = b_1  \  \delta.
  \label{eq:dens}
\end{equation}
In the linear bias model it is simply proportional to the density contrast $\delta$ in comoving gauge. 
The term $\delta_\text{rsd}$ includes the linear Kaiser RSD~\cite{Kaiser:1987qv} and other subdominant Doppler corrections,
\begin{equation}
\delta_\text{rsd} = \frac{1}{\mathcal{H}(z)} \partial_r (\mathbf{v}\cdot\mathbf{n}) + \Biggl(\frac{\mathcal{H'}}{\mathcal{H}^2} + \frac{2}{r\mathcal{H}} - f_\mathrm{evo}\Biggr)(\mathbf{v}\cdot \mathbf{n})
 + \left(f_\mathrm{evo}- 3\right)\HH V , \label{RSD-term}
\end{equation}
where $\mathbf{v} = -{\bf \nabla} V$ is the peculiar velocity in Poisson gauge (i.e.\ $V$ is the velocity potential), $r = \tau_o -\tau$ is the conformal distance,
$\mathcal{H}=a'/a$ is the conformal Hubble rate and a prime denotes a derivative with respect to the conformal time $\tau$. The evolution bias $f_\mathrm{evo}$ describes the redshift evolution of the halo number density in a comoving volume. Since $f_\mathrm{evo}$ affects only subdominant contributions to the number counts, we set it to zero in the analysis. 
The lensing magnification term~\cite{Matsubara:2004fr, Scranton:2005ci, Liu:2021gbm} is denoted by $\delta_{\kappa}$,
\begin{equation}
\delta_{\kappa} = -\int_0^{r(z)} \frac{r(z) - r}{r(z) r} \Delta_{\Omega} (\phi + \psi) dr,
\label{eq:kappa}
\end{equation}
where $\Delta_{\Omega}$
is the Laplace operator on the sphere.
The last term in Eq.~\eqref{eq:nc-full}, $ \delta_{\rm gr},$ includes
local and integrated combinations of the Bardeen potentials. 
These terms introduce a very small correction to the observable on the scales considered here and therefore we neglect them in the theoretical model. 
While we build our  model from the linear computation of the number counts, nonlinearities
in the matter density contrast, $\de$, are included using the \texttt{HMCODE}~\cite{Mead:2016zqy}.

The linear bias $b_1$ is the only free parameter.
We estimate it using the following procedures:
\begin{enumerate}
 \item We analyse the maps of number counts built from the comoving position of the halos and we fit their spectra to the theoretical prediction with only the density term in the number counts, assuming that the linear bias is constant within the redshift bin. Therefore, our model is given by
 \begin{equation}
C^\mathrm{th}_\ell = b^2_1 C^\mathrm{\delta}_\ell,
\end{equation}
where $C^\mathrm{\delta}_\ell$ is the angular power spectrum sourced by only density perturbations
for an unbiased tracer.  
\item We use the fully relativistic maps, and we fit their spectra to the theoretical prediction which includes all relevant relativistic effects. The theoretical prediction is constructed computing the separate contributions of density $C^\mathrm{\delta}_\ell$, the summed contributions of RSD and lensing magnification $C^{(\mathrm{rsd},\,\delta_{\kappa} )}_\ell$, and their cross-correlation with density for an unbiased tracer $C^{\delta \times (\mathrm{rsd},\,\delta_{\kappa})}_\ell$.  From those separate terms, our model is given by
\begin{equation}
C^\mathrm{th}_\ell = b^2_1 C^\mathrm{\delta}_\ell + b_1 C^{\delta \times (\mathrm{rsd},\,\delta_{\kappa})}_\ell +  C^{(\mathrm{rsd},\,\delta_{\kappa} )}_\ell. \label{eq:model-rel}
\end{equation}
\end{enumerate}

We fit the model to the data using the Levenberg-Marquardt algorithm, as implemented in the python routine 
\texttt{curve\_fit}, that minimizes the quantity
\begin{equation}
\chi^2 = \sum^{\ell = \ell_\mathrm{max}}_{\ell = \ell_\mathrm{min}} \frac{(C^\mathrm{th}_\ell - C^\mathrm{sim}_\ell)^2}{\sigma^2_\ell},
\end{equation}
where $\sigma_\ell$ are the errors on our data points estimated as described in Sec.~\ref{sec:method-spectra-cov}.

We fit the simulated data for different values of $\ell_\mathrm{max}$ in the range 
$\ell_\mathrm{max} \in [50, 1000]$. For each of these fits, we quantify the quality of the fit from the best-fit Chi-Square $\chi^2_\mathrm{fit}$. We estimate the probability to find a $\chi^2 > \chi^2_\mathrm{fit}$, under the hypothesis that
our model is adequate:
\begin{equation}
P(\chi^2 > \chi^2_\mathrm{fit}) = \int_{\chi^2_\mathrm{fit}}^\infty \mathrm{PDF}_{\chi^2}(x, n_\mathrm{dof}) \,dx,
\end{equation}
where $\mathrm{PDF}_{\chi^2}(x, n_\mathrm{dof})$ is the $\chi^2$ probability distribution function for $n_\mathrm{dof}$
degrees of freedom, having $n_\mathrm{dof} = \ell_\mathrm{max} - \ell_\mathrm{min} +1 -  n_\mathrm{bias}) =  (\mbox{``number of points included in the fit''} - \mbox{``number of free parameters''}$). As there is only one free bias parameter to be fitted, we have here $n_\mathrm{bias} = 1$.
The best-fit linear bias is chosen to be the value that, in the
range of $\ell_\mathrm{max}$ considered, gives the largest $P(\chi^2 > \chi^2_\mathrm{fit})$. 

        \begin{figure*}[ht]
        \centering
        \begin{subfigure}[b]{0.48\textwidth}
            \centering
            \includegraphics[width=\textwidth]{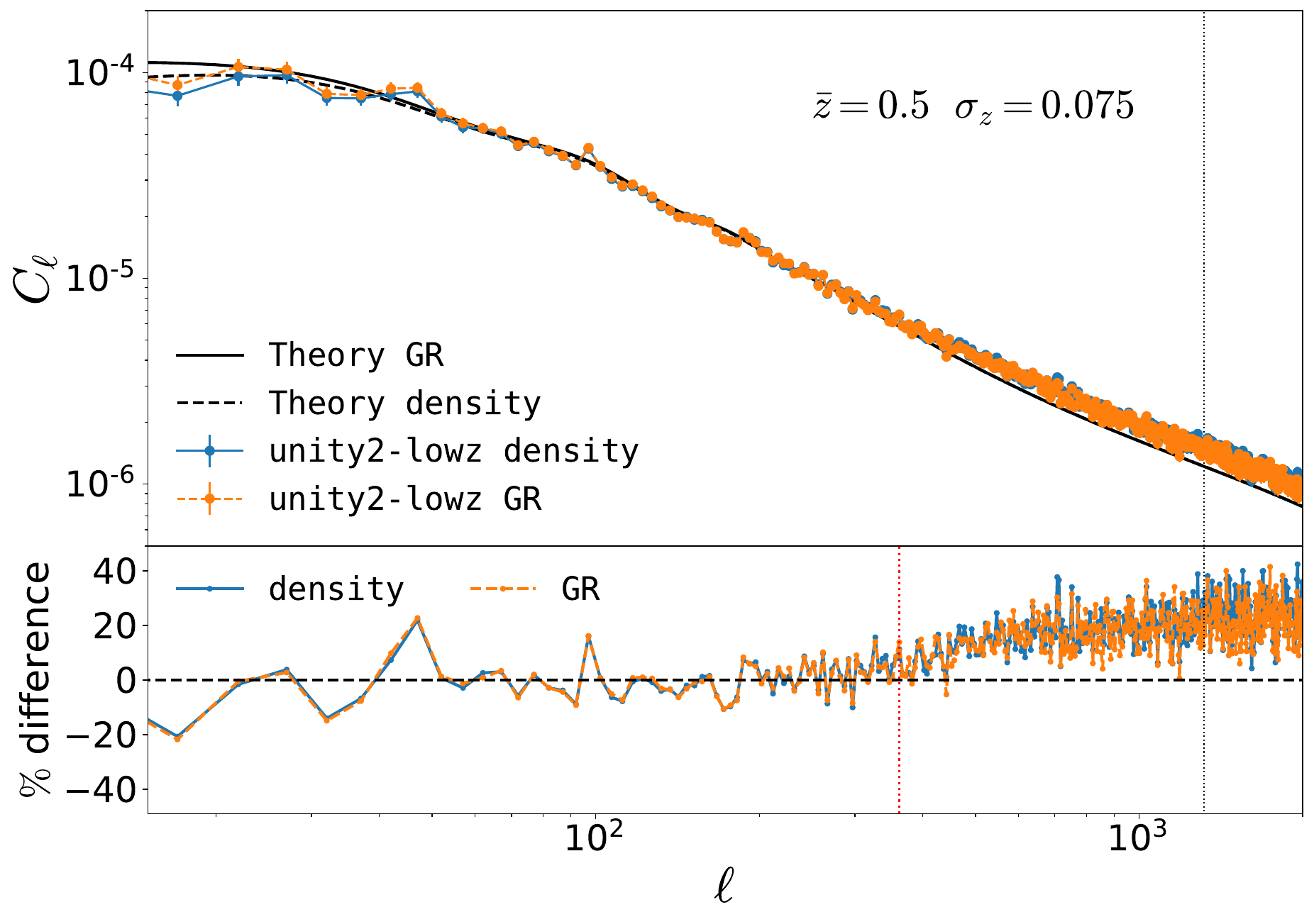}
        \end{subfigure}
        \hfill
        \begin{subfigure}[b]{0.48\textwidth}
            \centering 
            \includegraphics[width=\textwidth]{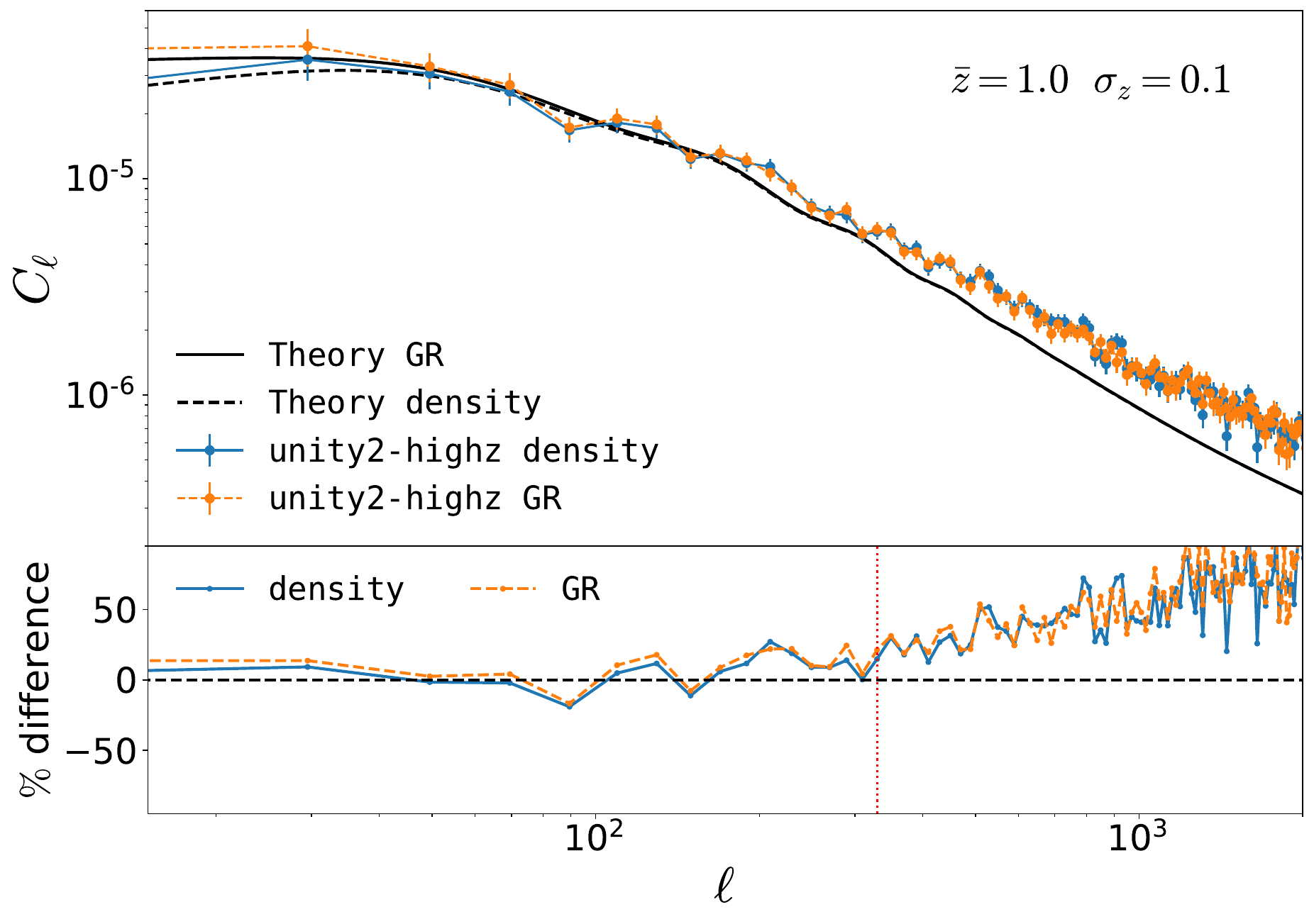}
        \end{subfigure} \\
        \begin{subfigure}[b]{0.48\textwidth}  
            \centering 
            \includegraphics[width=\textwidth]{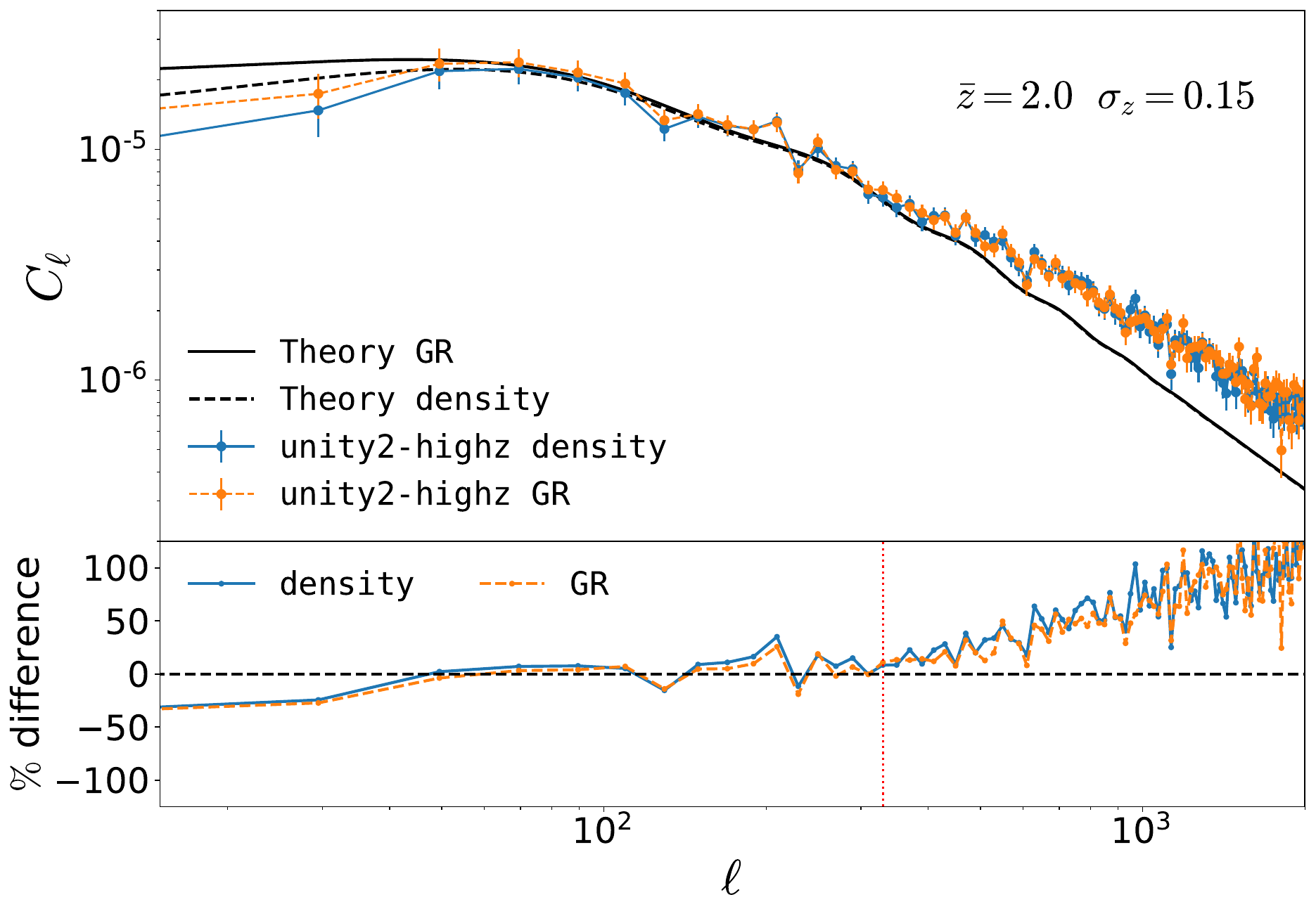}
        \end{subfigure}
        \hfill
        \begin{subfigure}[b]{0.48\textwidth}  
            \centering 
            \includegraphics[width=\textwidth]{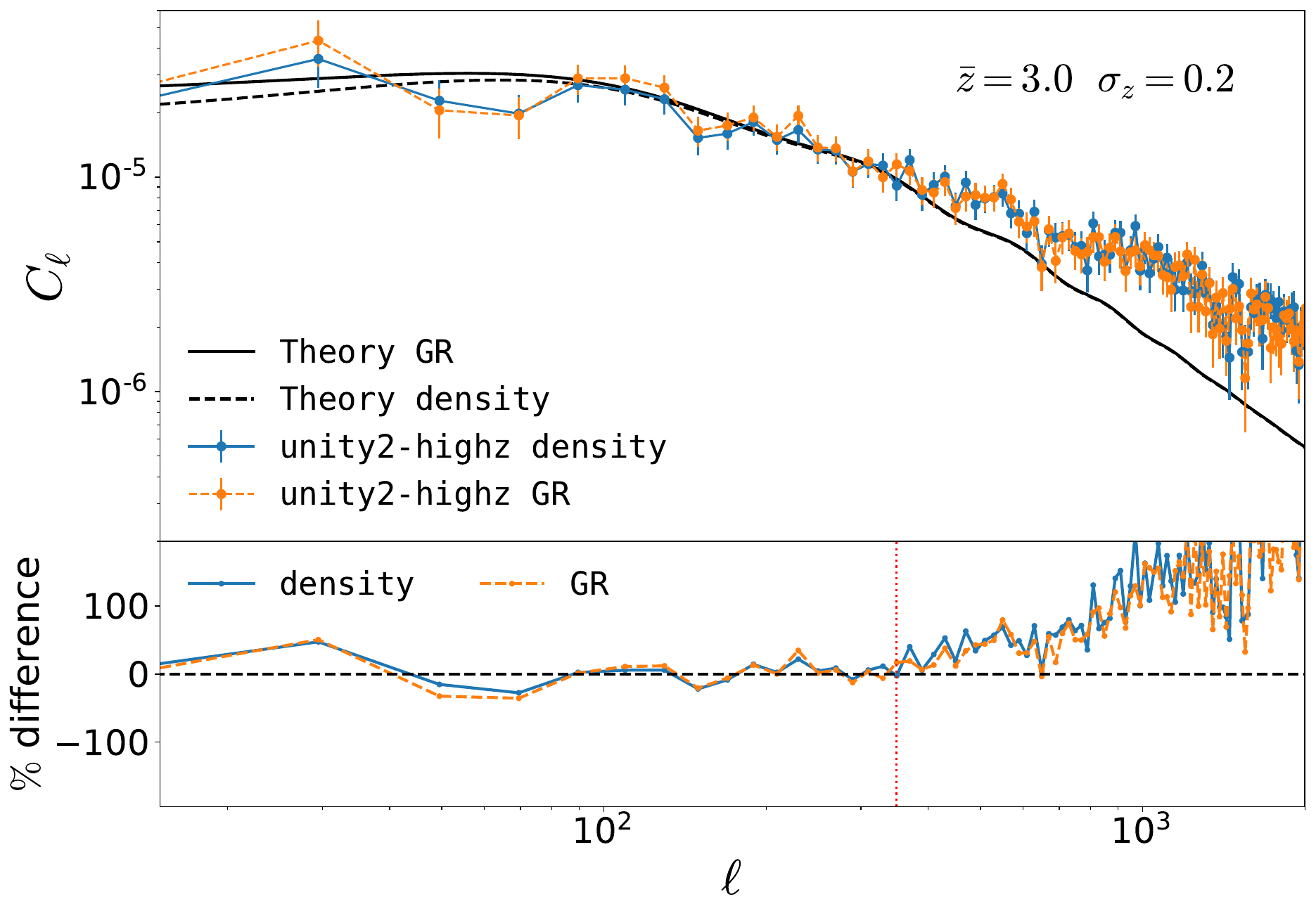}
        \end{subfigure}
        \caption[ ]
        {\small \underline{Top panels}: examples of halo angular power spectra at different redshifts, for a halo selection with $M_\mathrm{min} = 5\times 10^{12}\,M_\odot/h$. Blue data points represent maps constructed from the comoving positions of the halos and thus contain density perturbations only. Orange data points are the angular power spectra of the fully relativistic halo number counts. The best-fit theoretical models for the density and the relativistic spectra are displayed in black dashed and continuous lines, respectively. For the lowest redshift, $\bar{z} = 0.5$ (top left), a dotted vertical black line shows the multipole at which resolution effects become larger than $5\%$. The corresponding angular scale is beyond the range shown for the other three redshifts. \underline{Bottom panels}: relative difference between the best-fit theoretical model and the simulation data. A red vertical line shows the maximum multipole that is well modelled by a linear bias, using the $5\%$ deviation criterion described in the text. }
        \label{fig:lin-bias-fit-1}
    \end{figure*}

In Fig.~\ref{fig:lin-bias-fit-1} we show the angular 
power spectra extracted from our simulations for four representative redshifts. 
Simulation data are the blue (density maps) and orange (fully relativistic maps) points, while
the best-fit theoretical models are represented by dashed and continuous black lines, respectively. 
The best-fit values for the linear bias that we obtain for all the redshifts considered here are plotted in Fig.~\ref{fig:lin-bias-comp} as ``theory fit density'' (black line), and ``theory fit GR'' (grey dashed line). We find that the best-fit values obtained by fitting the halo angular power spectrum to the theoretical model are in good agreement with the values measured directly from the cross-correlation with the particle maps. On small scales, the linear bias prescription breaks down; actually, the deviations between
our simulation data and the best-fit theoretical model at a fixed scale $\ell \sim 1000$ increases with redshift (bottom panels of Fig.~\ref{fig:lin-bias-fit-1}).

In order to quantify the quality of our fit and to test up to which $\ell_\mathrm{nl}$  our model describes reasonably well the simulation data, we choose two criteria to assess at which multipole the linear bias model becomes unreliable:
\begin{enumerate}[label=(\alph*)]
\item $\chi^2$ criterion: the nonlinear cut $\ell_\mathrm{nl}$ is the minimum 
$\ell_\mathrm{max}$ at which $P(\chi^2 > \chi^2_\mathrm{fit})$ drops below the $5\%$ threshold. 
\item $5\%$ deviation criterion: $\ell_\mathrm{nl}$ 
is the minimum multipole 
such that the relative difference in percent, $100\% \times (C^\mathrm{sim}_\ell -C^\mathrm{th}_\ell)/C^\mathrm{th}_\ell$, becomes larger than $5\%$ for four and eight consecutive data points for \texttt{unity2-lowz} and \texttt{unity2-highz}, respectively. 
\end{enumerate}
The criterion a) takes into account the statistical uncertainty in the simulation data. On the other hand, the level of shot noise and its redshift dependence is a characteristic of the halo catalogue under consideration, and the cosmic variance error strongly depends on the survey geometry considered here. For these reasons, criterion b) gives a more general answer to the question of what are the smallest scales that can be modelled with the linear bias in the angular power spectrum.  

\begin{figure}
\begin{center}
  \includegraphics[width=0.55\textwidth]{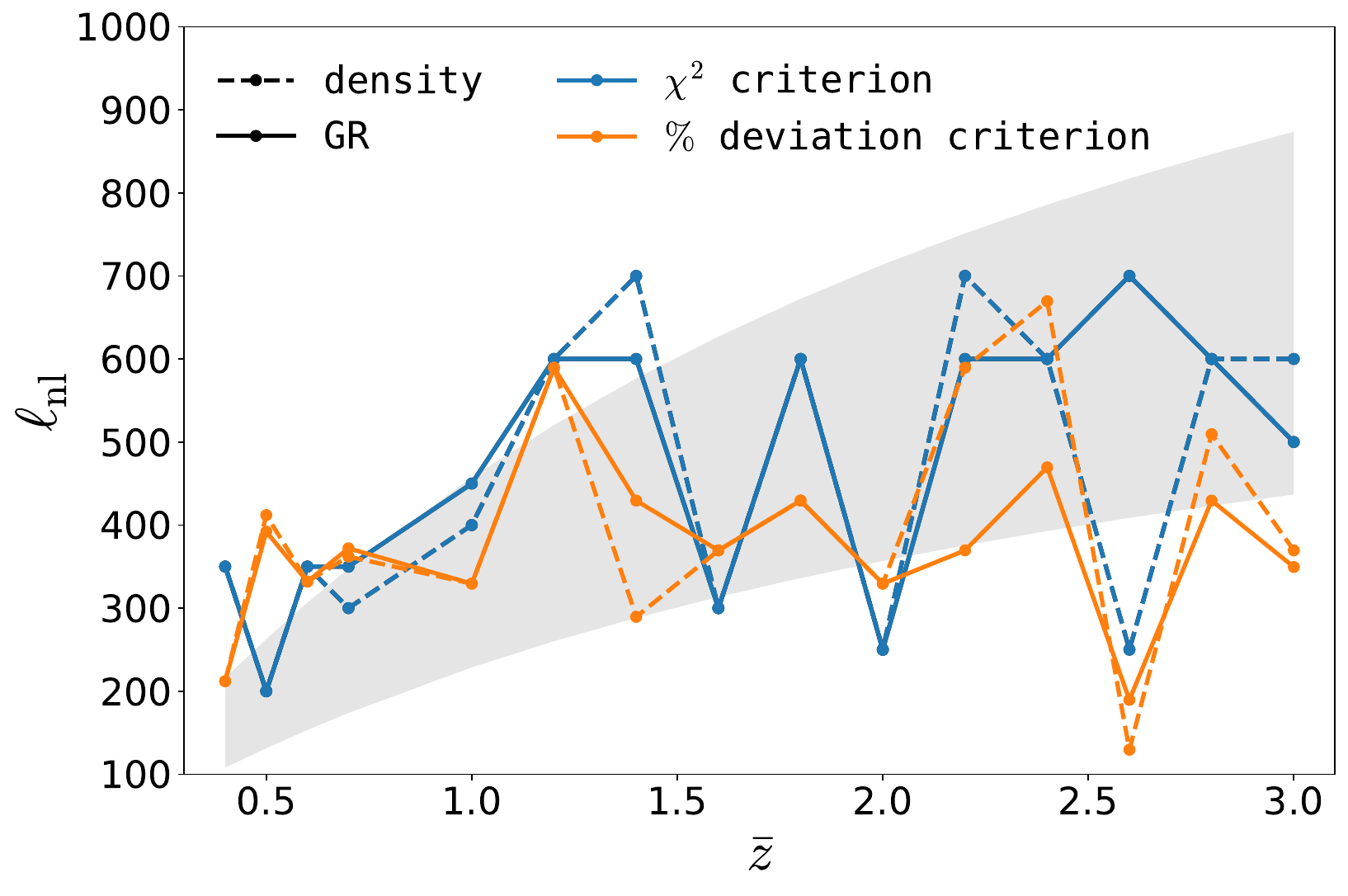}
  \end{center}
  \caption{Redshift dependence of the nonlinear angular scale $\ell_\mathrm{nl}$ from our simulation data.
  Blue points are estimated according to the $\chi^2$ criterion, while orange points are extracted applying the $5\%$ deviation criterion. These methods are described in the text. Continuous (dashed) lines mark the result obtained by fitting the fully relativistic (density) maps. A shaded region indicates the ansatz often used in the literature for the nonlinear scale $\ell_\mathrm{nl} = r(\bar{z}) k_\mathrm{max}$, with $0.1\,h/\mathrm{Mpc}  < k_\mathrm{max} < 0.2 \,h/\mathrm{Mpc}$. }
  \label{fig:lmax-fit}
\end{figure}

In Fig.~\ref{fig:lmax-fit} we show the nonlinear angular scale estimated from our simulation as a function of redshift, for the two criteria described above. 
The nonlinear angular scale estimated from the density and fully relativistic maps gives very similar results, while the $5\%$ deviation criterion gives a more conservative estimate of $\ell_\mathrm{nl}$ at large redshift. This is due to the large statistical uncertainty on the simulation measurements, the small fraction of sky available and the high level of shot noise at high redshift. 

We compare the nonlinear angular scale from our simulation to the ansatz often used in the literature $\ell_\mathrm{nl} = r(\bar{z}) k_\mathrm{max}$, see for example Refs.~\cite{Lorenz:2017iez, Tanidis:2021uxp}. Note that in our application $k_\mathrm{max}$ is the maximum wavenumber at which the linear bias model is valid in Fourier space, assuming that the nonlinearities of matter are well-modelled in the matter power spectrum by the \texttt{HMCODE}, while in Refs.~\cite{Lorenz:2017iez, Tanidis:2021uxp} $k_\mathrm{max}$ fixes the maximum wavenumber of validity for linear perturbation theory. 
In practice, for biased tracers, it is not realistic to separate the nonlinearities of matter from the effect of biasing. Nevertheless, this approximation
is often used in cosmological forecasts, see for example~\cite{Euclid:2019clj, Euclid:2021rez, Euclid:2021qvm}. Therefore, in this section, we assess up to which scales this approach is legitimate. 

In our simulation results we cannot identify a clear redshift dependence from both our criteria. In \texttt{unity2-lowz} ($\bar{z} < 0.85$) we see that
the linear bias prescription that we adopt here is not valid anymore for
$\ell > 200-400$, while the nonlinear angular scale in \texttt{unity2-highz} ($0.85 <  \bar{z} \le 3$) ranges roughly between $ 150 < \ell_\mathrm{nl} < 700$ with large fluctuations. 
The general trend of our numerical results tells us that at high redshift the linear bias description cannot be extended up to smaller scales, unlike the ansatz $\ell_\mathrm{nl} = r(\bar{z}) k_\mathrm{max}$, and that including scales above $\ell_\mathrm{nl} \sim 400$ in a cosmological analysis without a consistent treatment of nonlinear biasing is too optimistic. 
We understand this result as follows. Matter non-linearities and biasing have an opposite redshift dependence: nonlinear corrections to the clustering of matter grow in the late-time Universe, while biasing has a larger impact at early times. 
In the clustering statistics of a biased tracer these two effects are mixed, and they lead to the noisy redshift dependence of non-linearities displayed in Fig.~\ref{fig:lmax-fit}. At early times, the halo power spectrum is non-linear since halos are highly biased objects. At late times it is non-linear since matter fluctuations have entered the non-linear regime. This behaviour is certainly mass dependent and halos with much smaller masses will tend to be less biased, especially at higher redshift.

\section{Beyond linear bias}
\label{s:nl-bias}

In order to model the halo angular power spectrum accurately on scales $\ell > 400$, we need to go beyond the linear bias prescription.
A consistent formalism to describe nonlinear biasing has been developed in the past decade in the framework of perturbation theory, see for example Refs.~\cite{McDonald:2009dh, Assassi:2014fva, Senatore:2014eva, Mirbabayi:2014zca, Desjacques:2016bnm}.
The perturbative bias expansion has been tested with numerical simulations, see e.g.\ \cite{Nishimichi:2020tvu}, and recently applied to the analysis of cosmological surveys~\cite{Ivanov:2019pdj, DAmico:2019fhj, Zhang:2021yna}. Furthermore, there are now public codes that compute nonlinear power spectra of biased tracers at one-loop in cosmological perturbation theory, such as \textsc{class-pt}~\cite{Chudaykin:2020aoj}, based on the linear Boltzmann code \class, and \textsc{PyBird}~\cite{DAmico:2020kxu}.  
While the perturbative approach to biasing provides a robust framework to model galaxy clustering observables in the mildly nonlinear regime, in full generality it
also significantly increases the number of free parameters in the model. 
This is particularly problematic for a tomographic analysis where we need to marginalize 
over the extra bias parameters in each of the tomographic redshift bins. 
Therefore, it is crucial to find a balance between improving our modelling of biasing on small scales and limiting the number of nuisance parameters in order to preserve the information gain. For this reason, simulations have been used in the literature to investigate the relation between the linear bias and higher order bias parameters, see for example~\cite{Lazeyras:2015lgp,Barreira:2021ukk}. 

In this work, we test the accuracy of minimal extensions to the linear bias for the halo angular power spectrum. 
Following for example Ref.~\cite{Ivanov:2019pdj}, we consider a second-order bias expansion 
in the basis of bias operators $\{\delta, \delta^2, \mathcal{G}_2\}$, where $\delta$ 
is the matter density contrast, and $\mathcal{G}_2$ is the tidal field operator.
In comoving position space, i.e.\ neglecting light-cone effects, the perturbation of the halo number counts in this basis is 
\begin{equation}
\Delta_\mathrm{h} = b_1 \delta +\frac{b_2}{2} \delta^2 + b_{\mathrm{G}2} \mathcal{G}_2.
\label{eq:bias-exp}
\end{equation}
Given this model, we compute the theoretical prediction for the angular power spectrum in two steps.
First, we estimate the biased power spectrum in Fourier space $P_\mathrm{hh}(z, k, \mu)$ using the code \textsc{class-pt} (see Ref.~\cite{Chudaykin:2020aoj} for the details on the implementation). Then, we compute the projected statistics, that is the angular power spectrum, using 
the flat-sky approximation~\cite{Matthewson:2020rdt, Matthewson:2021rmb}.
For autocorrelations and tophat bins with half-bin width $\sigma_z$, we
need to compute\footnote{Note that here $\sigma_z$ is the half-bin width, while in~\cite{Matthewson:2021rmb}, $\sigma_z$ denotes the full width.}
\begin{equation}
C_\ell(\bar{z}, \sigma_z) = \frac{1}{2\pi r(\bar{z})} \int_{-\infty}^{\infty} \mathrm{d} k_\parallel 
\,\, P_\mathrm{hh}(\bar{z}, k, \mu)\,\,j^2_0 \left(\frac{k_\parallel \sigma_z}{H(\bar{z})}\right), \label{eq:flat-sky}
\end{equation}
where $j_0$ is the spherical Bessel function, $k = \sqrt{k^2_\parallel + \ell^2/r^2(\bar{z})}$ and $\mu = k_\parallel/k$. 

Our theoretical model, based on Eq.~\ref{eq:flat-sky}, neglects wide-angle effects 
and the non-uniform redshift distribution of sources inside the bin. 
In Appendix~\ref{ap:flat-sky} we test the impact of these two approximations on our study, and we find that up to $\bar{z} = 2$ our approximation agrees within $5\%$ with the full computation down to $\ell = 10$.
The largest discrepancies are found on large scales, and they are within the statistical uncertainties of our simulation measurements. 

The simulation data are fitted to these models with the Markov Chain Monte Carlo (MCMC)
method, making use of the Goodman \& Weare’s affine MCMC invariant ensemble sampler
implemented in the python package \textsc{emcee}~\cite{Foreman-Mackey:2012any}.
In Sec.~\ref{sec:pt-bias-density} we report the results for the density maps, 
and in Sec.~\ref{sec:pt-bias-rel} we discuss the analysis of the maps from  fully relativistic
number counts. 

\subsection{Results from density maps}
\label{sec:pt-bias-density}

We first consider the maps of number counts in comoving position space. 
Neglecting RSD and relativistic effects, the halo power spectrum in Fourier space is only a function of $\bar{z}$ and $k$. 
The full model for the Fourier space power spectrum, at fixed redshift, is then given by 
\begin{multline}
    P_\mathrm{hh}(\bar{z}, k) = b^2_1(\bar{z}) \Bigl[P_\mathrm{lin}(\bar{z}, k) + P_{1\text{-loop}, \text{SPT}} (\bar{z}, k) \Bigr] + b_1(\bar{z}) b_2(\bar{z}) \mathcal{I}_{\delta^2}
    (\bar{z}, k) 
    + 2 b_1(\bar{z})b_{\mathrm{G}2}(\bar{z}) \mathcal{I}_{\mathcal{G}_2} (\bar{z}, k)
     \label{eq:one-loop-real}
     \\
    + 2 b_1(\bar{z}) b_{\mathrm{G}2}(\bar{z}) \mathcal{F}_{\mathcal{G}_2} (\bar{z}, k)
    + \frac{1}{4} b^2_2(\bar{z}) \mathcal{I}_{\delta^2 \delta^2}  (\bar{z}, k) + b^2_{\mathrm{G}2} (\bar{z}) \mathcal{I}_{\mathcal{G}_2 \mathcal{G}_2} 
    (\bar{z}, k) + b_2(\bar{z}) b_{\mathrm{G}2} (\bar{z}) \mathcal{I}_{\delta^2 \mathcal{G}_2} 
    (\bar{z}, k)\,.
\end{multline}
We point the reader to Ref.~\cite{Chudaykin:2020aoj} for the exact form of the terms
$P_{1\text{-loop}, \text{SPT}}$, $\mathcal{I}_{\delta^2}$, $ \mathcal{I}_{\mathcal{G}_2}$, 
$\mathcal{F}_{\mathcal{G}_2}$,  $\mathcal{I}_{\delta^2 \delta^2}$, $\mathcal{I}_{\mathcal{G}_2 \mathcal{G}_2}$, $\mathcal{I}_{\delta^2 \mathcal{G}_2}$ appearing in Eq.~\eqref{eq:one-loop-real}.

We test four bias models against our simulation data:
a 1-parameter model $\{b_1\}$, that applies a linear bias on top of 
the linear power spectrum plus one-loop corrections from standard perturbation
theory (SPT) $P_{1\text{-loop}, \text{SPT}}$; a 2-parameter model $\{b_1, b_2\}$ which also includes as free parameter the quadratic bias, but neglects the tidal bias
$b_{\mathrm{G}2}$; a 2-parameter model $\{b_1, b_{\mathrm{G}2}\}$ that instead includes as
free parameter the tidal bias, but neglects the quadratic bias $b_2$;
and a 3-parameter model  $\{b_1, b_2, b_{\mathrm{G}2}\}$ that incorporates the full second-order
bias expansion in Eq.~\ref{eq:bias-exp}.
These models are obtained from Eq.~\eqref{eq:one-loop-real}, simply setting to zero the values of all parameters that we neglect in the corresponding bias expansion.
 
In our analysis, we fix the cosmological parameters to the input values in the simulation. This makes the fitting procedure very fast, as we can estimate the contributions to
the angular power spectrum individually and only once for each redshift bin, solving integrals of the form
\begin{equation}
C^\mathcal{\mathcal{T}_\text{X}}_\ell(\bar{z}, \sigma_z) = \frac{1}{2\pi r(\bar{z})} \int_{-\infty}^{\infty} \mathrm{d} k_\parallel \,\, \mathcal{T}_{X}(\bar{z}, k)\,\,j^2_0 \left(\frac{k_\parallel \sigma_z}{H(\bar{z})}\right), \label{eq:flat-sky-2}
\end{equation}
with $\mathcal{T}_\text{X} = \{P_\mathrm{lin} + P_{1\text{-loop}, \text{SPT}}, \mathcal{I}_{\delta^2}, \mathcal{I}_{\mathcal{G}_2}, \mathcal{F}_{\mathcal{G}_2}, \mathcal{I}_{\mathcal{G}_2 \mathcal{G}_2}, \mathcal{I}_{\delta^2 \mathcal{G}_2}\}$.
Once these terms are pre-computed, the model for the angular power spectrum and arbitrary 
values of the bias parameters can be quickly evaluated:
\begin{equation}
    C_\ell = b^2_1 C^{\text{lin} + 1\text{-loop}}_\ell + b_1 b_2 C^{\mathcal{I}_{\delta^2}}_\ell + 2 b_1 b_{\mathrm{G}2} \Bigl(C^{ \mathcal{I}_{\mathcal{G}_2}}_\ell + C^{\mathcal{F}_{\mathcal{G}_2}}_\ell \Bigr)
     + \frac{1}{4} b^2_2  C^{\mathcal{I}_{\delta^2 \delta^2}}_\ell
    + b^2_{\mathrm{G}2}   C^{\mathcal{I}_{\mathcal{G}_2 \mathcal{G}_2}}_\ell+ b_2 b_{\mathrm{G}2}   C^{  \mathcal{I}_{\delta^2 \mathcal{G}_2}}_\ell,
    \label{eq-pt-cell-model-real}
\end{equation}
where we omit the dependence on redshift of the bias parameters, and the $(\bar{z}, \sigma_z)$-dependence of the spectra for the sake of brevity.  We fit this model to our simulation data for all redshift bins considered in Sec.~\ref{s:lin-bias}, up to $\bar{z} = 2$. 
For the extended bias models, we include in the fit a range of multipoles between $\ell_\mathrm{min} = 20$ and $\ell_\mathrm{max} = 900$. Furthermore, we use flat uninformative priors on the bias parameters, with one exception that will be discussed later.

\begin{figure*}
        \centering
        \begin{subfigure}[b]{0.48\textwidth}
            \centering
            \includegraphics[width=\textwidth]{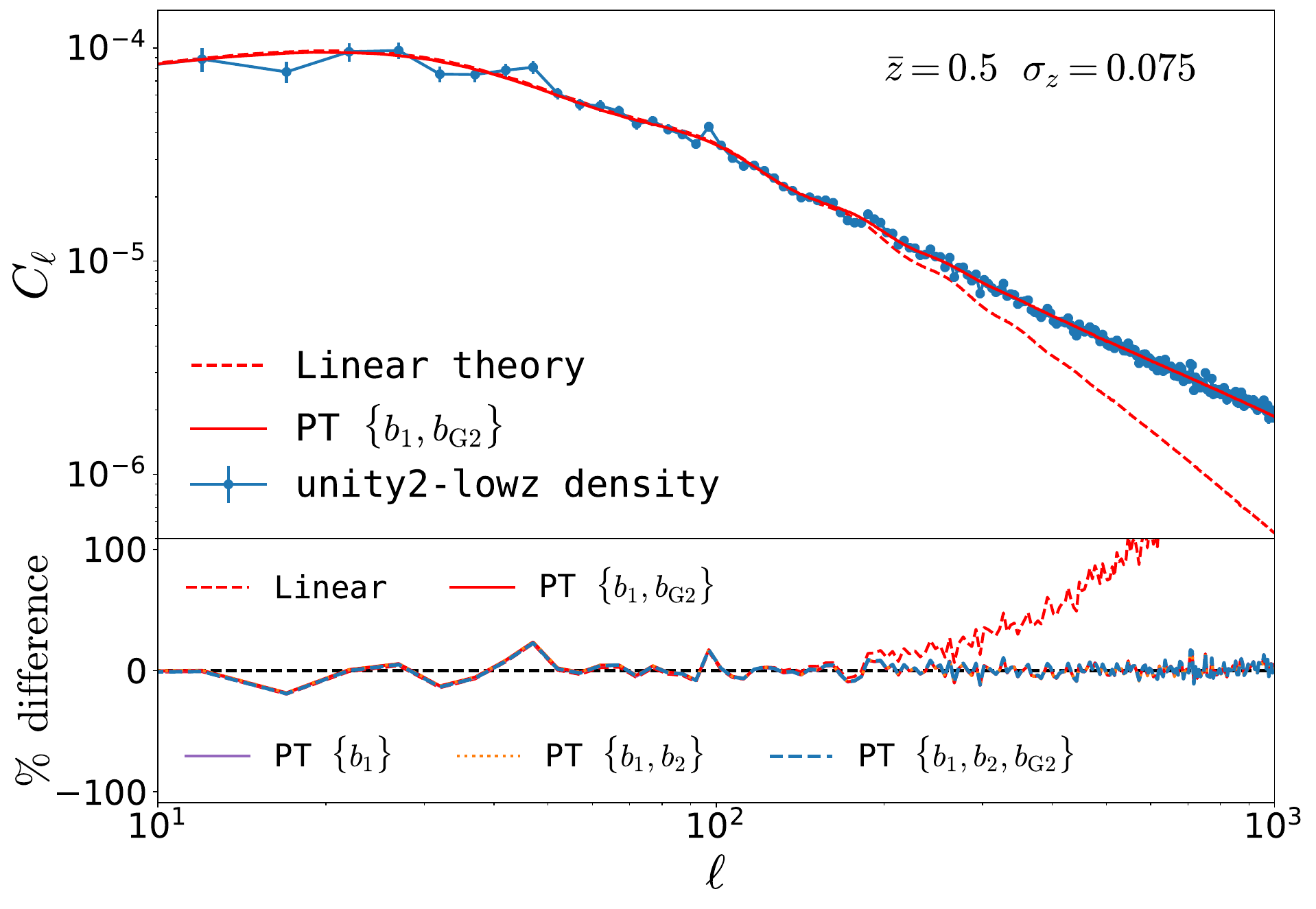}
        \end{subfigure}
        \hfill
        \begin{subfigure}[b]{0.48\textwidth}  
            \centering 
            \includegraphics[width=\textwidth]{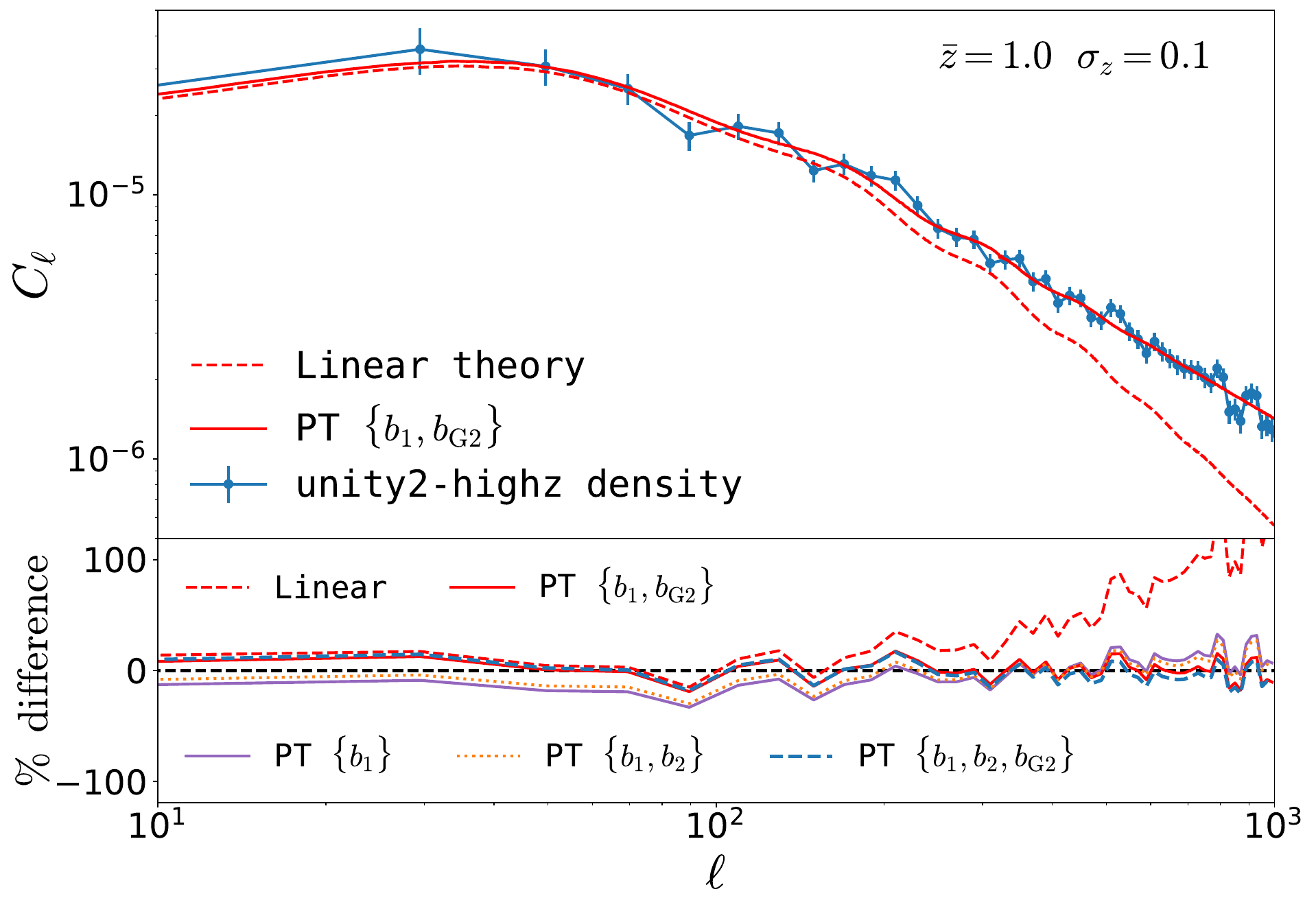}
        \end{subfigure} \\
        \begin{subfigure}[b]{0.48\textwidth}  
            \centering 
            \includegraphics[width=\textwidth]{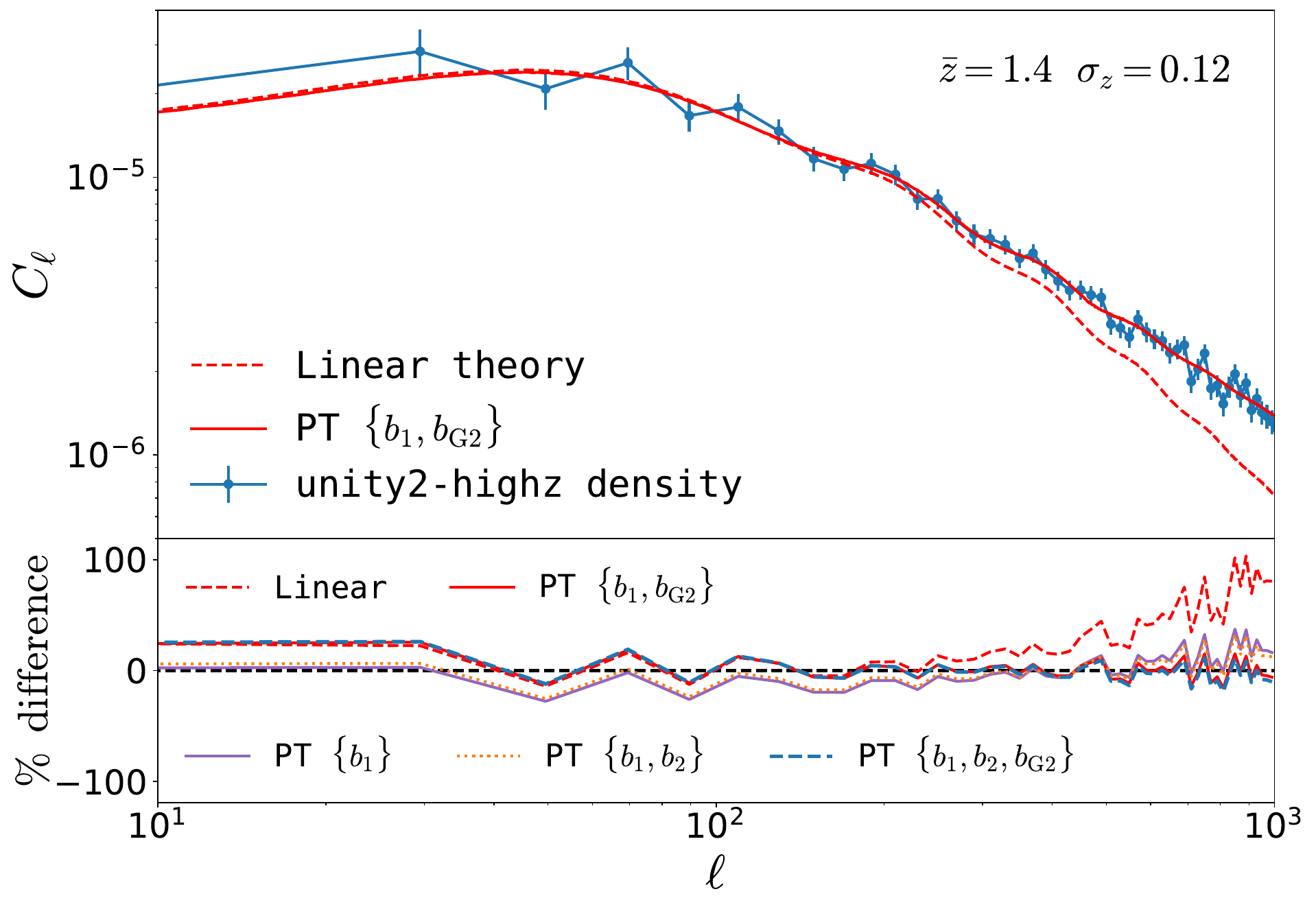}
        \end{subfigure}
        \hfill
        \begin{subfigure}[b]{0.48\textwidth}  
            \centering 
            \includegraphics[width=\textwidth]{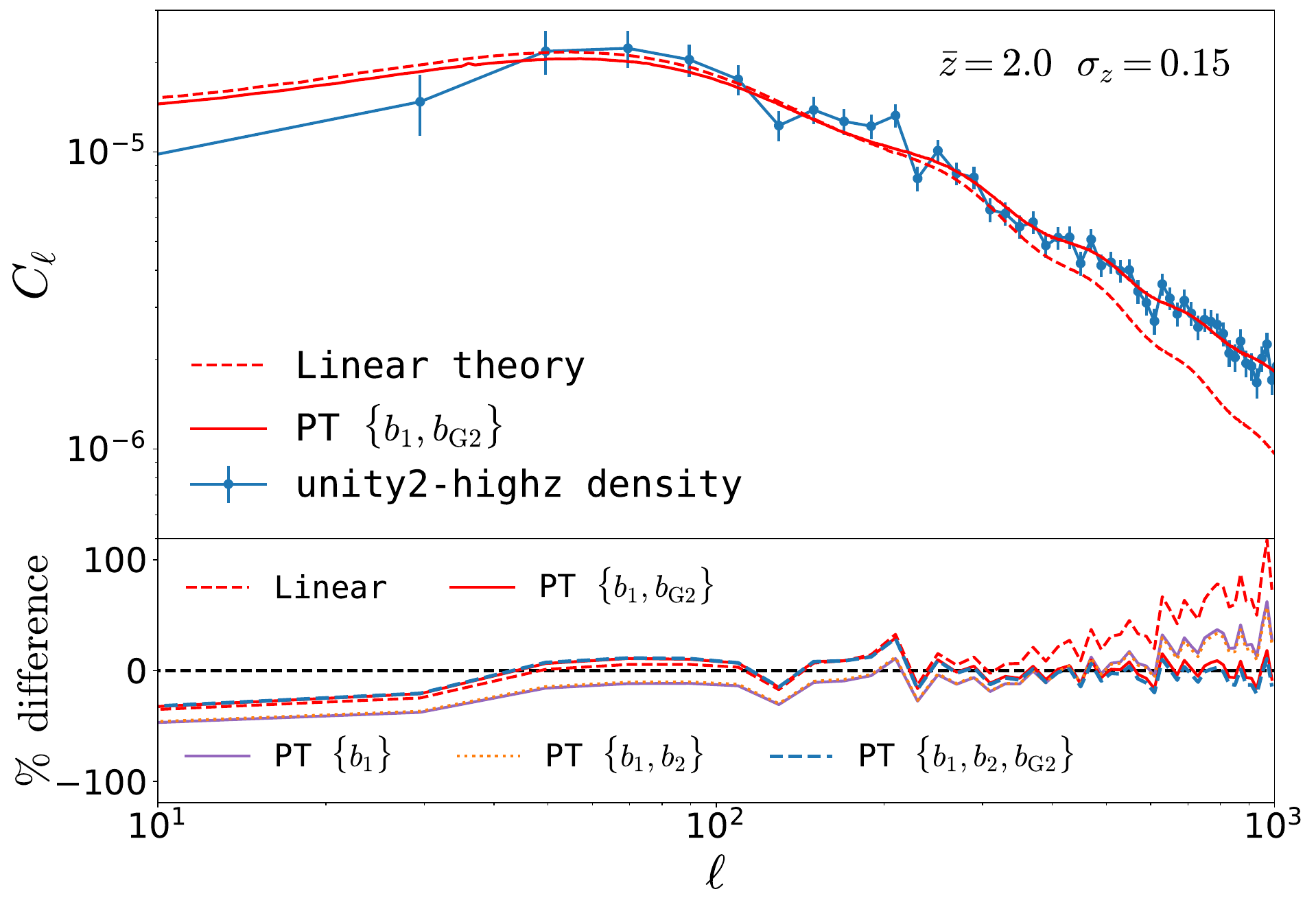}
        \end{subfigure}
        \caption[]
        {\small Outcome of our MCMC fit for the density angular power spectra, at four representative redshift. 
        \underline{Top panels:} Blue data points represent the angular power spectra extracted from the simulated density maps. Red dashed lines are the theoretical from linear theory, while continuous red lines represents the best-fit for the $\{b_1, b_{\mathrm{G}2}\}$ model. \underline{Bottom panels:} Percentage difference between the simulation data and the best-fit theoretical prediction, for linear theory and the four models described in Sec.~\ref{sec:pt-bias-density}. }
        \label{fig:lin-bias-pt-fit-1}
    \end{figure*}

\begin{table}[ht!]
\caption{Accuracy of the best-fit ($50$th percentile value) for the four bias models described in the text. We report the reduced $\chi^2$, that is $\chi^2/n_\mathrm{dof}$, and
the probability $P(\chi^2 > \chi^2_\mathrm{fit})$ to find a $\chi^2 > \chi^2_\mathrm{fit}$, under the null hypothesis that we are using the correct model.
Fits with $P(\chi^2 > \chi^2_\mathrm{fit}) < 5\%$ are considered not accurate, and highlighted in red. All the MCMC chains are run using uninformative flat priors, except for the model $\{b_1, b_2, b_{\mathrm{G}2}\}$ at $\bar{z} = 1$, where we force $b_2$ being positive (negative) and report the accuracy of the fit for the two cases. 
\label{tab:pt-bias}
}
\begin{center}
\resizebox{\textwidth}{!}{%
\begin{tabular}{ |c|c||c|c|c|c|c|c|c|c|}
 \hline
\multicolumn{2}{|c||}{}  & \multicolumn{2}{|c|}{$\{b_1\}$} & \multicolumn{2}{|c|}{$\{b_1, b_2\}$} & \multicolumn{2}{|c|}{$\{b_1, b_{\mathrm{G}2}\}$} & \multicolumn{2}{|c|}{$\{b_1, b_2, b_{\mathrm{G}2}\}$}  \\
 \hline
$\bar{z}$ &  & density & full GR &  density & full GR &  density & full GR &  density & full GR  \\
 \hline
\multirow{2}{*}{$0.4$} & $\chi^2/n_\mathrm{dof}$ & \cellcolor{green!10} $1.08$ & \cellcolor{green!10} $1.17$ & \cellcolor{green!10} $1.02$ & \cellcolor{green!10} $1.08$ & \cellcolor{green!10} $1.04$ & \cellcolor{green!10} $1.09$ & \cellcolor{green!5} $1.03$ & \cellcolor{green!5} $1.08$ \\
          & $P(\chi^2 > \chi^2_\mathrm{fit})$  & \cellcolor{green!10}$22\%$ & \cellcolor{green!10} $7\%$ & \cellcolor{green!10} $40\%$ & \cellcolor{green!10} $26\%$ & \cellcolor{green!10} $36\%$ & \cellcolor{green!10} $24\%$ & \cellcolor{green!5} $36\%$ & \cellcolor{green!5} $23\%$ 
          \\ 
 \hline
\multirow{2}{*}{$0.5$} & $\chi^2/n_\mathrm{dof}$ & \cellcolor{green!10} $1.07$ & \cellcolor{green!10} $1.24$ & \cellcolor{green!10} $1.10$ & \cellcolor{green!10} $1.22$ & \cellcolor{green!10} $1.09$ & \cellcolor{green!10} $1.22$ & \cellcolor{green!5} $1.11$ & \cellcolor{green!5} $1.23$\\
          & $P(\chi^2 > \chi^2_\mathrm{fit})$  & \cellcolor{green!10} $32\%$ & \cellcolor{green!10} $13\%$ & \cellcolor{green!10} $24\%$ & \cellcolor{green!10}  $16\%$ & \cellcolor{green!10} $26\%$ & \cellcolor{green!10} $16\%$  & \cellcolor{green!5} $24\%$ & \cellcolor{green!5} $15\%$\\ 
 \hline
 \multirow{2}{*}{$0.6$} & $\chi^2/n_\mathrm{dof}$ & \cellcolor{red!10} $1.48$ &  \cellcolor{red!10} $2$& \cellcolor{green!10} $1.24$ & \cellcolor{green!10} $1.24$ & \cellcolor{green!10} $1.16$  & \cellcolor{green!10} $1.3$& \cellcolor{green!5} $1.15$
 & \cellcolor{green!5} $1.3$\\
          & $P(\chi^2 > \chi^2_\mathrm{fit})$  &  \cellcolor{red!10} $\sim10^{-3}\%$ &  \cellcolor{red!10} $\sim10^{-3}\%$ & \cellcolor{green!10} $11\%$ & \cellcolor{green!10} $14\%$ & \cellcolor{green!10}$8\%$ & \cellcolor{green!10} $7\%$ & \cellcolor{green!5} $9\%$ & \cellcolor{green!5} $9\%$\\ 
 \hline
 \multirow{2}{*}{$0.7$} & $\chi^2/n_\mathrm{dof}$ &  \cellcolor{red!10} $2.02$ &  \cellcolor{red!10} $2.12$& \cellcolor{orange!10} $1.16$ & \cellcolor{orange!10} $1.4$&  \cellcolor{green!10} $1.08$ & \cellcolor{green!10} $1.3$ & \cellcolor{green!5} $0.95$ \cellcolor{green!5} & \cellcolor{green!5} $1.36$\\
          & $P(\chi^2 > \chi^2_\mathrm{fit})$  &  \cellcolor{red!10} $\sim 10^{-14}\,\,\%$ &  \cellcolor{red!10} $\sim 10^{-12}\,\,\%$ & \cellcolor{orange!10} $7\%$ & \cellcolor{orange!10} $0.6\%$ & \cellcolor{green!10} $22\%$ & \cellcolor{green!10} $2\%$  &  \cellcolor{green!5}$22\%$ & \cellcolor{green!5} $2\%$ \\ 
 \hline
 \multirow{2}{*}{$1$} & $\chi^2/n_\mathrm{dof}$ &  \cellcolor{red!10} $2.42$ &  \cellcolor{red!10} $1.92$ & \cellcolor{orange!10} $1.71$ & \cellcolor{orange!10} $1.32$ & \cellcolor{green!10} $0.99$ & \cellcolor{green!10} $0.79$& $1.28\,(1.15)$ \cellcolor{green!5} & \cellcolor{green!5} $0.97\,(0.87)$\\
          & $P(\chi^2 > \chi^2_\mathrm{fit})$  &  \cellcolor{red!10} $\sim10^{-4}\,\,\%$ &  \cellcolor{red!10} $\sim10^{-2}\,\,\%$ & \cellcolor{orange!10} $0.3\%$ & \cellcolor{orange!10} $8\%$ & \cellcolor{green!10} $50\%$ & \cellcolor{green!10} $83\%$& \cellcolor{green!5} $11\,(23)\%$ & \cellcolor{green!5} $52\,(70)\%$\\ 
 \hline
 \multirow{2}{*}{$1.2$} & $\chi^2/n_\mathrm{dof}$ &  \cellcolor{red!10} $1.84$ &  \cellcolor{red!10} $1.78$ & \cellcolor{orange!10} $1.43$ & \cellcolor{orange!10} $1.35$ & \cellcolor{green!10} $1.12$ & \cellcolor{green!10} $1.01$ & \cellcolor{green!5} $1.18$  & \cellcolor{green!5} $1.06$\\
          & $P(\chi^2 > \chi^2_\mathrm{fit})$  &  \cellcolor{red!10} $0.06\,\,\%$ &  \cellcolor{red!10} $0.13\,\,\%$ & \cellcolor{orange!10} $3\%$ & \cellcolor{orange!10} $7\%$ & \cellcolor{green!10} $27\%$ & \cellcolor{green!10} $46\%$  & \cellcolor{green!5} $20\%$ & \cellcolor{green!5} $37\%$\\ 
 \hline
 \multirow{2}{*}{$1.4$} & $\chi^2/n_\mathrm{dof}$ &  \cellcolor{red!10} $2.41$ &  \cellcolor{red!10} $2.32$& \cellcolor{red!10} $1.86$ & \cellcolor{red!10} $1.79$ & \cellcolor{green!10} $0.86$ & \cellcolor{green!10} $1.00$ & \cellcolor{green!5} $0.95$  & \cellcolor{green!5} $1.15$\\
          & $P(\chi^2 > \chi^2_\mathrm{fit})$  &  \cellcolor{red!10} $\sim10^{-4}\,\,\%$ &  \cellcolor{red!10} $\sim10^{-4}\,\,\%$& \cellcolor{red!10} $0.06\%$ & \cellcolor{red!10} $0.13\%$ & \cellcolor{green!10} $73\%$ & \cellcolor{green!10} $48\%$ & \cellcolor{green!5} $57\%$  & \cellcolor{green!5} $23\%$\\ 
 \hline
 \multirow{2}{*}{$1.6$} & $\chi^2/n_\mathrm{dof}$ &  \cellcolor{red!10} $2.46$ &  \cellcolor{red!10} $2.7$ & \cellcolor{red!10} $2.07$ & \cellcolor{red!10} $2.24$ & \cellcolor{green!10} $1.22$ & \cellcolor{green!10} $1.24$ & \cellcolor{green!5} $1.40$  & \cellcolor{green!5} $1.43$\\
          & $P(\chi^2 > \chi^2_\mathrm{fit})$  &  \cellcolor{red!10} $\sim10^{-5}\,\,\%$ &  \cellcolor{red!10} $\sim10^{-6}\,\,\%$ & \cellcolor{red!10} $0.006\%$ & \cellcolor{red!10} $\sim10^{-3}\,\,\%$ & \cellcolor{green!10} $15\%$ & \cellcolor{green!10} $13\%$ & \cellcolor{green!5} $4.5\%$ & \cellcolor{green!5} $4\%$ \\ 
 \hline
 \multirow{2}{*}{$1.8$} & $\chi^2/n_\mathrm{dof}$ &  \cellcolor{red!10} $3.5$ &  \cellcolor{red!10} $3.24$ & \cellcolor{red!10} $3.07$ & \cellcolor{red!10} $2.77$ & \cellcolor{green!10} $0.96$ & \cellcolor{green!10} $0.94$ & \cellcolor{green!5} $1.09$  & \cellcolor{green!5} $1.08$ \\
          & $P(\chi^2 > \chi^2_\mathrm{fit})$  &  \cellcolor{red!10} $\sim10^{-12}\,\,\%$ &  \cellcolor{red!10} $\sim10^{-10}\,\,\%$ & \cellcolor{red!10} $\sim10^{-8}\,\,\%$ & \cellcolor{red!10} $\sim10^{-6}\,\,\%$ & \cellcolor{green!10} $54\%$ & \cellcolor{green!10} $58\%$  & \cellcolor{green!5} $32\%$  & \cellcolor{green!5} $33\%$\\ 
 \hline
 \multirow{2}{*}{$2$} & $\chi^2/n_\mathrm{dof}$ &  \cellcolor{red!10} $3.36$ &  \cellcolor{red!10} $3.11$ & \cellcolor{red!10} $3.0$ &\cellcolor{red!10}  $2.76$ & \cellcolor{green!10} $0.97$ & \cellcolor{green!10} $0.94$ & \cellcolor{green!5} $1.19$  & \cellcolor{green!5} $1.12$\\
          & $P(\chi^2 > \chi^2_\mathrm{fit})$  &  \cellcolor{red!10} $\sim10^{-10}\,\,\%$ &  \cellcolor{red!10} $\sim10^{-9}\,\,\%$ & \cellcolor{red!10} $\sim10^{-8}\,\,\%$ & \cellcolor{red!10} $\sim10^{-6}\,\,\%$ & \cellcolor{green!10} $53\%$ & \cellcolor{green!10} $59\%$ & \cellcolor{green!5} $19\%$  & \cellcolor{green!5} $27\%$\\ 
 \hline
\end{tabular} 
}
\end{center}
\label{tab:gc-const-dens}
\end{table}

In Fig.~\ref{fig:lin-bias-pt-fit-1} we show the outcome of the MCMC fit at four
representative redshifts, while in Table~\ref{tab:pt-bias} (columns denoted by ``density'') we quantitatively compare the quality of the fit for the four bias models.
As estimators for the quality of the fit, we consider the reduced $\chi^2$ ($\chi^2/n_\text{dof}$) and the probability $P(\chi^2 > \chi^2_\mathrm{fit})$ to find a $\chi^2 > \chi^2_\mathrm{fit}$, under the null hypothesis that we are using the correct model. 
We consider the best-fit value of the parameters to be the value at the $50$th percentile.

At low redshift, that is $\bar{z} = 0.4, 0.5$, the best-fit of the four models gives a very similar prediction for the angular power spectrum, and they all reproduce accurately the simulation data in the entire multipole range used for the fit. 
Surprisingly, at low redshift, even the minimal model $\{b_1\}$ provides a reasonably good fit to our data. 
Therefore, the one-loop corrections together with a linear bias provide a better model to the halo angular power spectrum than the prescription used in the previous section, that is a linear bias applied to the \texttt{HMCODE} matter power spectrum. 
The matter power spectrum used in the $\{b_1\}$ model does not include what is called the \lq counterterm\rq in the literature~\cite{Chudaykin:2020aoj}, needed to model accurately the clustering of matter in perturbation theory. Therefore, 
the fact that the model with linear bias and one-loop corrections in standard perturbation theory works fairly well in the two lowest redshift bins could be simply a coincidence. At these redshifts the nonlinear bias has a similar amplitude but opposite sign as the contribution from the counterterm, and the two contributions roughly cancel. 
This cancellation does not occur anymore at higher redshift, starting from $\bar{z} = 0.6$, the quality of the fit for the 1-parameter model
$\{b_1\}$ quickly degrades. At higher redshift, this model is not able to reproduce the shape of the power spectrum on a wide range of scales. In order to accommodate the 
data on small scales it loses accuracy in the large-scale part of the spectrum, as shown in the bottom panels of Fig.~\ref{fig:lin-bias-pt-fit-1} for $\bar{z} = 1, 1.4, 2$. 
A similar trend emerges for the model $\{b_1, b_2\}$ starting from $\bar{z} \sim 0.7$. 
Overall, we find that both the 2-parameter model $\{b_1, b_{\mathrm{G}2}\}$ and 
the 3-parameter model  $\{b_1, b_2, b_{\mathrm{G}2}\}$ can describe the simulation
data well in the whole range of redshifts and angular scales considered here, the
former having the clear advantage of requiring only one nonlinear bias parameter.

As mentioned previously, we use uninformative flat priors to run the MCMC chains. The only exception is given by the model $\{b_1, b_2, b_{\mathrm{G}2}\}$ at $\bar{z} = 1$, where we 
find that, for uninformative priors, the posterior distribution of $b_2$ has two local 
maxima, one at $b_2 < 0$ and one at $b_2 > 0$. The resulting $50$th percentile value for $b_2$ does not provide a good fit to the data. Therefore, in the results we present here for this model and redshift bin we impose a flat prior $b_2 > 0$, which instead gives a good fit to the data. 
In Table~\ref{tab:pt-bias} we report the information on the quality of the fit for both 
prior choices,  $b_2 > 0$ and  $b_2 < 0$. 

\begin{figure}
\begin{center}
  \includegraphics[width=0.8\textwidth]{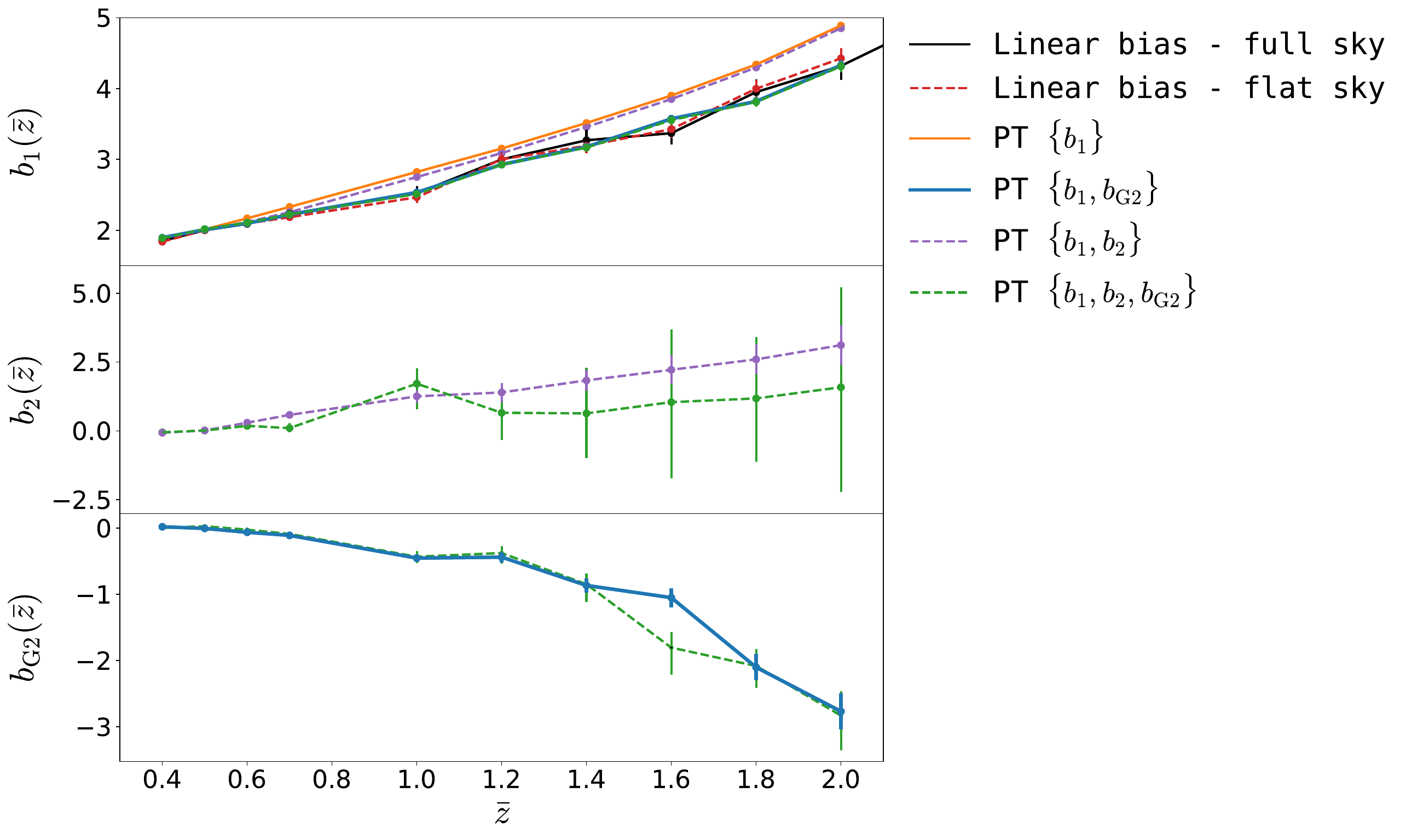}
  \end{center}
  \caption{Best-fit values of the bias parameters, obtained from the maps of the number counts in comoving space, for the models considered in Sec.~\ref{s:nl-bias}. The error bars are based on 
  the $16$th (lower bounds) and $84$th (upper bounds) percentiles of the samples in the marginalized distributions. The black line and dashed red line respectively refer to the best-fit values of the linear bias from the full sky and flat sky model when only the large scales are included in the fit.}
  \label{fig:pt-biases-z}
\end{figure}

In Fig.~\ref{fig:pt-biases-z} we show the redshift dependence of the best-fit values for $b_1$ (top panel), $b_2$ (central panel) and $b_{\mathrm{G}2}$ (bottom panel). For $b_1$, we compare the values obtained from models that use a perturbative bias expansion and the results of a linear fit, which only includes data on large scales.   
We see that the perturbative models providing a good fit on all scales,
that is $\{b_1, b_{\mathrm{G}2}\}$ and $\{b_1, b_2, b_{\mathrm{G}2}\}$, find 
best-fit values of $b_1$ fully consistent with the measurements on large scales,
while both $\{b_1\}$ and $\{b_1, b_2\}$ give larger values of the linear bias. This means that these two models inaccurately predict a lower amplitude
for the angular power spectra at low multipoles, which results in a larger value of the bias and an inaccurate fit.  
We also see from Fig.~\ref{fig:pt-biases-z} that the models $\{b_1, b_{\mathrm{G}2}\}$ and $\{b_1, b_2, b_{\mathrm{G}2}\}$ consistently find negative values for the parameter
$b_{\mathrm{G}2}$, with amplitude monotonically increasing with redshift.  
On the other hand, we also notice that in the 3-parameter model $b_2$ is mostly unconstrained at $\bar{z} > 1$, and the best-fit values are consistent with zero within the error bars.
It seems that the quadratic contributions from the density are already well modelled by one-loop perturbation theory and introducing tidal bias is sufficient to model the data.

\subsection{Results from relativistic maps}
\label{sec:pt-bias-rel}

In this section, we apply the same methodology discussed in Sec.~\ref{sec:pt-bias-density} to test the perturbative bias expansion against the fully relativistic number counts, and thus assess the robustness of our results
when light-cone effects are taken into account. 
In our model, we neglect the effect of magnification and other relativistic effects. Using the {\sc class} code we tested the impact of magnification in the autocorrelation of a linearly biased tracer, with bias given by the linear 
bias of our halo population. We find that for $\bar{z} \le 2$ the impact of magnification is always below $1\%$ and thus it would not affect the analysis
in a significant way if included. 
RSD are included in the model using the simple linear Kaiser prescription~\cite{Kaiser:1987qv}. The motivation for this choice is twofold. First, adding nonlinear RSD would extend the set of free parameters. Second, we know that RSD affects the angular power spectrum on large scales, where our simulation data have larger statistical errors due to cosmic variance, and that the effect is highly suppressed for wide redshift bins. While 
it was pointed out in Refs.~\cite{Gebhardt:2020imr,Matthewson:2021rmb,Lepori:2021lck} that nonlinear velocities affect the angular power spectrum on large scales for thin redshift bins, in our study we use relatively wide bins where
the linear prescription for RSD is expected to work well. Therefore,
to include the linear RSD model, we simply add the following two extra terms to Eq.~\eqref{eq-pt-cell-model-real},
\begin{equation}
    C_\ell \rightarrow  C_\ell + b_1 C^{\mathrm{rsd}\times \delta_\mathrm{lin}}_\ell + C^{\mathrm{rsd} \times \mathrm{rsd}}_\ell,
    \label{eq-pt-cell-model-rsd}
\end{equation}
with
\begin{equation}
   C^{\mathrm{rsd}\times \delta_\mathrm{lin}}_\ell (\bar{z}, \sigma_z) = \frac{1}{\pi r(\bar{z})} \int_{-\infty}^{\infty} \mathrm{d} k_\parallel \,\, f(\bar{z})\,\mu^2\, P_\mathrm{lin}(k, \bar{z})\,\,j^2_0 \left(\frac{k_\parallel \sigma_z}{H(\bar{z})}\right), 
\end{equation}
and 
\begin{equation}
   C^{\mathrm{rsd} \times \mathrm{rsd}}_\ell (\bar{z}, \sigma_z) = \frac{1}{2 \pi r(\bar{z})} \int_{-\infty}^{\infty} \mathrm{d} k_\parallel \,\, f^2(\bar{z})\,\mu^4\, P_\mathrm{lin}(k, \bar{z})\,\,j^2_0 \left(\frac{k_\parallel \sigma_z}{H(\bar{z})}\right), 
    \label{eq-pt-cell-model-rsd-2}
\end{equation}
where $f(\bar{z})$ is the growth rate. 

        \begin{figure*}
        \centering
        \begin{subfigure}[b]{0.48\textwidth}
            \centering
            \includegraphics[width=\textwidth]{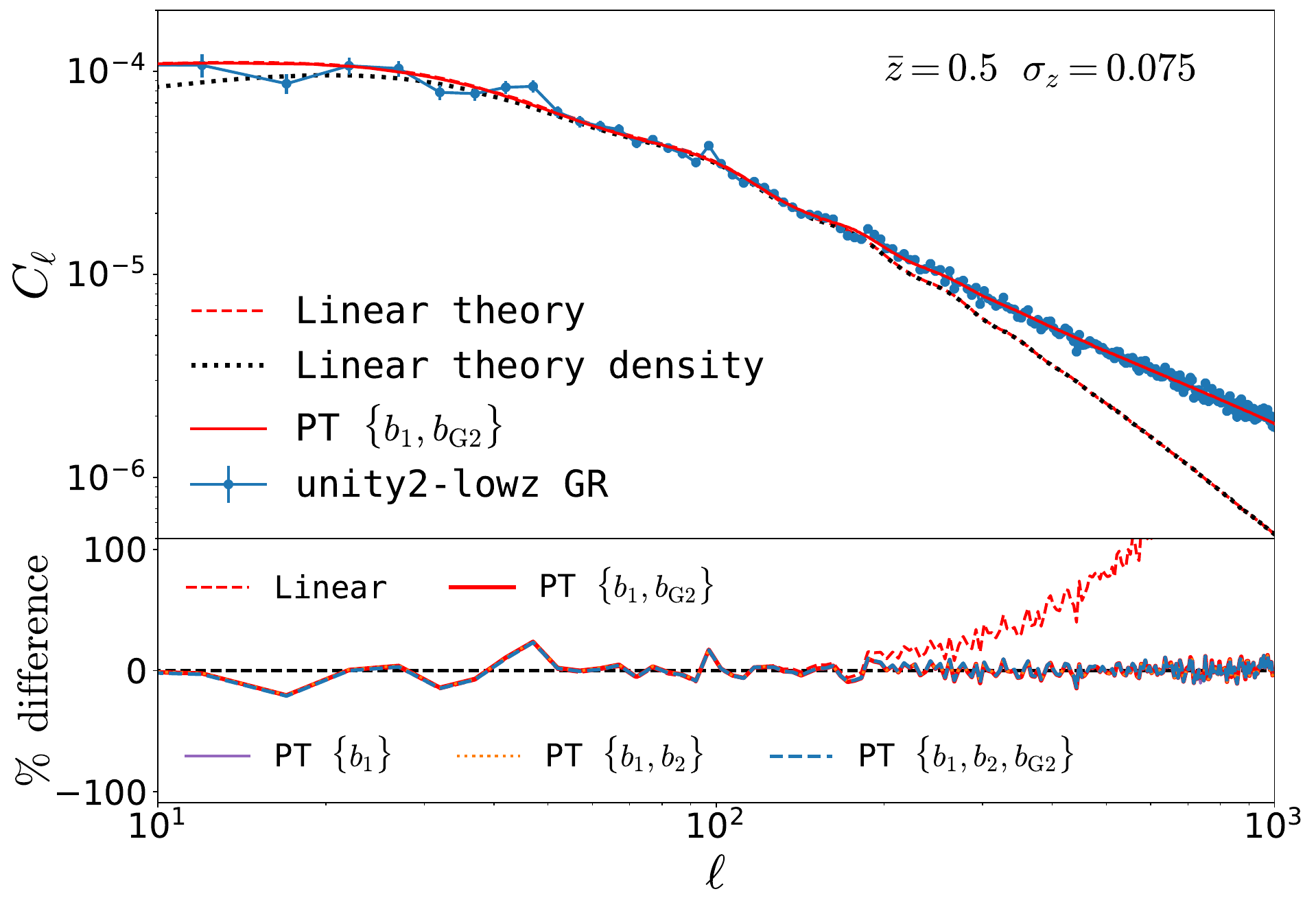}
            \caption[]%
            {{\small  }}    
            \label{fig:pt-bias-gr-1}
        \end{subfigure}
        \hfill
        \begin{subfigure}[b]{0.48\textwidth}  
            \centering 
            \includegraphics[width=\textwidth]{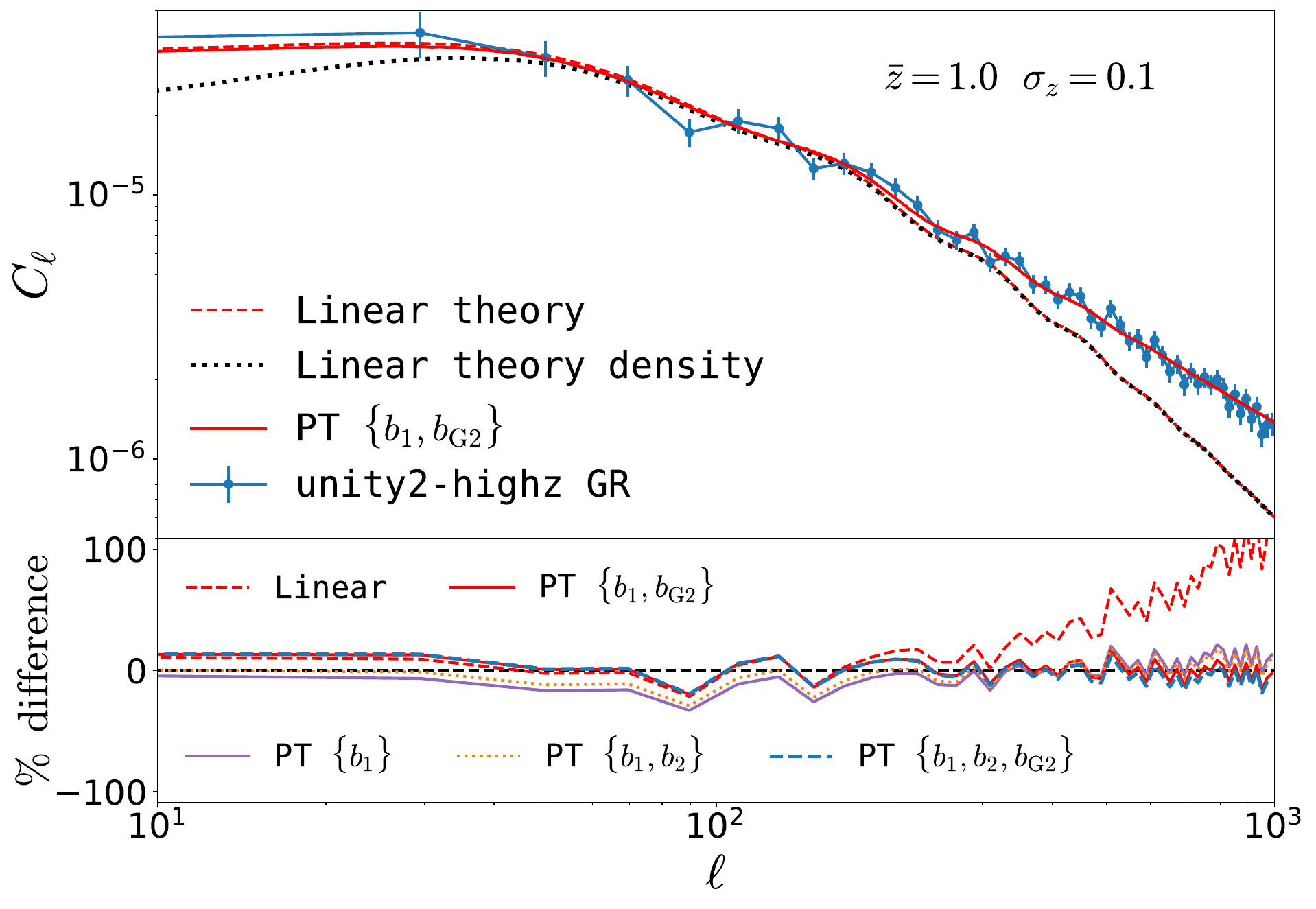}
            \caption[]%
            {{\small }}    
            \label{fig:pt-bias-gr-2}
        \end{subfigure} \\
        \begin{subfigure}[b]{0.48\textwidth}  
            \centering 
            \includegraphics[width=\textwidth]{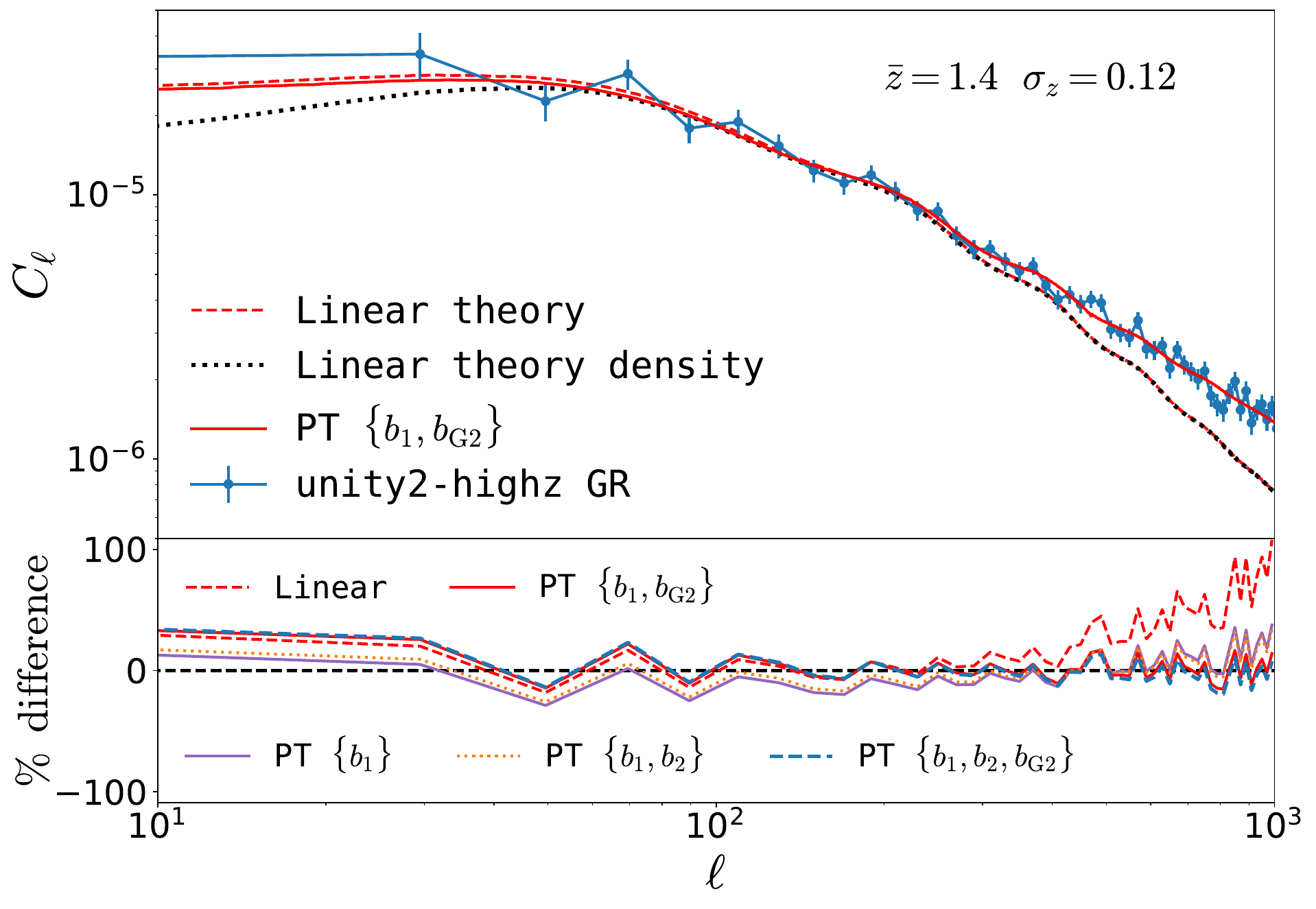}
            \caption[]%
            {{\small }}    
            \label{fig:pt-bias-gr-3}
        \end{subfigure}
        \hfill
        \begin{subfigure}[b]{0.48\textwidth}  
            \centering 
            \includegraphics[width=\textwidth]{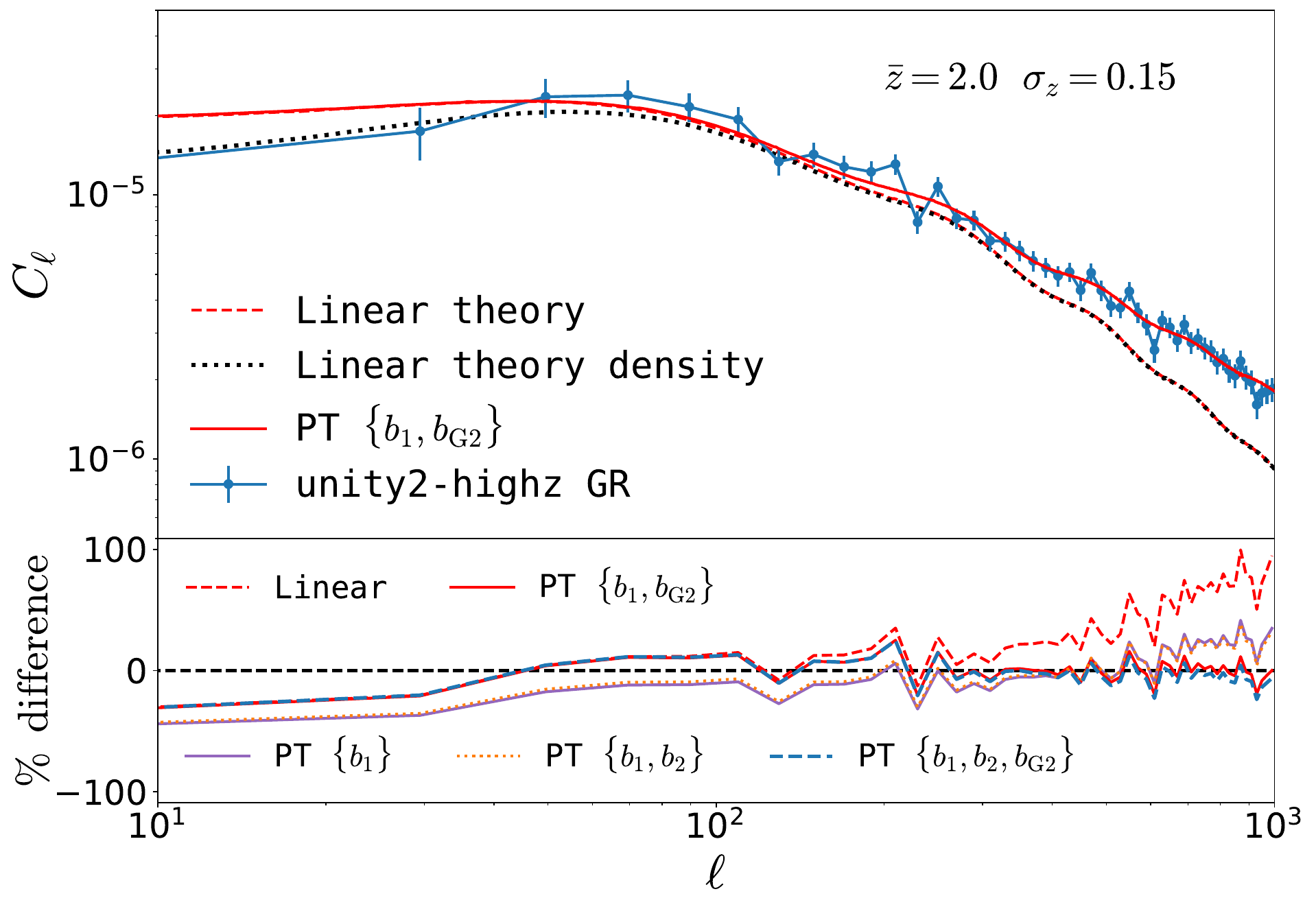}
            \caption[]%
            {{\small }}    
            \label{fig:pt-bias-gr-4}
        \end{subfigure}
        \caption[ ]
        {\small Outcome of our MCMC fit for the fully relativistic angular power spectra, at four representative redshifts. 
        \underline{Top panels:} Blue data points represent the angular power spectra extracted from the simulated maps. Red dashed (black dotted) lines are the theoretical prediction from linear theory including (neglecting) RSD, while continuous red lines represents the best fit for the 
        $\{b_1, b_{\mathrm{G}2}\}$ model. \underline{Bottom panels:} Percentage difference between the simulation data and the best-fit theoretical prediction, for linear theory and the four models described in Sec.~\ref{sec:pt-bias-rel}.  }
        \label{fig:lin-bias-pt-fit-2}
    \end{figure*}

\begin{figure}
\begin{center}
  \includegraphics[width=0.5\textwidth]{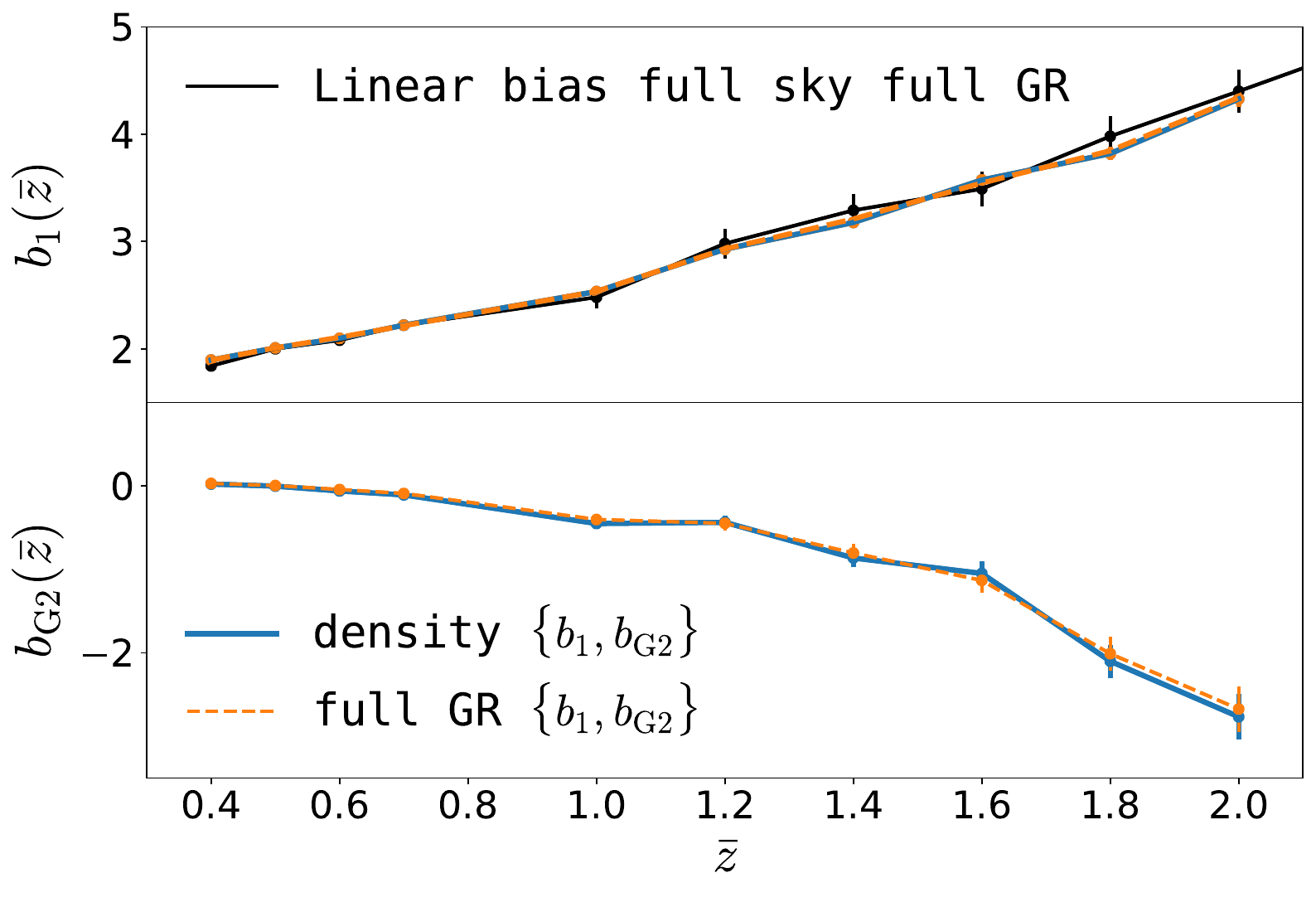}
  \end{center}
  \caption{Best-fit bias parameters for the model $\{b_1, b_{\mathrm{G}2}\}$. 
  Blue data refer to analysis of the density maps, orange data refers to the 
  analysis of the fully relativistic maps. For comparison, the black line shows
  the bias in linear theory, estimated only from the large scales in the angular power spectrum.}
  \label{fig:pt-biases-z-comp}
\end{figure}

We apply the same pipeline as described in Sec.~\ref{sec:pt-bias-density}
to the relativistic angular power spectra, using the updated model
described above. 
In Fig.~\ref{fig:lin-bias-pt-fit-2} we show the outcome of the fit 
for four representative redshifts, while the accuracy of the fit for the different models is reported in Table~\ref{tab:pt-bias} (columns denoted by ``full GR''). 
The analysis of the relativistic maps leads to the same results highlighted 
in the previous section: $\{b_1\}$ and $\{b_1, b_2\}$ are inaccurate at redshift $\bar{z} > 0.6, 0.7$, as they are not able to fit the large-scale and small-scale parts of the spectra simultaneously; on the other hand, 
$\{b_1, b_{\mathrm{G}2}\}$ and $\{b_1, b_2, b_{\mathrm{G}2}\}$ provide a good
model up to scales $\ell \sim 1000$.
We conclude that the model $\{b_1, b_{\mathrm{G}2}\}$ is preferable for providing an accurate prediction with the minimum number of free parameters. 

In Fig.~\ref{fig:pt-biases-z-comp} we compare the best-fit values
of the bias parameters from the analysis of the density maps and the fully relativistic maps for the model $\{b_1, b_{\mathrm{G}2}\}$.
We find that the bias measurements from the relativistic maps are fully consistent with the measurements from the density maps, and the linear bias agrees with the large scale bias estimated in Sec.~\ref{sec:3.2}.
The agreement between the results with/without relativistic effects is not surprising: as we already mentioned, radial correlations are smeared over a wide bin. Therefore, the contribution of RSD is highly suppressed for the binning chosen in our work, and confined to scales with large cosmic variance. We expect this result not to hold for thin bins. Since this requires a separate analysis, we will address this aspect in future work.

\section{Bias from cross-correlations}
\label{s:bias-cross}

In the previous section, we analysed the correlations of number counts at equal redshift $\bar{z}$. 
Here we consider the cross-correlation between halo number counts and lensing convergence $\kappa$. 
This cross-correlation provides an alternative way to estimate the halo bias
when we correlate maps at unequal redshift: the images of objects at $z_2$ are lensed by the halo and matter overdensities at $z_1 < z_2$. Therefore, the convergence field at $z_2$ is correlated with the  number counts at $z_1$. By measuring the cross-correlation of the convergence field in our simulation with the halo number counts in comoving position space,
$C^{\delta_\mathrm{h} \kappa}_\ell(z_1, z_2)$, and similarly the cross-correlation with the particle number counts, $C^{\delta \kappa}_\ell(z_1, z_2)$, we can estimate the linear bias $b_1$ at $z_1$ as
\begin{equation}
b_1 (z_1) = \frac{C^{\delta_\mathrm{h} \kappa}_\ell(z_1, z_2)}{C^{\delta \kappa}_\ell(z_1, z_2)}. 
\end{equation}

When considering the fully relativistic number counts, that is what we observe in 
galaxy surveys, there are extra terms that contribute to the halo and particle number counts.
On top of RSD, also lensing contributes to the number counts through magnification $\delta_{\kappa}$, and this effect is highly correlated with the lensing convergence field because in linear theory $\delta_{\kappa} = - 2 \kappa$.
Since the convergence is an integrated effect, the convergence maps at different redshifts are correlated. 
In this section, we study the impact of the magnification in the cross-correlation of number counts and 
convergence, and the effect on the linear bias estimation in entails.

We extract the convergence maps from our fully relativistic simulations for the \texttt{unity2-highz} catalogue and the binning in redshift considered in the previous sections. To this end we use the particle catalogue in order to have a good sampling of the field. Following the methodology described in detail in Ref.~\cite{Lepori_2020b}, we first use ray tracing to compute the area distance at the position of each particle. Then, we estimate the convergence as fluctuation of the area distance compared to its background value. Finally, we construct the map by taking in each pixel the average of the convergence over all the particles that fall into the pixel. 

We compute the cross-correlation of the convergence maps with 
the halo number counts, $C^{\Delta_\mathrm{h} \kappa}_\ell(z_1, z_2)$, and particle number counts, $C^{\Delta \kappa}_\ell(z_1, z_2)$, including all relativistic effects. We also estimate the correlation of the convergence field $C^{\kappa \kappa}_\ell(z_1, z_2)$.

        \begin{figure*}
        \centering
        \begin{subfigure}[b]{0.48\textwidth}
            \centering
            \includegraphics[width=\textwidth]{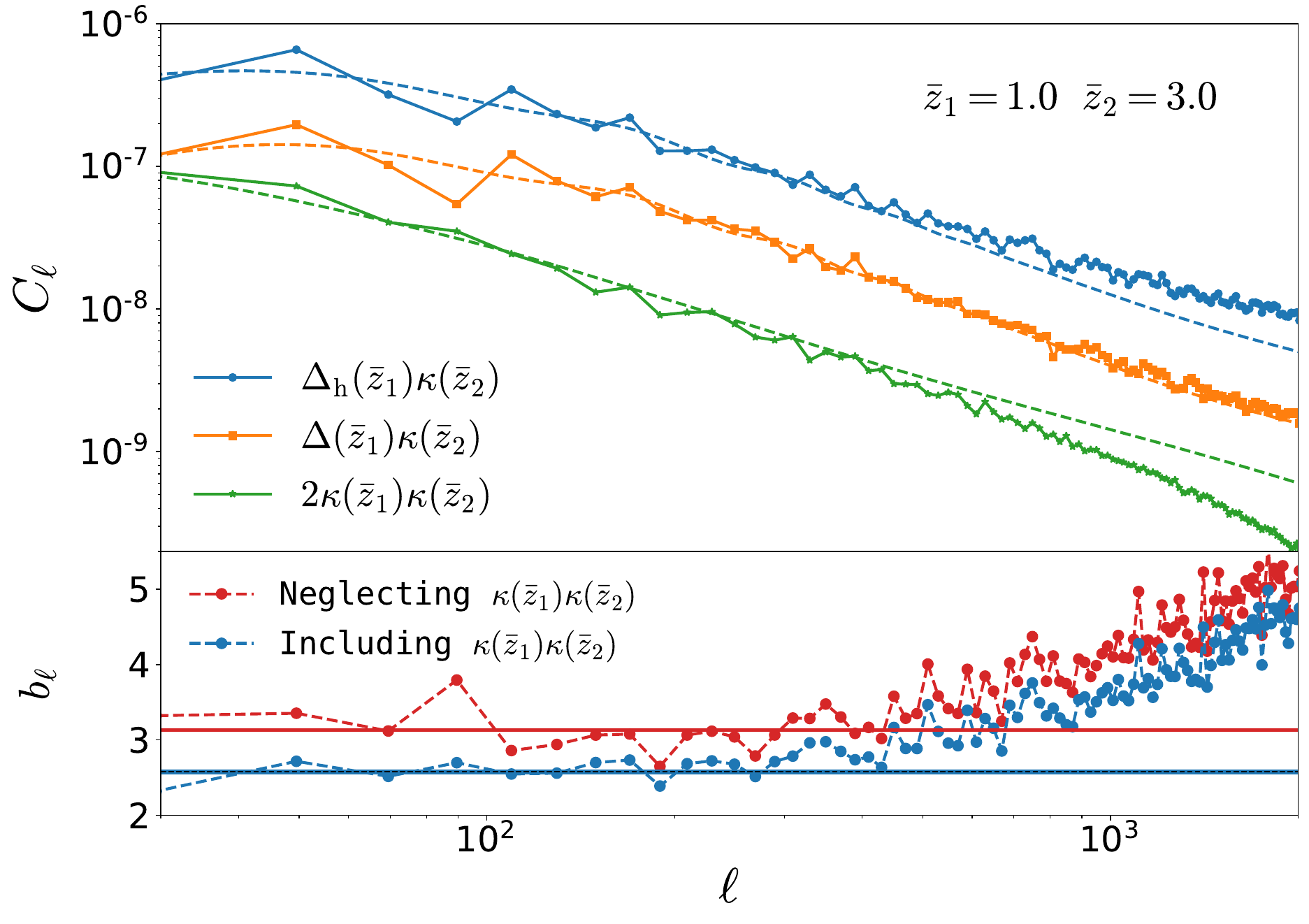}
        \end{subfigure} 
        \hfill
        \begin{subfigure}[b]{0.48\textwidth}  
            \centering 
            \includegraphics[width=\textwidth]{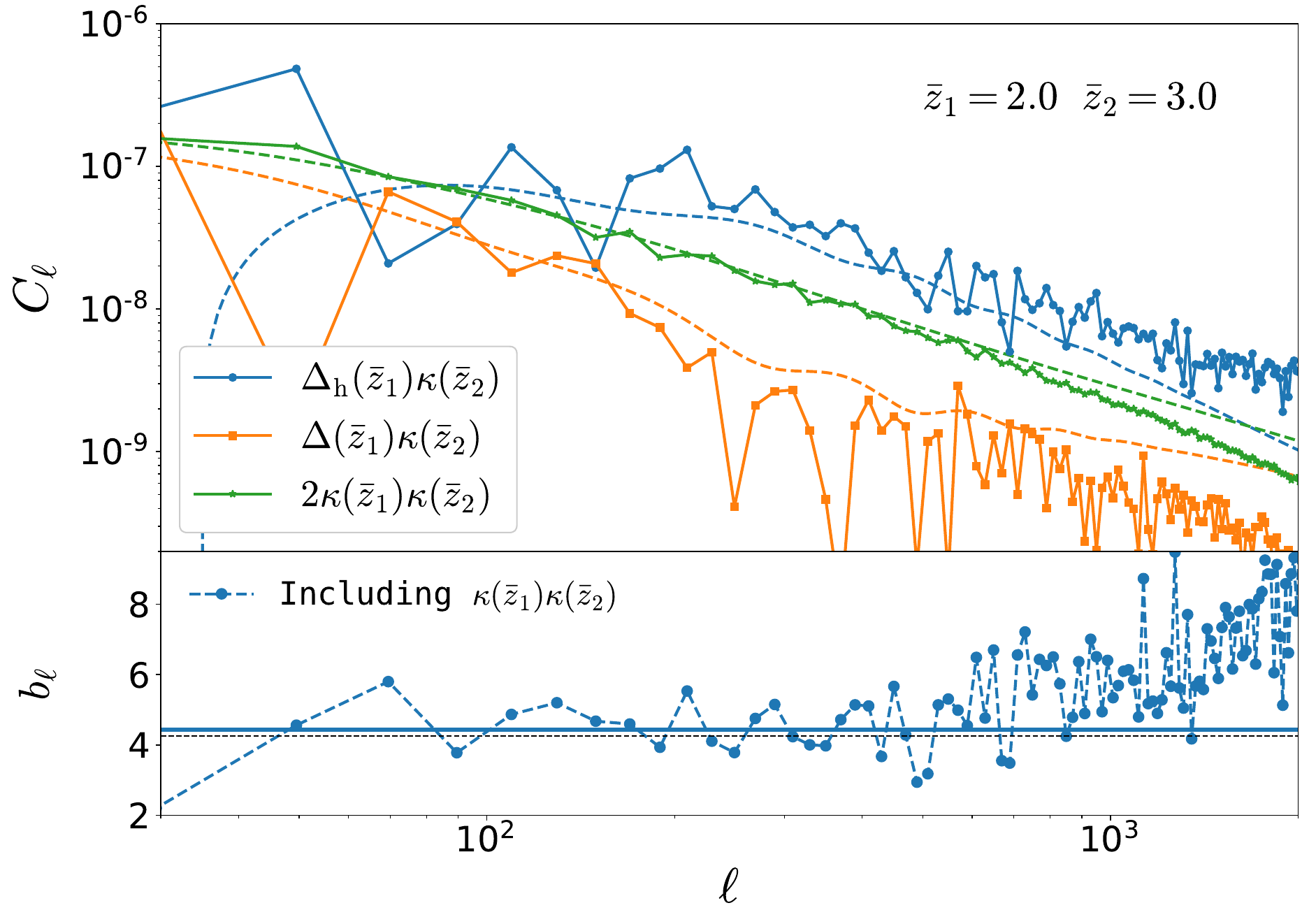}
        \end{subfigure}
        \caption[ ]
        {\small Cross-correlation of number counts at $z_1= 1$ and $z_1= 2$ (left and right plots, respectively) with the convergence map at $z_2 = 3$. \underline{Top panels}:
        angular power spectra $C^{\Delta_\mathrm{h} \kappa}_\ell(z_1, z_2)$ (blue), $C^{\Delta \kappa}_\ell(z_1, z_2)$ (orange), and the amplitude of the contribution from magnification to these spectra, that is $2 C^{\kappa \kappa}_\ell(z_1, z_2)$ (green). Data points are the spectra extracted from the simulations, the dashed lines represent the linear theory prediction from the {\class} code. \underline{Bottom panels}:
        Bias estimation at $z_1$ from this cross-correlation, in a case where
        we ignore the contribution of magnification (red lines) and in a case where we subtract that contribution (blue lines). The data points are computed from Eq.~\eqref{eq:kappa-bias-1} and Eq.~\eqref{eq:kappa-bias-2}, respectively. The horizontal solid red/blue lines mark the respective linear bias estimated from the cross-correlation. A dashed black lines shows the linear bias estimated previously in Sec.~\ref{sec:3.1} from the autocorrelation. For $z_1 = 2$ (right plot) the contribution from magnification is comparable to the one from density on all scales in the spectrum $C^{\Delta \kappa}_\ell(z_1, z_2)$, and thus it is not
        possible to estimate the linear bias in a meaningful way when neglecting this effect. 
        }
        \label{fig:lin-bias-cross-kappa-1}
    \end{figure*}

In Fig.~\ref{fig:lin-bias-cross-kappa-1} (top panel on the left plot), we
show the cross-correlation power spectra $C^{\Delta_\mathrm{h} \kappa}_\ell(z_1, z_2)$ (blue), $C^{\Delta \kappa}_\ell(z_1, z_2)$ (orange), and the amplitude of the contribution from magnification to these spectra, that is $2 C^{\kappa \kappa}_\ell(z_1, z_2)$ (green), for $z_1 = 1$ and $z_2 = 3$. We estimate the bias at $z_1$ for a case where we neglect the presence of relativistic effects, that is
\begin{equation}
b_\ell (z_1) = \frac{C^{\Delta_\mathrm{h} \kappa}_\ell(z_1, z_2)}{C^{\Delta \kappa}_\ell(z_1, z_2)}\,, 
\label{eq:kappa-bias-1}
\end{equation}
and also for a case where we subtract the effect of lensing magnification, i.e.\
\begin{equation}
b_\ell (z_1) = \frac{C^{\Delta_\mathrm{h} \kappa}_\ell(z_1, z_2) + 2 C^{\kappa \kappa}_\ell(z_1, z_2)}{C^{\Delta \kappa}_\ell(z_1, z_2)+ 2 C^{\kappa \kappa}_\ell(z_1, z_2)}\,. 
\label{eq:kappa-bias-2}
\end{equation}
The results are shown in Fig.~\ref{fig:lin-bias-cross-kappa-1} (bottom panel on the left plot). In the case where we neglect the presence of relativistic effects, the linear bias fitted to $b_\ell$, including scales below $\ell = 150$, overestimates the value obtained in the previous sections from the autocorrelation. 
However, in the case where we subtract the contribution of magnification, we are able to recover the value of the linear bias accurately.

For the configuration $z_1 = 1$ and $z_2 = 3$, 
the contribution of magnification to
the particle angular power spectrum ranges between $\sim 20-70\%$ on scales $\ell \sim 30-1000$, and the shift in the bias estimate is $\sim 20\%$. When we cross-correlate number counts at $z_1 > 1$, the contribution of magnification increases, while the density term becomes smaller. At $z_1 = 2$, magnification constitutes a significant fraction of both the spectra  $C^{\Delta_\mathrm{h} \kappa}_\ell(z_1, z_2)$ and $C^{\Delta \kappa}_\ell(z_1, z_2)$, as shown in Fig.~\ref{fig:lin-bias-cross-kappa-1} (right plot). Therefore, for these configurations, it is not possible to obtain a reasonable estimate of the linear bias neglecting the contribution of magnification. However, when magnification is taken into account in the bias measurement, we can recover the linear bias obtained from the autocorrelation at $z_1 = 2$. 

\begin{figure*}
        \centering
        \begin{subfigure}[b]{0.48\textwidth}  
            \centering 
            \includegraphics[width=\textwidth]{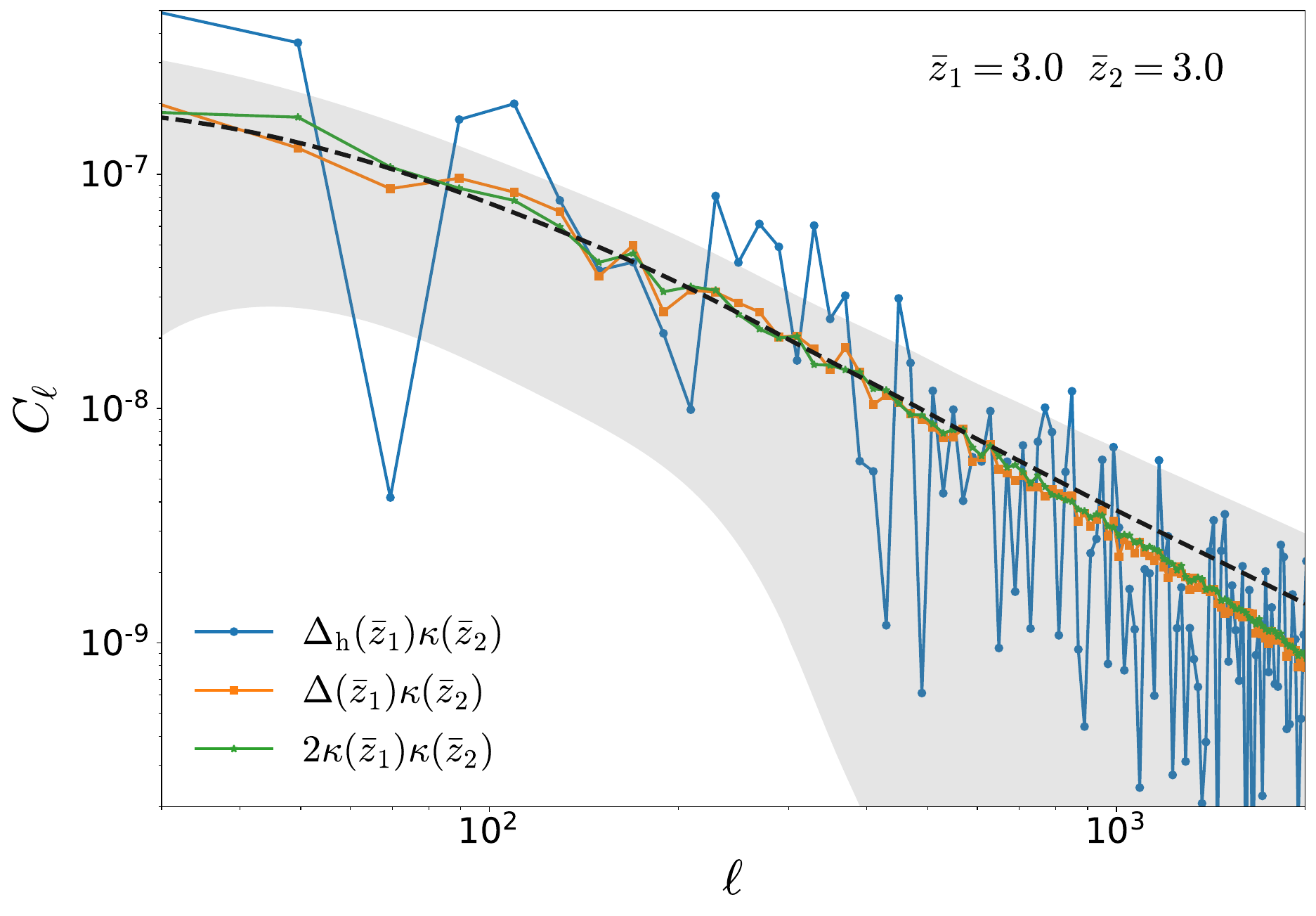}
        \end{subfigure}
         \hfill
        \begin{subfigure}[b]{0.48\textwidth}  
            \centering 
            \includegraphics[width=\textwidth]{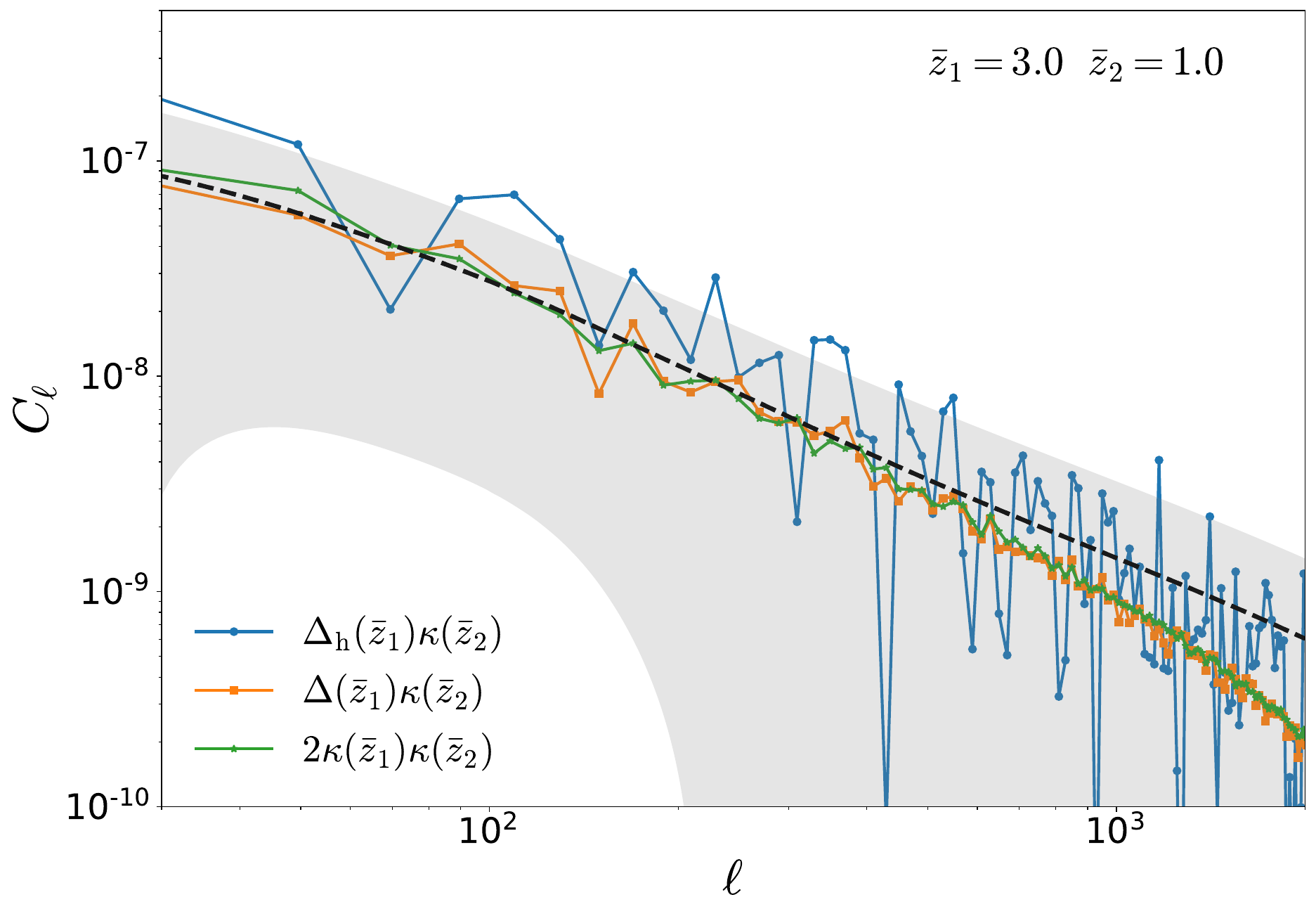}
        \end{subfigure}
        \caption[ ]
        {\small Cross-correlation of number counts at $z_1= 3$ and the convergence map at $z_2 = 3$ (left plot), and $z_2 = 1$ (right plot). The data points show the measurements from simulation for $C^{\Delta_\mathrm{h} \kappa}_\ell(z_1, z_2)$ (blue), $C^{\Delta \kappa}_\ell(z_1, z_2)$ (orange), and the amplitude of the contribution from magnification to these spectra, that is $2 C^{\kappa \kappa}_\ell(z_1, z_2)$ (green). These three lines coincide within statistical fluctuations because magnification is the dominant contribution
        to both $C^{\Delta_\mathrm{h} \kappa}_\ell(z_1, z_2)$ and $C^{\Delta \kappa}_\ell(z_1, z_2)$, for either of the configurations. The dashed black lines show the predictions from linear theory for $C^{\Delta \kappa}_\ell(z_1, z_2)$, while the grey shaded region shows the $1\sigma$ statistical error for $C^{\Delta_\mathrm{h} \kappa}_\ell(z_1, z_2)$ around the theory prediction. This error is dominated by the shot noise of halos at $\bar{z}_1 = 3$.  
        }
        \label{fig:lin-bias-cross-kappa-2}
    \end{figure*}

In Fig.~\ref{fig:lin-bias-cross-kappa-2} we show the same cross-correlations, for number counts at $z_1 = 3$ and convergence at $z_2 = 3$ (left plot) and $z_2 = 1$ (right plot). 
For $z_1 \ge z_2$, the contribution from magnification is by far the dominant one in the cross-correlations $C^{\Delta_\mathrm{h} \kappa}_\ell(z_1, z_2)$ and $C^{\Delta \kappa}_\ell(z_1, z_2)$, and there is no information on the halo bias. 
While we cannot measure the halo bias from these cross-correlations, they are a clean observable to detect cosmic magnification. A first detection of this effect from galaxy shear measurements cross-correlated with background galaxy counts has been presented in Ref.~\cite{Liu:2021gbm}. In the near future, galaxy surveys like Euclid~\cite{EUCLID:2011zbd, Amendola:2016saw} and the Vera C.\ Rubin Observatory's Legacy Survey of Space and Time (LSST)~\cite{LSSTScience:2009jmu,LSSTDarkEnergyScience:2012kar} will be able to detect this effect with high significance and extract cosmological
information from it~\cite{Montanari:2015rga}.

In Fig.~\ref{fig:lin-bias-cross-kappa-1} and Fig.~\ref{fig:lin-bias-cross-kappa-2}, we notice that
the correlations dominated by magnification exhibit 
a power suppression on small scales when compared to the theoretical prediction. This is an artifact of finite resolution that occurs because magnification and lensing convergence are integrated effects and, therefore, the large error at low redshifts propagates into the estimated spectra at all redshifts. A detailed study of this issue can be found in Ref.~\cite{Lepori_2020b}, Appendix C. 

\section{Conclusions}\label{s:con} 

In this paper we have computed for the first time the halo number counts on the light cone of a relativistic N-body simulation and the model independent two-point statistics, the angular power
spectrum. 
Since halos are biased tracers of the LSS, 
this study is a first step towards building a robust model for the galaxy bias in the angular power spectrum of galaxies, the primary observable in the galaxy clustering analysis of photometric surveys. 

We have measured the redshift dependence of  halo bias for a halo population with masses above 
$M_\mathrm{min} = 5\times 10^{12}\,M_\odot/h$, using different methods and estimators.
We have computed the halo bias from the snapshots using  cross-correlation of the density field for halos and particles, together with the matter power spectrum in Fourier space.
We have compared these results to the bias values measured on the light cone using the angular power spectrum. The bias on the light cone has been extracted in tomographic redshift bins a) by comparing the correlations of halo density maps, particle density maps and their cross-correlation; b) by fitting the halo spectra to a theoretical model, both with and without relativistic effects.
We have found that these different estimates give consistent results for the linear halo bias, i.e.\
the halo bias on large scales. On small scales, 
high order corrections become significant and the 
bias from the halo autocorrelation differs from
the one obtained using the cross-correlation of halos and particles.

We have investigated the limitations of the linear bias model for our halo population. In particular, we have established at which multipoles a theoretical model with linear bias, and nonlinearities modelled in the matter power spectrum with the \texttt{HMCODE}, breaks down. This nonlinear multipole scale
estimated in our simulation ranges between $\ell_\mathrm{nl} = 200-600$, and it does not show a clear redshift dependence. The values of the nonlinear multipole scale that we obtain from our simulations are roughly compatible with the ansatz often used in the literature $\ell_\mathrm{nl} = r(\bar{z}) k_\mathrm{max}$, with $k_\mathrm{max} = 0.1-0.2\,h/$Mpc. However, unlike this simple prediction, we find that at high redshift the linear bias description cannot be extended up to smaller scales. This reflects the fact that the bias becomes larger at high redshift,  while nonlinearities of matter decay at early time. Nonlinear biasing mixes these two effects, thus it is not expected to decay at high redshift in the same way as the nonlinear clustering of matter does. 

We have tested minimal extensions of the linear bias model, including second-order terms in the bias expansion. These extra terms are known in the literature as quadratic bias and tidal bias. We have found that a two-parameter model with linear bias and tidal bias (and quadratic bias set to zero) is able to fit the halo angular power spectra well at equal redshifts on a wide range of scales, up to $\ell \sim 1000$. This represents a great improvement compared to the linear bias model. We have also shown that this model is significantly preferred compared to adding a quadratic bias. A quadratic bias addition
is not able to reproduce nonlinearities in our simulated angular power spectra at $\bar{z} \ge 0.7$. 

Finally, we have shown that it is possible to estimate the linear bias from the cross-correlation of halo number counts and convergence, provided that the convergence field is at higher redshift than the number counts, and that the contribution of magnification is properly accounted for.  

This work can be extended in several directions. We have tested the bias expansion against the halo angular power spectra at equal redshift. However, a significant amount of information can be extracted from the correlations at unequal redshift, see for example Ref.~\cite{DiDio:2013sea}, and thus it is crucial to assess the robustness of a model against the full set of auto and cross-correlations. A detailed study of the cross-correlation beyond the linear bias model might require going beyond the flat-sky approximation that we have employed in our analysis as it provides a fast computation of our observable. We will address this issue in future work.
We have found that RSD can be modelled reasonably well using Kaiser's linear prescription for the binning considered in our work. We expect this picture to change for thinner redshift bins, as several works have pointed out that nonlinear velocities contribute to RSD on large scales in the angular statistics for a typical bin size $\sigma_z \sim 0.01$ or smaller, see e.g.\ Refs.~\cite{Jalilvand:2019brk,Gebhardt:2020imr,Lepori:2021lck, Matthewson:2021rmb}. A detailed analysis of nonlinear RSD in the angular statistics for biased tracers is left for future research.  

\section*{Acknowledgements}
FL thanks Andrej Obuljen for useful discussions. 
This work is supported by the Swiss National Science Foundation. We used high-performance computing resources provided by the Swiss National Super\-computing Centre (CSCS) under pay-per-use agreements (project IDs ``go25'' and ``uzh34'').
{We aknowledge the use of the following \texttt{python} packages to perform our analysis: \texttt{Ipython}~\cite{ipythonref},  \texttt{numpy}~\cite{numpyref, numpyref2}, \texttt{scipy}~\cite{scipyref}, \texttt{matplotlib}~\cite{matplotlibref}.

{\small\paragraph{Carbon footprint} 
This work re-used existing simulation data which drastically limits its carbon footprint. We estimate that additional post-processing consumed up to $250~\mathrm{kWh}$ of electrical energy from the Swiss power grid. Using a conversion factor of $0.119\,\mathrm{kg\,CO_2\,kWh}^{-1}$ (taken from \href{https://co2.myclimate.org/en}{myclimate.org}\footnote{Retrieved 13.\ September 2022}) this caused emissions of up to $30\,\mathrm{kg\,CO_2}$. Emissions are being fully offset through the carbon offsetting initiative of the Institute for Computational Science at University of Zurich, partnering with the myclimate Foundation.}

\appendix

\section{Jackknife resampling method}
\label{a:jk}

In the analysis presented in the main body of the paper, we adopt the jackknife method to suppress the shot noise from the estimated angular power spectra at fixed mean redshift $\bar{z}$. As explained in Sec.~\ref{sec:method-spectra-cov}, the method consists of
randomly splitting the number of objects (particles or halos) inside the redshift bins in two populations of equal size, generating the maps of the number counts from each of them, and estimating the angular power spectrum by cross-correlating these two maps.
Since the shot noise of the two maps is uncorrelated, it does not contribute to the expectation value of the estimated spectrum. 

    \begin{figure*}
        \centering
        \begin{subfigure}[b]{0.475\textwidth}
            \centering
            \includegraphics[width=\textwidth]{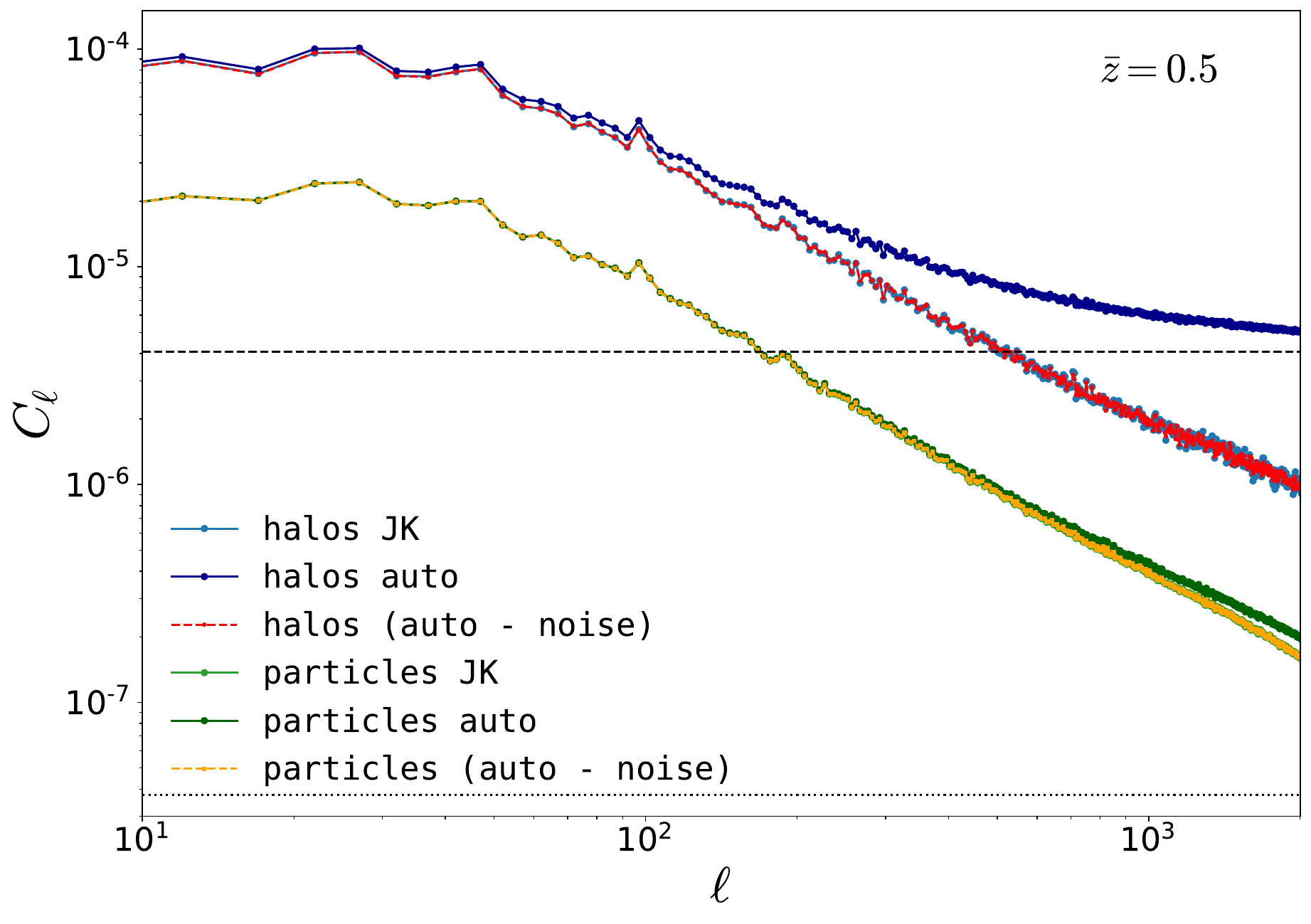}
            \caption[]%
            {{\small \texttt{unity2-lowz}}}    
            \label{fig:2a}
        \end{subfigure}
        \hfill
        \begin{subfigure}[b]{0.475\textwidth}  
            \centering 
            \includegraphics[width=\textwidth]{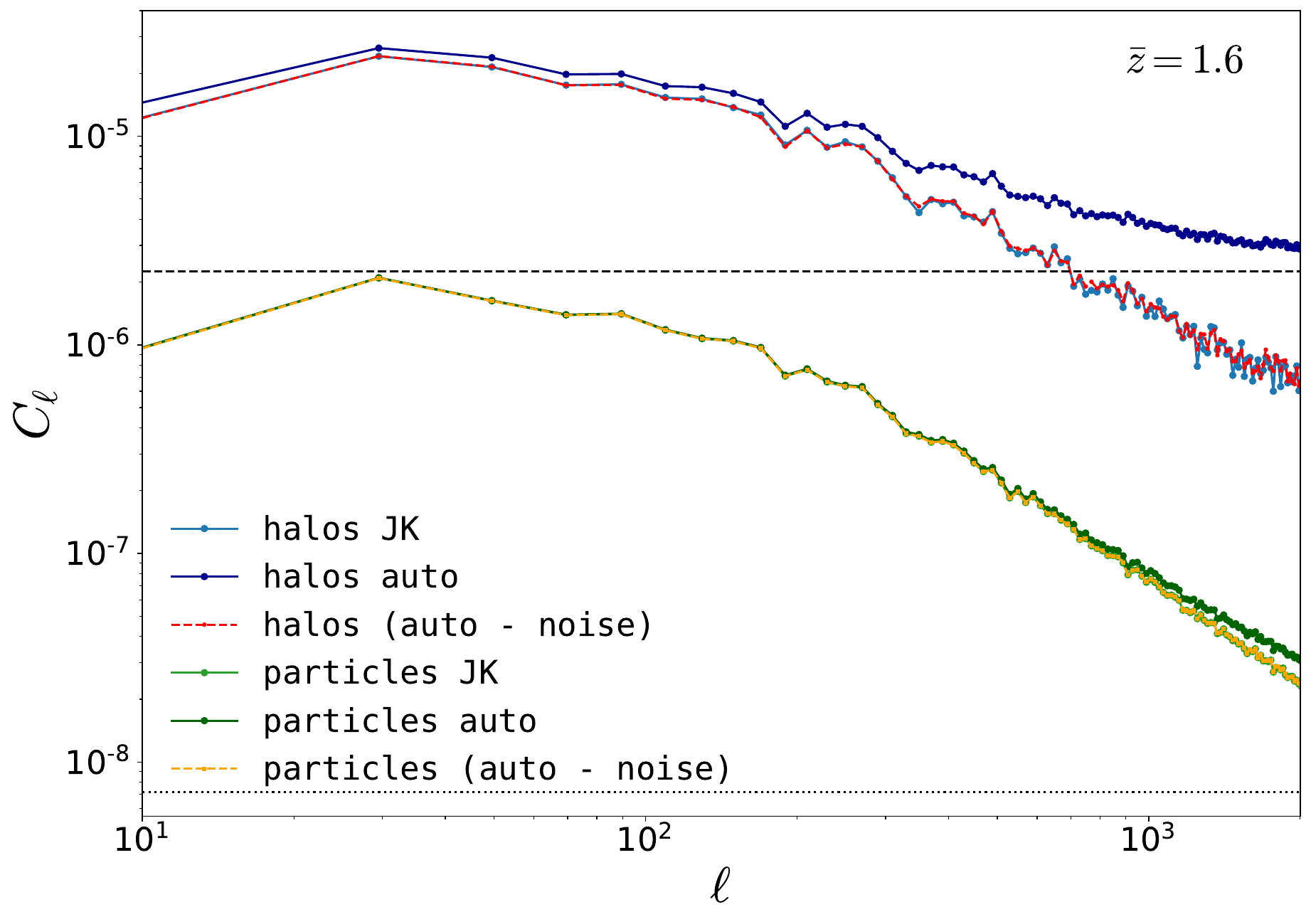}
            \caption[]%
            {{\small \texttt{unity2-highz}}}    
            \label{fig:2b}
        \end{subfigure}
        \caption[]
        {\small Angular power spectra at fixed mean redshift $\bar{z} = 0.5$ (left plot) and $\bar{z} = 1.6$ (right plot), for the halo and particle number counts. The lines denoted by \texttt{auto} refer to the standard estimator, where the angular power spectra are computed from the autocorrelation of the number counts and, therefore, are affected by shot noise. The lines denoted by \texttt{(auto - noise)} are obtained by explicitly subtracting the expected noise level $1/N_i$ from the \texttt{auto} spectra. 
        The lines denoted by \texttt{JK} refer to the jackknife estimator which effectively suppresses the shot noise. Black dashed (dotted) lines show the expected level of shot noise in the bin for halos (particles).
        The halos have been selected applying a mass selection with $M_\text{min} = 5\times 10^{12}\,M_\odot/h$.
        } 
        \label{fig:a-jk}
    \end{figure*}
    
In Fig.~\ref{fig:a-jk} we compare the jackknife estimator (\texttt{JK}) to 
the standard estimator of the spectrum (\texttt{auto}) that simply uses the autocorrelation of the number counts as is, for both halos and particles. 
We see that the standard estimator is affected by shot noise on all scales for halos, while the spectra of particles are visibly affected only on small scales. 
This is fully consistent with the level of Poisson noise, given by  $1/N_i$ for a redshift bin with $N_i$ galaxies per steradian, that is shown in the plot with a dashed horizontal line for halos, and a dotted horizontal line for particles. 
The noise level is $\sim 2$ orders of magnitudes smaller for particles due to their higher average number density. 
The \texttt{JK} method provides an effective way to remove the shot noise contribution from the spectra, without assuming a prior knowledge of the noise. 
Fig.~\ref{fig:a-jk} shows that the \texttt{JK} method gives consistent results with the angular power spectrum extracted from the autocorrelation if the Poisson noise is manually subtracted from the spectrum. 

\section{Linear bias on comoving slices}
\label{ap:Fourier-bias}
In order to measure the linear bias from the Fourier modes on equal-time hypersurfaces we run a separate simulation with the same parameters and mass resolution as \texttt{unity2} but with a smaller volume of $L^3_\mathrm{box} = (1344\,\mathrm{Mpc}/h)^3$. The snapshots then have $1920^3$ particles which is easier to handle and allows us to store them at a large number of redshift values $z \in \{0, 0.1, 0.2, 0.3, 0.4, 0.5, 0.6, 0.7, 0.8, 0.9, 1, 1.2, 1.4, 1.6, 1.8, 2.2, 2.6, 3\}$.
The halo-matter cross-power spectrum and the matter auto-power spectrum in each snapshot is estimated using the publicly available code \textsc{Pylians3}.\footnote{\url{https://pylians3.readthedocs.io}}
This code implements a routine that deposits particles into a regular 3D grid with $N^3$ voxels using the cloud-in-cell mass-assignment scheme. Here, we have used a grid with $N = 1000$, which is sufficient for the intended measurements of the linear bias well below its Nyquist wavenumber of $k_\mathrm{Nyq} = \pi N / L_\mathrm{box} \approx 2.33\,h/\mathrm{Mpc}$. We also note that $L_\mathrm{box}$ is much smaller than the horizon scale and the gauge dependence of the measured power spectra is therefore expected to have a minuscule effect on our bias measurements.
The auto-power spectra and cross-power spectra are then estimated from the density fields in Fourier space. 
The linear bias is finally estimated by 
dividing the halo-matter cross-power spectrum by the matter auto-power spectrum, see Eq.~\eqref{eq:bias_fourier},
and fitting a constant $b_\mathrm{fit}(\bar{z})$ in the range between $k_\mathrm{min} = 0.02\,h/\mathrm{Mpc}$ and $k_\mathrm{max} = 0.08\,h/\mathrm{Mpc}$ using least-squares regression. The uncertainty in the bias measurement is estimated from the standard deviation of the residuals of the data points of $b(k, \bar{z})$ with respect to $b_\mathrm{fit}(\bar{z})$,
\begin{equation}
    \sigma^2_{b} = \frac{1}{N_\mathrm{bins}-1} \sum_{k_i} \left[b(k_i, \bar{z}) - b_\mathrm{fit}(\bar{z})\right]^2\,,
\label{eq:sigma_bias}
\end{equation}
where $N_\mathrm{bins}$ is the total number of bins which contain the individual estimates at the wavenumbers $k_i$ within the fitting range. The residuals for the bias measurement in each snapshot are shown in Fig.~\ref{fig:snapshot_bias_residuals} and the best-fit bias value is shown in Fig.~\ref{fig:lin-bias-comp} over the entire redshift range.

\begin{figure}
\begin{center}
  \includegraphics[width=0.6\textwidth]{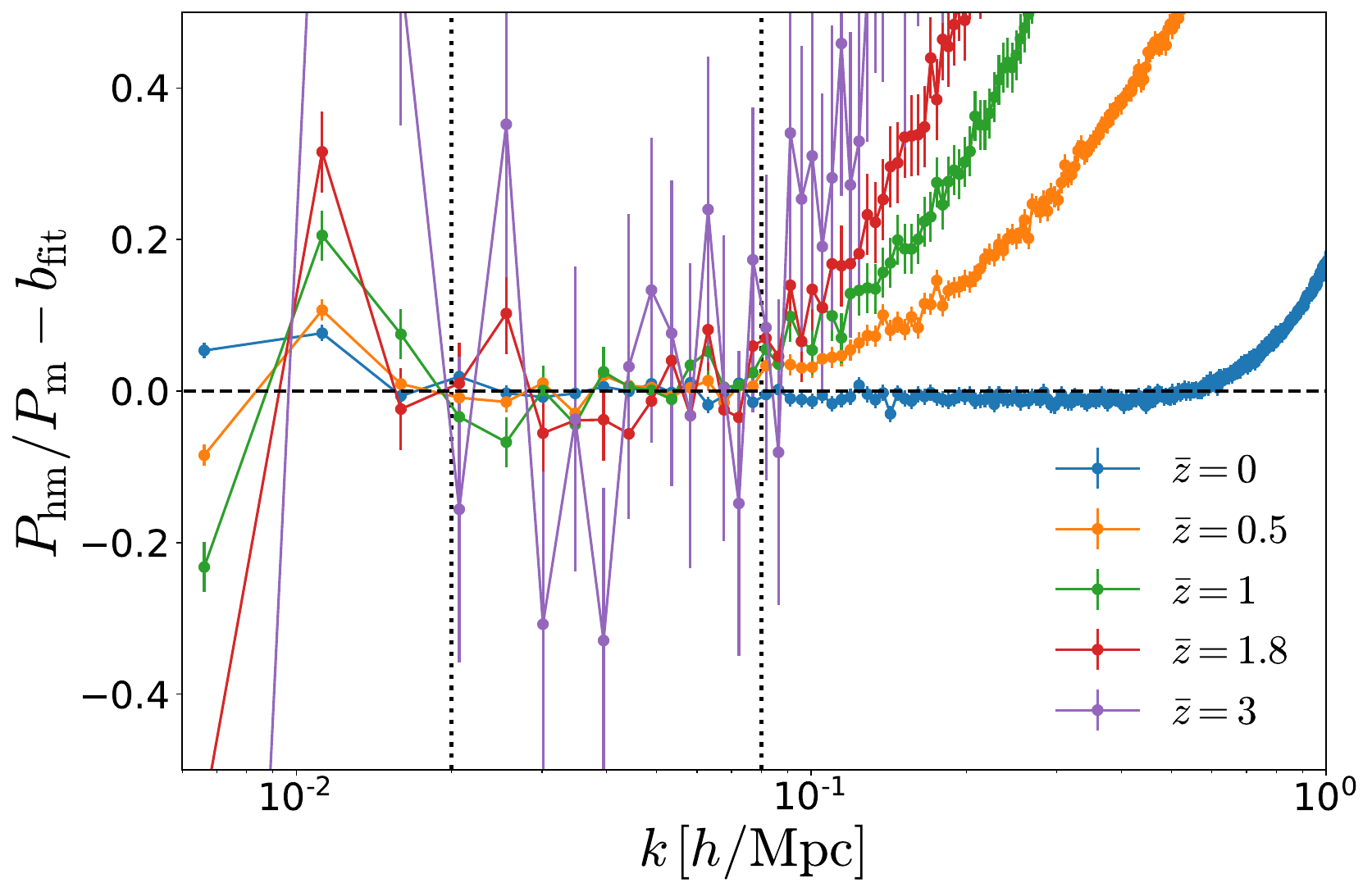}
  \end{center}
  \caption{Residuals for constant bias fits for a subset of the 17 snapshots that have been used for the estimation of the linear bias in Fourier space, with halo masses $M > 5 \times 10^{12}\,M_\odot/h$. The error bars are estimated from Eq.~\eqref{eq:sigma_bias}. The fit was performed in the range between the black dotted vertical lines.}
  \label{fig:snapshot_bias_residuals}
 \end{figure}

\section{Validation of the flat-sky approximation}
\label{ap:flat-sky}

        \begin{figure*}
        \centering
        \begin{subfigure}[b]{0.48\textwidth}
            \centering
            \includegraphics[width=\textwidth]{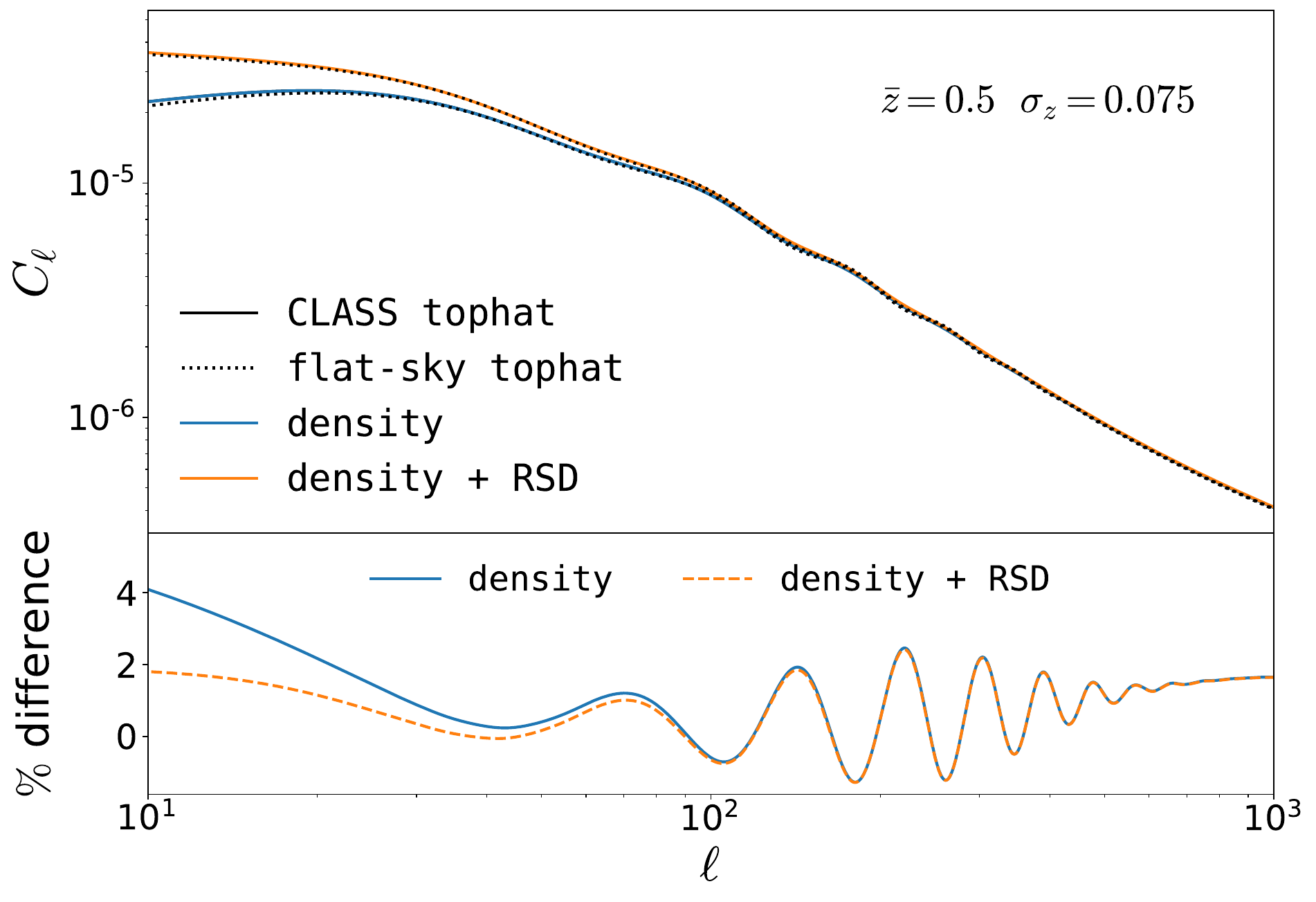}
        \end{subfigure}
        \hfill
        \begin{subfigure}[b]{0.48\textwidth}  
            \centering 
            \includegraphics[width=\textwidth]{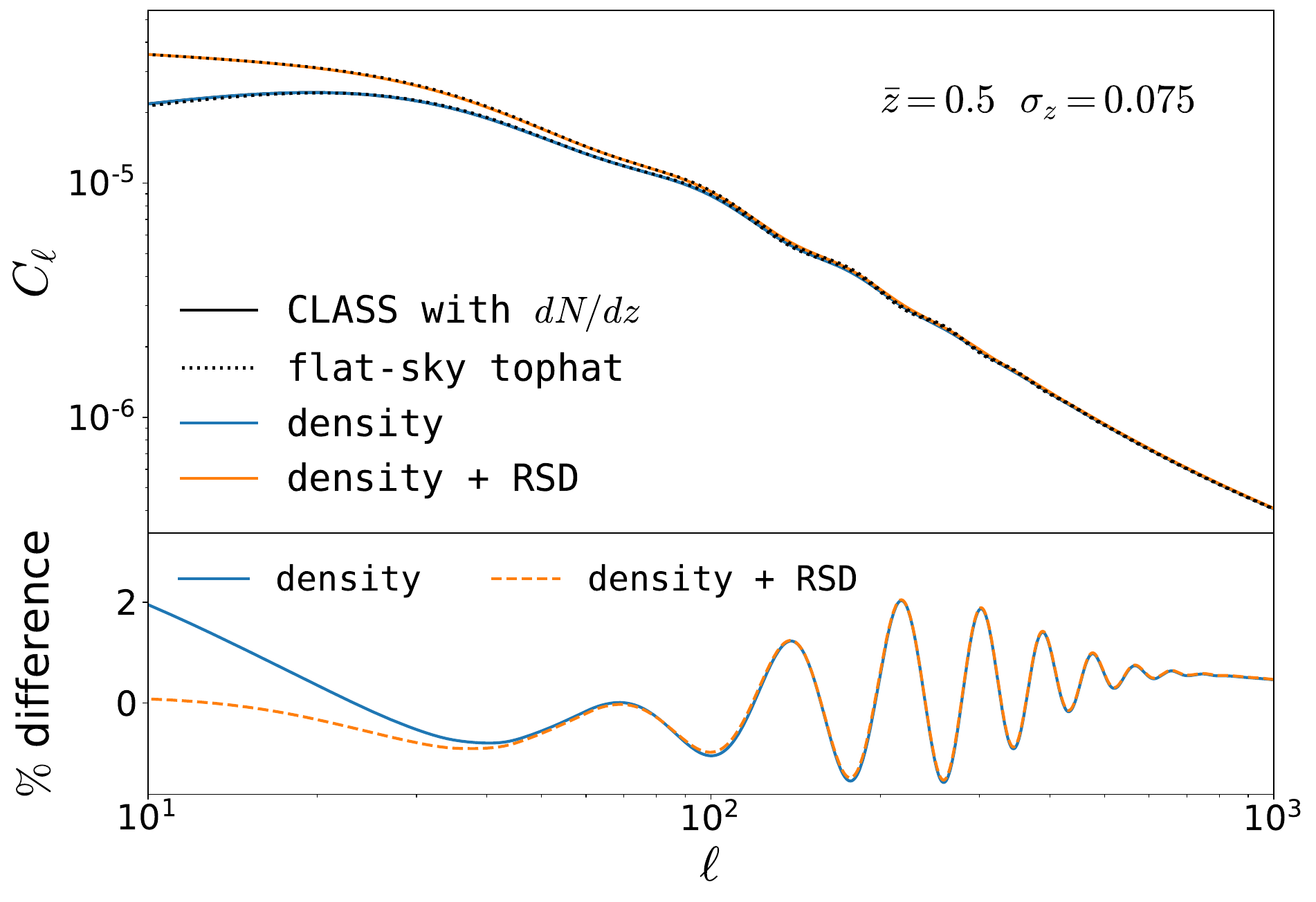}
        \end{subfigure} \\
        \begin{subfigure}[b]{0.48\textwidth}  
            \centering 
            \includegraphics[width=\textwidth]{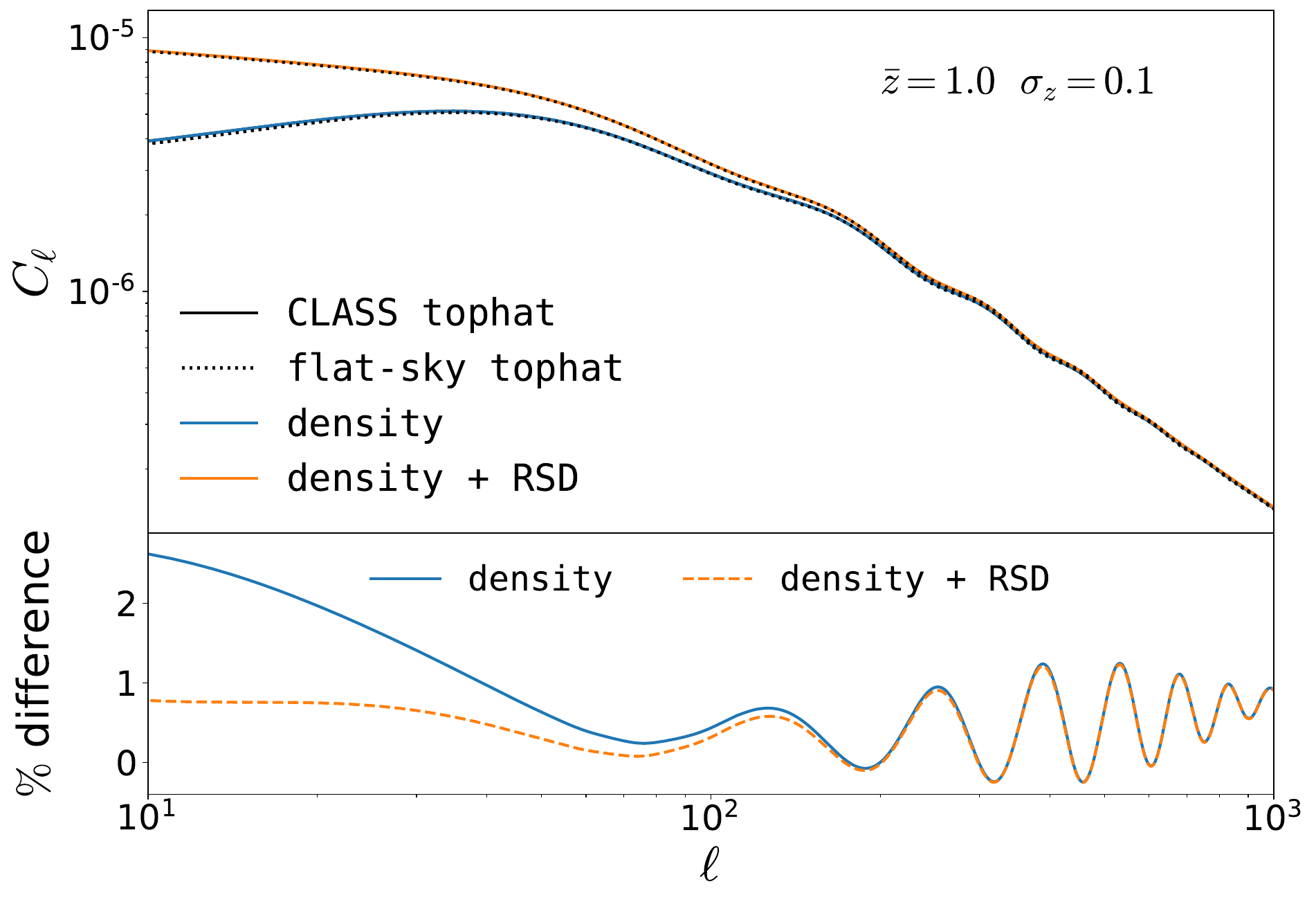}
        \end{subfigure}
        \hfill
        \begin{subfigure}[b]{0.48\textwidth}  
            \centering 
            \includegraphics[width=\textwidth]{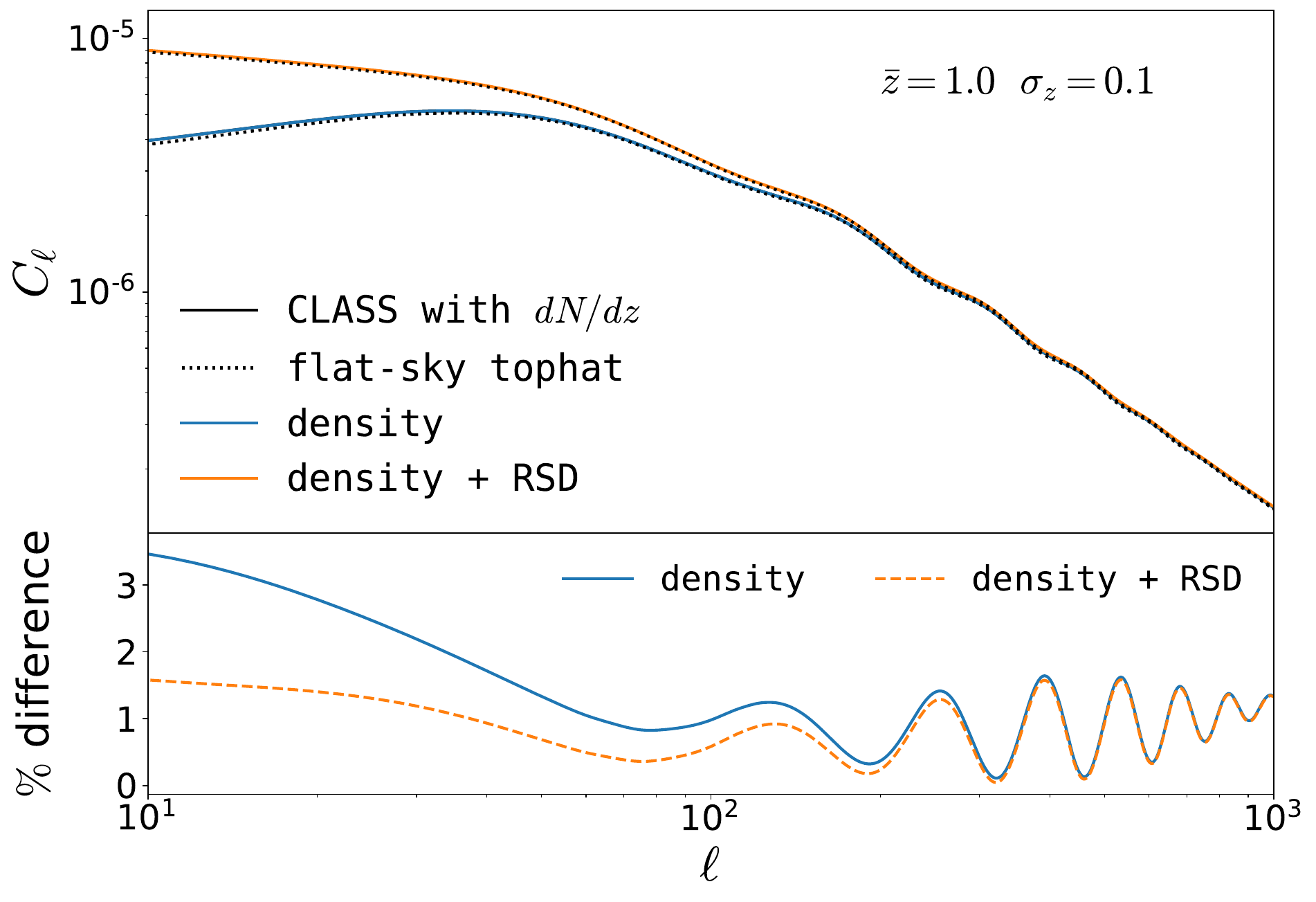}
        \end{subfigure}
        \begin{subfigure}[b]{0.48\textwidth}  
            \centering 
            \includegraphics[width=\textwidth]{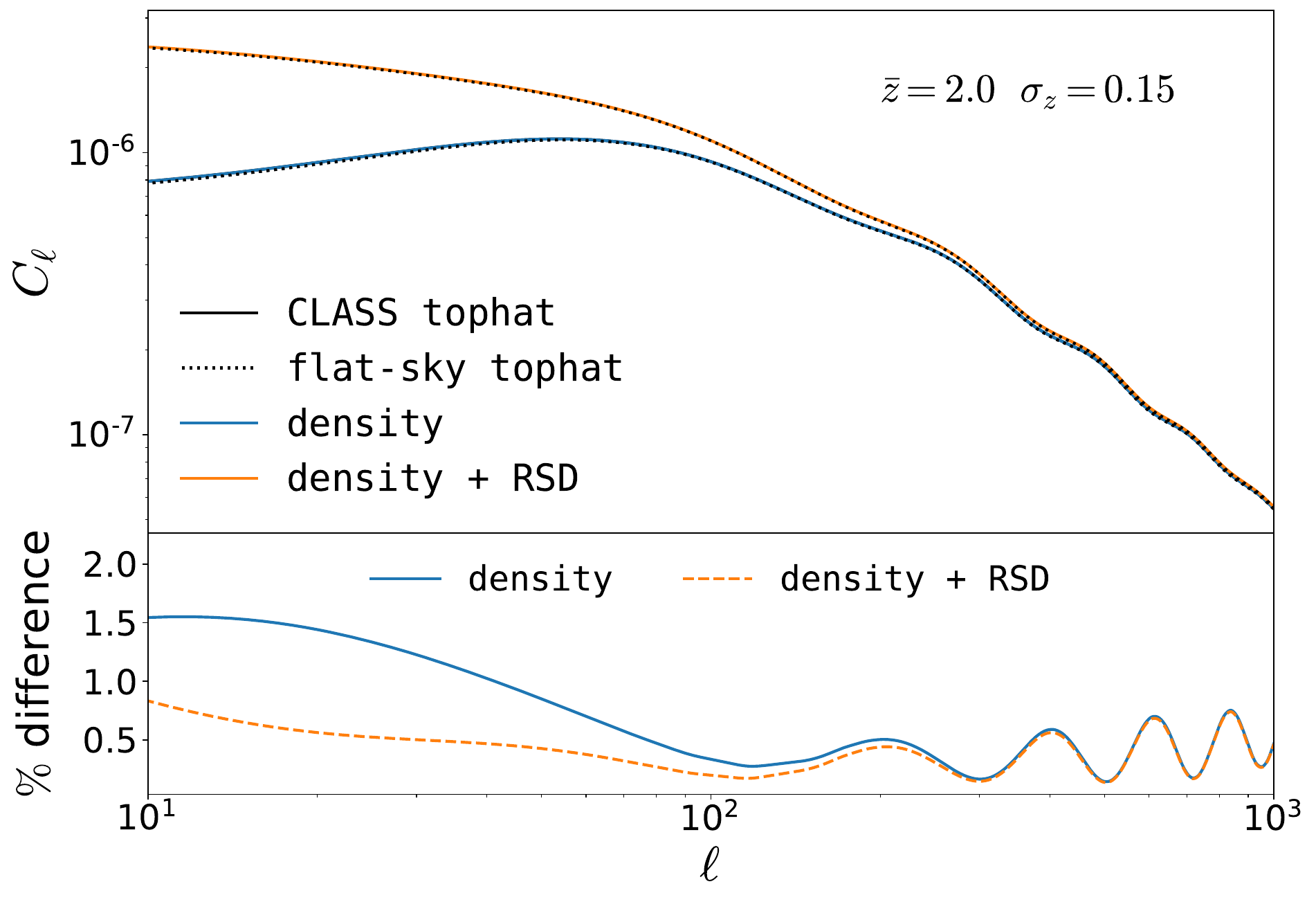}
        \end{subfigure}
        \hfill
        \begin{subfigure}[b]{0.48\textwidth}  
            \centering 
            \includegraphics[width=\textwidth]{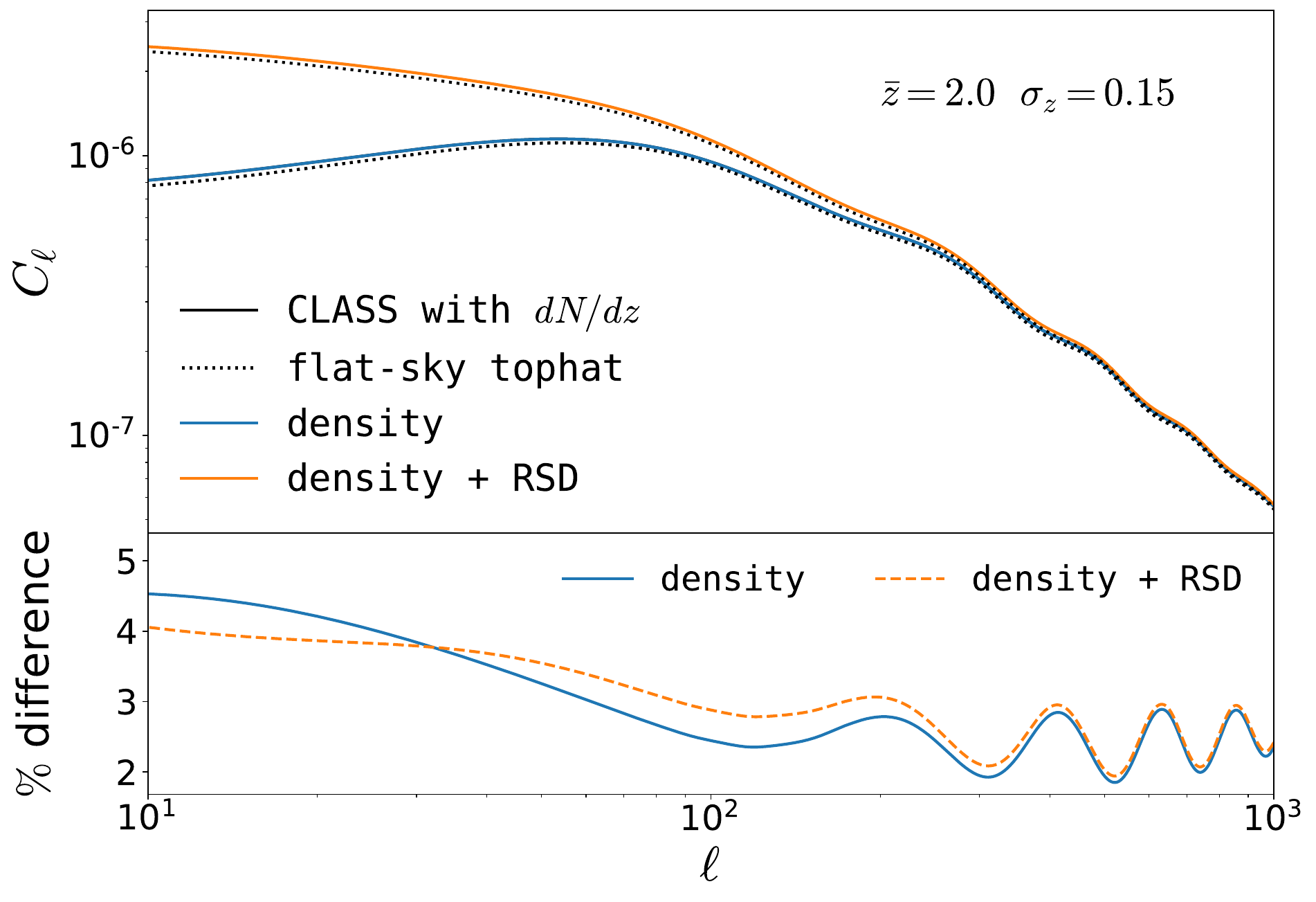}
        \end{subfigure}
        \caption[ ]
        {\small Comparison between the angular power spectrum in the flat-sky approximation with tophat bins and a) the full-sky computation with a uniform source distribution $dN/dz$ (left column);
        b) the full-sky computation with the $dN/dz$ of the halos in our simulation with masses above $M_\mathrm{min} = 5\times 10^{12}\,M_\odot/h$ (right column). In each plot, the top panel displays the angular power spectra, while the bottom panel shows the difference between the full-sky and the flat-sky spectra. Blue lines include only density contribution to the number counts, while orange lines take into account both density and RSD. The angular power spectra in the flat-sky approximation are shown as black dotted lines. 
        }
        \label{fig:flat-sky}
    \end{figure*}

In Sec.~\ref{s:nl-bias} we test a model of the angular power spectrum that includes second-order terms in the bias expansion.
In order to evaluate this model and run our MCMC chains efficiently, we neglect wide-angle effects and we assume a uniform redshift distribution of the sources in our tophat bin.
In this appendix, we study the impact of these approximations
on the angular power spectra of the number counts.
For this purpose, we compare the angular
power spectra for unbiased tracers 
in the flat-sky approximation to a) the angular power spectra estimated in full-sky with the code \class, using a uniform 
redshift distribution of the sources within the tophat bins; b) the angular power spectra estimated in full-sky with the code \class, for tophat bins and the redshift distribution of the halo population used in our analysis. 
In this test, we use the same bin width adopted in our analysis in Sec.~\ref{s:nl-bias}, and nonlinear corrections are included in the power spectrum using
the \texttt{HMCODE} recipe implemented in \class.

In Fig.~\ref{fig:flat-sky} we present this comparison at three representative 
redshifts: $\bar{z} = 0.5$ (top row), $\bar{z} = 1$ (middle row) $\bar{z} = 2$ (bottom row). The left column shows the comparison of the flat-sky approximation with case a), while  the right column shows the comparison with case b). 
We find that wide angle effects are below $4\%$ at $z = 0.5$, and their impact decreases at higher redshift. On the other hand, the difference between the full computation and our approximation (which also includes the effect of a non-uniform $dN/dz$ in the bin) increases with redshift. At $z = 2$ it is still below $5\%$ on large scales, and it is still below the statistical errors of our simulated data. Therefore, we can safely use this approximation in our analysis for bins $\bar{z} \le 2$. 

\bibliographystyle{JHEP}
\bibliography{refs}

\providecommand{\href}[2]{#2}\begingroup\raggedright\begin{thebibliography}{100}

\bibitem{DES:2021wwk}
{\bf DES} Collaboration, T.~M.~C. Abbott et~al., {\it {Dark Energy Survey Year
  3 results: Cosmological constraints from galaxy clustering and weak
  lensing}},  {\em Phys. Rev. D} {\bf 105} (2022), no.~2 023520,
  [\href{http://arxiv.org/abs/2105.13549}{{\tt arXiv:2105.13549}}].

\bibitem{DESI:2016fyo}
{\bf DESI} Collaboration, A.~Aghamousa et~al., {\it {The DESI Experiment Part
  I: Science,Targeting, and Survey Design}},
  \href{http://arxiv.org/abs/1611.00036}{{\tt arXiv:1611.00036}}.

\bibitem{LSSTScience:2009jmu}
{\bf LSST Science, LSST Project} Collaboration, P.~A. Abell et~al., {\it {LSST
  Science Book, Version 2.0}},  \href{http://arxiv.org/abs/0912.0201}{{\tt
  arXiv:0912.0201}}.

\bibitem{LSSTDarkEnergyScience:2012kar}
{\bf LSST Dark Energy Science} Collaboration, A.~Abate et~al., {\it {Large
  Synoptic Survey Telescope: Dark Energy Science Collaboration}},
  \href{http://arxiv.org/abs/1211.0310}{{\tt arXiv:1211.0310}}.

\bibitem{EUCLID:2011zbd}
{\bf EUCLID} Collaboration, R.~Laureijs et~al., {\it {Euclid Definition Study
  Report}},  \href{http://arxiv.org/abs/1110.3193}{{\tt arXiv:1110.3193}}.

\bibitem{Amendola:2016saw}
L.~Amendola et~al., {\it {Cosmology and fundamental physics with the Euclid
  satellite}},  {\em Living Rev. Rel.} {\bf 21} (2018), no.~1 2,
  [\href{http://arxiv.org/abs/1606.00180}{{\tt arXiv:1606.00180}}].

\bibitem{Dore:2014cca}
O.~Dor\'e et~al., {\it {Cosmology with the SPHEREX All-Sky Spectral Survey}},
  \href{http://arxiv.org/abs/1412.4872}{{\tt arXiv:1412.4872}}.

\bibitem{Maartens:2015mra}
{\bf SKA Cosmology SWG} Collaboration, R.~Maartens, F.~B. Abdalla, M.~Jarvis,
  and M.~G. Santos, {\it {Overview of Cosmology with the SKA}},  {\em PoS} {\bf
  AASKA14} (2015) 016, [\href{http://arxiv.org/abs/1501.04076}{{\tt
  arXiv:1501.04076}}].

\bibitem{RuthBook}
R.~Durrer, {\em The Cosmic Microwave Background}.
\newblock Cambridge University Press, 2008.

\bibitem{Bernardeau:2009bm}
F.~Bernardeau, C.~Bonvin, and F.~Vernizzi, {\it {Full-sky lensing shear at
  second order}},  {\em Phys. Rev. D} {\bf 81} (2010) 083002,
  [\href{http://arxiv.org/abs/0911.2244}{{\tt arXiv:0911.2244}}].

\bibitem{Baumann:2010tm}
D.~Baumann, A.~Nicolis, L.~Senatore, and M.~Zaldarriaga, {\it {Cosmological
  Non-Linearities as an Effective Fluid}},  {\em JCAP} {\bf 07} (2012) 051,
  [\href{http://arxiv.org/abs/1004.2488}{{\tt arXiv:1004.2488}}].

\bibitem{Carrasco:2012cv}
J.~J.~M. Carrasco, M.~P. Hertzberg, and L.~Senatore, {\it {The Effective Field
  Theory of Cosmological Large Scale Structures}},  {\em JHEP} {\bf 09} (2012)
  082, [\href{http://arxiv.org/abs/1206.2926}{{\tt arXiv:1206.2926}}].

\bibitem{Porto:2013qua}
R.~A. Porto, L.~Senatore, and M.~Zaldarriaga, {\it {The Lagrangian-space
  Effective Field Theory of Large Scale Structures}},  {\em JCAP} {\bf 05}
  (2014) 022, [\href{http://arxiv.org/abs/1311.2168}{{\tt arXiv:1311.2168}}].

\bibitem{Nbody:1988}
R.~W. {Hockney} and J.~W. {Eastwood}, {\em {Computer simulation using
  particles}}.
\newblock CRC Press, 1988.

\bibitem{Angulo:2021kes}
R.~E. Angulo and O.~Hahn, {\it {Large-scale dark matter simulations}},
  \href{http://arxiv.org/abs/2112.05165}{{\tt arXiv:2112.05165}}.

\bibitem{Desjacques:2016bnm}
V.~Desjacques, D.~Jeong, and F.~Schmidt, {\it {Large-Scale Galaxy Bias}},  {\em
  Phys. Rept.} {\bf 733} (2018) 1--193,
  [\href{http://arxiv.org/abs/1611.09787}{{\tt arXiv:1611.09787}}].

\bibitem{mo_van_den_bosch_white_2010}
H.~Mo, F.~van~den Bosch, and S.~White, {\em Galaxy Formation and Evolution}.
\newblock Cambridge University Press, 2010.

\bibitem{Kaiser:1987qv}
N.~Kaiser, {\it {Clustering in real space and in redshift space}},  {\em Mon.
  Not. Roy. Astron. Soc.} {\bf 227} (1987) 1--27.

\bibitem{Matsubara:2004fr}
T.~Matsubara, {\it {Correlation function in deep redshift space as a
  cosmological probe}},  {\em Astrophys. J.} {\bf 615} (2004) 573--585,
  [\href{http://arxiv.org/abs/astro-ph/0408349}{{\tt astro-ph/0408349}}].

\bibitem{Yoo:2009au}
J.~Yoo, A.~L. Fitzpatrick, and M.~Zaldarriaga, {\it {A New Perspective on
  Galaxy Clustering as a Cosmological Probe: General Relativistic Effects}},
  {\em Phys. Rev. D} {\bf 80} (2009) 083514,
  [\href{http://arxiv.org/abs/0907.0707}{{\tt arXiv:0907.0707}}].

\bibitem{Yoo:2010ni}
J.~Yoo, {\it {General Relativistic Description of the Observed Galaxy Power
  Spectrum: Do We Understand What We Measure?}},  {\em Phys. Rev. D} {\bf 82}
  (2010) 083508, [\href{http://arxiv.org/abs/1009.3021}{{\tt
  arXiv:1009.3021}}].

\bibitem{Bonvin:2011bg}
C.~Bonvin and R.~Durrer, {\it {What galaxy surveys really measure}},  {\em
  Phys. Rev. D} {\bf 84} (2011) 063505,
  [\href{http://arxiv.org/abs/1105.5280}{{\tt arXiv:1105.5280}}].

\bibitem{Challinor:2011bk}
A.~Challinor and A.~Lewis, {\it {The linear power spectrum of observed source
  number counts}},  {\em Phys. Rev. D} {\bf 84} (2011) 043516,
  [\href{http://arxiv.org/abs/1105.5292}{{\tt arXiv:1105.5292}}].

\bibitem{Jeong:2011as}
D.~Jeong, F.~Schmidt, and C.~M. Hirata, {\it {Large-scale clustering of
  galaxies in general relativity}},  {\em Phys. Rev. D} {\bf 85} (2012) 023504,
  [\href{http://arxiv.org/abs/1107.5427}{{\tt arXiv:1107.5427}}].

\bibitem{Yoo:2014sfa}
J.~Yoo and M.~Zaldarriaga, {\it {Beyond the Linear-Order Relativistic Effect in
  Galaxy Clustering: Second-Order Gauge-Invariant Formalism}},  {\em Phys. Rev.
  D} {\bf 90} (2014), no.~2 023513, [\href{http://arxiv.org/abs/1406.4140}{{\tt
  arXiv:1406.4140}}].

\bibitem{Umeh:2014ana}
O.~Umeh, C.~Clarkson, and R.~Maartens, {\it {Nonlinear relativistic corrections
  to cosmological distances, redshift and gravitational lensing magnification.
  II - Derivation}},  {\em Class. Quant. Grav.} {\bf 31} (2014) 205001,
  [\href{http://arxiv.org/abs/1402.1933}{{\tt arXiv:1402.1933}}].

\bibitem{Bertacca:2014hwa}
D.~Bertacca, {\it {Observed galaxy number counts on the light cone up to second
  order: III. Magnification bias}},  {\em Class. Quant. Grav.} {\bf 32} (2015),
  no.~19 195011, [\href{http://arxiv.org/abs/1409.2024}{{\tt
  arXiv:1409.2024}}].

\bibitem{DiDio:2014lka}
E.~Di~Dio, R.~Durrer, G.~Marozzi, and F.~Montanari, {\it {Galaxy number counts
  to second order and their bispectrum}},  {\em JCAP} {\bf 12} (2014) 017,
  [\href{http://arxiv.org/abs/1407.0376}{{\tt arXiv:1407.0376}}]. [Erratum:
  JCAP 06, E01 (2015)].

\bibitem{Nielsen:2016ldx}
J.~T. Nielsen and R.~Durrer, {\it {Higher order relativistic galaxy number
  counts: dominating terms}},  {\em JCAP} {\bf 03} (2017) 010,
  [\href{http://arxiv.org/abs/1606.02113}{{\tt arXiv:1606.02113}}].

\bibitem{Magi:2022nfy}
M.~Magi and J.~Yoo, {\it {Second-order gauge-invariant formalism for the
  cosmological observables: Complete verification of their gauge-invariance}},
  \href{http://arxiv.org/abs/2204.01751}{{\tt arXiv:2204.01751}}.

\bibitem{Durrer:2016jzq}
R.~Durrer and V.~Tansella, {\it {Vector perturbations of galaxy number
  counts}},  {\em JCAP} {\bf 07} (2016) 037,
  [\href{http://arxiv.org/abs/1605.05974}{{\tt arXiv:1605.05974}}].

\bibitem{DiDio:2016ykq}
E.~Di~Dio, F.~Montanari, A.~Raccanelli, R.~Durrer, M.~Kamionkowski, and
  J.~Lesgourgues, {\it {Curvature constraints from Large Scale Structure}},
  {\em JCAP} {\bf 06} (2016) 013, [\href{http://arxiv.org/abs/1603.09073}{{\tt
  arXiv:1603.09073}}].

\bibitem{Bonvin:2008ni}
C.~Bonvin, {\it {Effect of Peculiar Motion in Weak Lensing}},  {\em Phys. Rev.
  D} {\bf 78} (2008) 123530, [\href{http://arxiv.org/abs/0810.0180}{{\tt
  arXiv:0810.0180}}].

\bibitem{Hall:2012wd}
A.~Hall, C.~Bonvin, and A.~Challinor, {\it {Testing General Relativity with
  21-cm intensity mapping}},  {\em Phys. Rev. D} {\bf 87} (2013), no.~6 064026,
  [\href{http://arxiv.org/abs/1212.0728}{{\tt arXiv:1212.0728}}].

\bibitem{Irsic:2015nla}
V.~Ir\v{s}i\v{c}, E.~Di~Dio, and M.~Viel, {\it {Relativistic effects in
  Lyman-\ensuremath{\alpha} forest}},  {\em JCAP} {\bf 02} (2016) 051,
  [\href{http://arxiv.org/abs/1510.03436}{{\tt arXiv:1510.03436}}].

\bibitem{Alonso:2021obj}
D.~Alonso, {\it {Linear anisotropies in dispersion-measure-based cosmological
  observables}},  {\em Phys. Rev. D} {\bf 103} (2021), no.~12 123544,
  [\href{http://arxiv.org/abs/2103.14016}{{\tt arXiv:2103.14016}}].

\bibitem{DiDio:2013bqa}
E.~Di~Dio, F.~Montanari, J.~Lesgourgues, and R.~Durrer, {\it {The CLASSgal code
  for Relativistic Cosmological Large Scale Structure}},  {\em JCAP} {\bf 11}
  (2013) 044, [\href{http://arxiv.org/abs/1307.1459}{{\tt arXiv:1307.1459}}].

\bibitem{DiDio:2013sea}
E.~Di~Dio, F.~Montanari, R.~Durrer, and J.~Lesgourgues, {\it {Cosmological
  Parameter Estimation with Large Scale Structure Observations}},  {\em JCAP}
  {\bf 01} (2014) 042, [\href{http://arxiv.org/abs/1308.6186}{{\tt
  arXiv:1308.6186}}].

\bibitem{Lesgourgues:2011re}
J.~Lesgourgues, {\it {The Cosmic Linear Anisotropy Solving System (CLASS) I:
  Overview}},  \href{http://arxiv.org/abs/1104.2932}{{\tt arXiv:1104.2932}}.

\bibitem{Blas:2011rf}
D.~Blas, J.~Lesgourgues, and T.~Tram, {\it {The Cosmic Linear Anisotropy
  Solving System (CLASS) II: Approximation schemes}},  {\em JCAP} {\bf 07}
  (2011) 034, [\href{http://arxiv.org/abs/1104.2933}{{\tt arXiv:1104.2933}}].

\bibitem{Challinor:2011}
A.~{Challinor} and A.~{Lewis}, ``{CAMB Sources: Number Counts, Lensing \&
  Dark-age 21cm Power Spectra}.'' Astrophysics Source Code Library, record
  ascl:1105.013, May, 2011.

\bibitem{Lewis:1999bs}
A.~Lewis, A.~Challinor, and A.~Lasenby, {\it {Efficient computation of CMB
  anisotropies in closed FRW models}},  {\em Astrophys. J.} {\bf 538} (2000)
  473--476, [\href{http://arxiv.org/abs/astro-ph/9911177}{{\tt
  astro-ph/9911177}}].

\bibitem{Fosalba:2007mf}
P.~Fosalba, E.~Gaztanaga, F.~Castander, and M.~Manera, {\it {The onion
  universe: all sky light-cone simulations in shells}},  {\em Mon. Not. Roy.
  Astron. Soc.} {\bf 391} (2008) 435,
  [\href{http://arxiv.org/abs/0711.1540}{{\tt arXiv:0711.1540}}].

\bibitem{Hilbert:2008kb}
S.~Hilbert, J.~Hartlap, S.~D.~M. White, and P.~Schneider, {\it {Ray-tracing
  through the Millennium Simulation: Born corrections and lens-lens coupling in
  cosmic shear and galaxy-galaxy lensing}},  {\em Astron. Astrophys.} {\bf 499}
  (2009) 31, [\href{http://arxiv.org/abs/0809.5035}{{\tt arXiv:0809.5035}}].

\bibitem{Fosalba:2013mra}
P.~Fosalba, E.~Gazta\~naga, F.~J. Castander, and M.~Crocce, {\it {The MICE
  Grand Challenge light-cone simulation \textendash{} III. Galaxy lensing mocks
  from all-sky lensing maps}},  {\em Mon. Not. Roy. Astron. Soc.} {\bf 447}
  (2015), no.~2 1319--1332, [\href{http://arxiv.org/abs/1312.2947}{{\tt
  arXiv:1312.2947}}].

\bibitem{Fosalba:2013wxa}
P.~Fosalba, M.~Crocce, E.~Gazta\~naga, and F.~J. Castander, {\it {The MICE
  grand challenge lightcone simulation \textendash{} I. Dark matter
  clustering}},  {\em Mon. Not. Roy. Astron. Soc.} {\bf 448} (2015), no.~4
  2987--3000, [\href{http://arxiv.org/abs/1312.1707}{{\tt arXiv:1312.1707}}].

\bibitem{Bentivegna:2015flc}
E.~Bentivegna and M.~Bruni, {\it {Effects of nonlinear inhomogeneity on the
  cosmic expansion with numerical relativity}},  {\em Phys. Rev. Lett.} {\bf
  116} (2016), no.~25 251302, [\href{http://arxiv.org/abs/1511.05124}{{\tt
  arXiv:1511.05124}}].

\bibitem{Giblin:2015vwq}
J.~T. Giblin, J.~B. Mertens, and G.~D. Starkman, {\it {Departures from the
  Friedmann-Lemaitre-Robertston-Walker Cosmological Model in an Inhomogeneous
  Universe: A Numerical Examination}},  {\em Phys. Rev. Lett.} {\bf 116}
  (2016), no.~25 251301, [\href{http://arxiv.org/abs/1511.01105}{{\tt
  arXiv:1511.01105}}].

\bibitem{Adamek:2015eda}
J.~Adamek, D.~Daverio, R.~Durrer, and M.~Kunz, {\it {General relativity and
  cosmic structure formation}},  {\em Nature Phys.} {\bf 12} (2016) 346--349,
  [\href{http://arxiv.org/abs/1509.01699}{{\tt arXiv:1509.01699}}].

\bibitem{Adamek:2016zes}
J.~Adamek, D.~Daverio, R.~Durrer, and M.~Kunz, {\it {gevolution: a cosmological
  N-body code based on General Relativity}},  {\em JCAP} {\bf 07} (2016) 053,
  [\href{http://arxiv.org/abs/1604.06065}{{\tt arXiv:1604.06065}}].

\bibitem{Macpherson:2016ict}
H.~J. Macpherson, P.~D. Lasky, and D.~J. Price, {\it {Inhomogeneous Cosmology
  with Numerical Relativity}},  {\em Phys. Rev. D} {\bf 95} (2017), no.~6
  064028, [\href{http://arxiv.org/abs/1611.05447}{{\tt arXiv:1611.05447}}].

\bibitem{Barrera-Hinojosa:2019mzo}
C.~Barrera-Hinojosa and B.~Li, {\it {GRAMSES: a new route to general
  relativistic $N$-body simulations in cosmology. Part I. Methodology and code
  description}},  {\em JCAP} {\bf 01} (2020) 007,
  [\href{http://arxiv.org/abs/1905.08890}{{\tt arXiv:1905.08890}}].

\bibitem{Adamek:2020jmr}
J.~Adamek, C.~Barrera-Hinojosa, M.~Bruni, B.~Li, H.~J. Macpherson, and J.~B.
  Mertens, {\it {Numerical solutions to Einstein's equations in a shearing-dust
  Universe: a code comparison}},  {\em Class. Quant. Grav.} {\bf 37} (2020),
  no.~15 154001, [\href{http://arxiv.org/abs/2003.08014}{{\tt
  arXiv:2003.08014}}].

\bibitem{Giblin:2017ezj}
J.~T. Giblin, J.~B. Mertens, G.~D. Starkman, and A.~R. Zentner, {\it {General
  Relativistic Corrections to the Weak Lensing Convergence Power Spectrum}},
  {\em Phys. Rev. D} {\bf 96} (2017), no.~10 103530,
  [\href{http://arxiv.org/abs/1707.06640}{{\tt arXiv:1707.06640}}].

\bibitem{Borzyszkowski:2017ayl}
M.~Borzyszkowski, D.~Bertacca, and C.~Porciani, {\it {LIGER: mock relativistic
  light-cones from Newtonian simulations}},  {\em Mon. Not. Roy. Astron. Soc.}
  {\bf 471} (2017), no.~4 3899--3914,
  [\href{http://arxiv.org/abs/1703.03407}{{\tt arXiv:1703.03407}}].

\bibitem{Adamek:2018rru}
J.~Adamek, C.~Clarkson, L.~Coates, R.~Durrer, and M.~Kunz, {\it {Bias and
  scatter in the Hubble diagram from cosmological large-scale structure}},
  {\em Phys. Rev. D} {\bf 100} (2019), no.~2 021301,
  [\href{http://arxiv.org/abs/1812.04336}{{\tt arXiv:1812.04336}}].

\bibitem{Lepori_2020b}
F.~Lepori, J.~Adamek, R.~Durrer, C.~Clarkson, and L.~Coates, {\it {Weak-lensing
  observables in relativistic N-body simulations}},  {\em Mon. Not. Roy.
  Astron. Soc.} {\bf 497} (2020), no.~2 2078--2095,
  [\href{http://arxiv.org/abs/2002.04024}{{\tt arXiv:2002.04024}}].

\bibitem{Lepori:2021lck}
F.~Lepori, J.~Adamek, and R.~Durrer, {\it {Cosmological simulations of number
  counts}},  {\em JCAP} {\bf 12} (2021), no.~12 021,
  [\href{http://arxiv.org/abs/2106.01347}{{\tt arXiv:2106.01347}}].

\bibitem{Barrera-Hinojosa:2021msx}
C.~Barrera-Hinojosa, B.~Li, and Y.-C. Cai, {\it {Looking for a twist: probing
  the cosmological gravitomagnetic effect via weak lensing-kSZ
  cross-correlations}},  {\em Mon. Not. Roy. Astron. Soc.} {\bf 510} (2022),
  no.~3 3589--3604, [\href{http://arxiv.org/abs/2109.02632}{{\tt
  arXiv:2109.02632}}].

\bibitem{Macpherson:2021gbh}
H.~J. Macpherson and A.~Heinesen, {\it {Luminosity distance and anisotropic
  sky-sampling at low redshifts: A numerical relativity study}},  {\em Phys.
  Rev. D} {\bf 104} (2021) 023525, [\href{http://arxiv.org/abs/2103.11918}{{\tt
  arXiv:2103.11918}}]. [Erratum: Phys.Rev.D 104, 109901 (2021)].

\bibitem{Rasera:2021mvk}
Y.~Rasera et~al., {\it {The RayGalGroupSims cosmological simulation suite for
  the study of relativistic effects: An application to lensing-matter
  clustering statistics}},  {\em Astron. Astrophys.} {\bf 661} (2022) A90,
  [\href{http://arxiv.org/abs/2111.08745}{{\tt arXiv:2111.08745}}].

\bibitem{Tian:2021qgg}
C.~Tian, M.~F. Carney, J.~B. Mertens, and G.~Starkman, {\it {Accurate
  relativistic observables from postprocessing light cone catalogs}},  {\em
  Phys. Rev. D} {\bf 105} (2022), no.~6 063511,
  [\href{http://arxiv.org/abs/2110.00893}{{\tt arXiv:2110.00893}}].

\bibitem{Hui:2007cu}
L.~Hui, E.~Gaztanaga, and M.~LoVerde, {\it {Anisotropic Magnification
  Distortion of the 3D Galaxy Correlation. 1. Real Space}},  {\em Phys. Rev. D}
  {\bf 76} (2007) 103502, [\href{http://arxiv.org/abs/0706.1071}{{\tt
  arXiv:0706.1071}}].

\bibitem{Schmidt:2009ri}
F.~Schmidt, E.~Rozo, S.~Dodelson, L.~Hui, and E.~Sheldon, {\it {Lensing Bias in
  Cosmic Shear}},  {\em Astrophys. J.} {\bf 702} (2009) 593--602,
  [\href{http://arxiv.org/abs/0904.4703}{{\tt arXiv:0904.4703}}].

\bibitem{Schmidt:2009b}
F.~Schmidt, E.~Rozo, S.~Dodelson, L.~Hui, and E.~Sheldon, {\it Size bias in
  galaxy surveys},  {\em Phys. Rev. Lett.} {\bf 103} (Jul, 2009) 051301.

\bibitem{Euclid:2019clj}
{\bf Euclid} Collaboration, A.~Blanchard et~al., {\it {Euclid preparation: VII.
  Forecast validation for Euclid cosmological probes}},  {\em Astron.
  Astrophys.} {\bf 642} (2020) A191,
  [\href{http://arxiv.org/abs/1910.09273}{{\tt arXiv:1910.09273}}].

\bibitem{Euclid:2021qvm}
{\bf Euclid} Collaboration, S.~Ili\'c et~al., {\it {Euclid preparation - XV.
  Forecasting cosmological constraints for the Euclid and CMB joint analysis}},
   {\em Astron. Astrophys.} {\bf 657} (2022) A91,
  [\href{http://arxiv.org/abs/2106.08346}{{\tt arXiv:2106.08346}}].

\bibitem{Euclid:2021rez}
{\bf Euclid} Collaboration, F.~Lepori et~al., {\it {Euclid preparation - XIX.
  Impact of magnification on photometric galaxy clustering}},  {\em Astron.
  Astrophys.} {\bf 662} (2022) A93,
  [\href{http://arxiv.org/abs/2110.05435}{{\tt arXiv:2110.05435}}].

\bibitem{Planck:2013pxb}
{\bf Planck} Collaboration, P.~A.~R. Ade et~al., {\it {Planck 2013 results.
  XVI. Cosmological parameters}},  {\em Astron. Astrophys.} {\bf 571} (2014)
  A16, [\href{http://arxiv.org/abs/1303.5076}{{\tt arXiv:1303.5076}}].

\bibitem{Gorski:2004by}
K.~M. G{\'o}rski, E.~Hivon, A.~J. Banday, B.~D. Wandelt, F.~K. Hansen,
  M.~Reinecke, and M.~Bartelman, {\it {HEALPix - A Framework for high
  resolution discretization, and fast analysis of data distributed on the
  sphere}},  {\em Astrophys. J.} {\bf 622} (2005) 759--771,
  [\href{http://arxiv.org/abs/astro-ph/0409513}{{\tt astro-ph/0409513}}].

\bibitem{Adamek:2019aad}
J.~Adamek and C.~Fidler, {\it {The large-scale general-relativistic correction
  for Newtonian mocks}},  {\em JCAP} {\bf 09} (2019) 026,
  [\href{http://arxiv.org/abs/1905.11721}{{\tt arXiv:1905.11721}}].

\bibitem{Behroozi_2012}
P.~S. Behroozi, R.~H. Wechsler, and H.-Y. Wu, {\it {The Rockstar Phase-Space
  Temporal Halo Finder and the Velocity Offsets of Cluster Cores}},  {\em
  Astrophys. J.} {\bf 762} (2013) 109,
  [\href{http://arxiv.org/abs/1110.4372}{{\tt arXiv:1110.4372}}].

\bibitem{Szapudi:2000xj}
I.~Szapudi, S.~Prunet, D.~Pogosyan, A.~S. Szalay, and J.~R. Bond, {\it {Fast
  CMB analyses via correlation functions}},
  \href{http://arxiv.org/abs/astro-ph/0010256}{{\tt astro-ph/0010256}}.

\bibitem{Chon:2003gx}
G.~Chon, A.~Challinor, S.~Prunet, E.~Hivon, and I.~Szapudi, {\it {Fast
  estimation of polarization power spectra using correlation functions}},  {\em
  Mon. Not. Roy. Astron. Soc.} {\bf 350} (2004) 914,
  [\href{http://arxiv.org/abs/astro-ph/0303414}{{\tt astro-ph/0303414}}].

\bibitem{Inman:2015pfa}
D.~Inman, J.~D. Emberson, U.-L. Pen, A.~Farchi, H.-R. Yu, and
  J.~Harnois-D\'eraps, {\it {Precision reconstruction of the cold dark
  matter-neutrino relative velocity from $N$-body simulations}},  {\em Phys.
  Rev. D} {\bf 92} (2015), no.~2 023502,
  [\href{http://arxiv.org/abs/1503.07480}{{\tt arXiv:1503.07480}}].

\bibitem{Fidler:2018geb}
C.~Fidler, N.~Sujata, and C.~Rampf, {\it {A Relativistic Interpretation of Bias
  in Newtonian Simulations}},  {\em JCAP} {\bf 02} (2019) 049,
  [\href{http://arxiv.org/abs/1810.10835}{{\tt arXiv:1810.10835}}].

\bibitem{Mead:2016zqy}
A.~Mead, C.~Heymans, L.~Lombriser, J.~Peacock, O.~Steele, and H.~Winther, {\it
  {Accurate halo-model matter power spectra with dark energy, massive neutrinos
  and modified gravitational forces}},  {\em Mon. Not. Roy. Astron. Soc.} {\bf
  459} (2016), no.~2 1468--1488, [\href{http://arxiv.org/abs/1602.02154}{{\tt
  arXiv:1602.02154}}].

\bibitem{Heitmann:2013bra}
K.~Heitmann, E.~Lawrence, J.~Kwan, S.~Habib, and D.~Higdon, {\it {The Coyote
  Universe Extended: Precision Emulation of the Matter Power Spectrum}},  {\em
  Astrophys. J.} {\bf 780} (2014) 111,
  [\href{http://arxiv.org/abs/1304.7849}{{\tt arXiv:1304.7849}}].

\bibitem{Scranton:2005ci}
{\bf SDSS} Collaboration, R.~Scranton et~al., {\it {Detection of cosmic
  magnification with the Sloan Digital Sky Survey}},  {\em Astrophys. J.} {\bf
  633} (2005) 589--602, [\href{http://arxiv.org/abs/astro-ph/0504510}{{\tt
  astro-ph/0504510}}].

\bibitem{Liu:2021gbm}
X.~Liu, D.~Liu, Z.~Gao, C.~Wei, G.~Li, L.~Fu, T.~Futamase, and Z.~Fan, {\it
  {Detection of Cosmic Magnification via Galaxy Shear -- Galaxy Number Density
  Correlation from HSC Survey Data}},  {\em Phys. Rev. D} {\bf 103} (2021),
  no.~12 123504, [\href{http://arxiv.org/abs/2104.13595}{{\tt
  arXiv:2104.13595}}].

\bibitem{Lorenz:2017iez}
C.~S. Lorenz, D.~Alonso, and P.~G. Ferreira, {\it {Impact of relativistic
  effects on cosmological parameter estimation}},  {\em Phys. Rev. D} {\bf 97}
  (2018), no.~2 023537, [\href{http://arxiv.org/abs/1710.02477}{{\tt
  arXiv:1710.02477}}].

\bibitem{Tanidis:2021uxp}
K.~Tanidis and S.~Camera, {\it {Model-independent constraints on clustering and
  growth of cosmic structures from BOSS DR12 galaxies in harmonic space}},
  \href{http://arxiv.org/abs/2107.00026}{{\tt arXiv:2107.00026}}.

\bibitem{McDonald:2009dh}
P.~McDonald and A.~Roy, {\it {Clustering of dark matter tracers: generalizing
  bias for the coming era of precision LSS}},  {\em JCAP} {\bf 08} (2009) 020,
  [\href{http://arxiv.org/abs/0902.0991}{{\tt arXiv:0902.0991}}].

\bibitem{Assassi:2014fva}
V.~Assassi, D.~Baumann, D.~Green, and M.~Zaldarriaga, {\it {Renormalized Halo
  Bias}},  {\em JCAP} {\bf 08} (2014) 056,
  [\href{http://arxiv.org/abs/1402.5916}{{\tt arXiv:1402.5916}}].

\bibitem{Senatore:2014eva}
L.~Senatore, {\it {Bias in the Effective Field Theory of Large Scale
  Structures}},  {\em JCAP} {\bf 11} (2015) 007,
  [\href{http://arxiv.org/abs/1406.7843}{{\tt arXiv:1406.7843}}].

\bibitem{Mirbabayi:2014zca}
M.~Mirbabayi, F.~Schmidt, and M.~Zaldarriaga, {\it {Biased Tracers and Time
  Evolution}},  {\em JCAP} {\bf 07} (2015) 030,
  [\href{http://arxiv.org/abs/1412.5169}{{\tt arXiv:1412.5169}}].

\bibitem{Nishimichi:2020tvu}
T.~Nishimichi, G.~D'Amico, M.~M. Ivanov, L.~Senatore, M.~Simonovi\'c,
  M.~Takada, M.~Zaldarriaga, and P.~Zhang, {\it {Blinded challenge for
  precision cosmology with large-scale structure: results from effective field
  theory for the redshift-space galaxy power spectrum}},  {\em Phys. Rev. D}
  {\bf 102} (2020), no.~12 123541, [\href{http://arxiv.org/abs/2003.08277}{{\tt
  arXiv:2003.08277}}].

\bibitem{Ivanov:2019pdj}
M.~M. Ivanov, M.~Simonovi\'c, and M.~Zaldarriaga, {\it {Cosmological Parameters
  from the BOSS Galaxy Power Spectrum}},  {\em JCAP} {\bf 05} (2020) 042,
  [\href{http://arxiv.org/abs/1909.05277}{{\tt arXiv:1909.05277}}].

\bibitem{DAmico:2019fhj}
G.~D'Amico, J.~Gleyzes, N.~Kokron, K.~Markovic, L.~Senatore, P.~Zhang,
  F.~Beutler, and H.~Gil-Mar\'\i{}n, {\it {The Cosmological Analysis of the
  SDSS/BOSS data from the Effective Field Theory of Large-Scale Structure}},
  {\em JCAP} {\bf 05} (2020) 005, [\href{http://arxiv.org/abs/1909.05271}{{\tt
  arXiv:1909.05271}}].

\bibitem{Zhang:2021yna}
P.~Zhang, G.~D'Amico, L.~Senatore, C.~Zhao, and Y.~Cai, {\it {BOSS Correlation
  Function analysis from the Effective Field Theory of Large-Scale Structure}},
   {\em JCAP} {\bf 02} (2022), no.~02 036,
  [\href{http://arxiv.org/abs/2110.07539}{{\tt arXiv:2110.07539}}].

\bibitem{Chudaykin:2020aoj}
A.~Chudaykin, M.~M. Ivanov, O.~H.~E. Philcox, and M.~Simonovi\'c, {\it
  {Nonlinear perturbation theory extension of the Boltzmann code CLASS}},  {\em
  Phys. Rev. D} {\bf 102} (2020), no.~6 063533,
  [\href{http://arxiv.org/abs/2004.10607}{{\tt arXiv:2004.10607}}].

\bibitem{DAmico:2020kxu}
G.~D'Amico, L.~Senatore, and P.~Zhang, {\it {Limits on $w$CDM from the EFTofLSS
  with the PyBird code}},  {\em JCAP} {\bf 01} (2021) 006,
  [\href{http://arxiv.org/abs/2003.07956}{{\tt arXiv:2003.07956}}].

\bibitem{Lazeyras:2015lgp}
T.~Lazeyras, C.~Wagner, T.~Baldauf, and F.~Schmidt, {\it {Precision measurement
  of the local bias of dark matter halos}},  {\em JCAP} {\bf 02} (2016) 018,
  [\href{http://arxiv.org/abs/1511.01096}{{\tt arXiv:1511.01096}}].

\bibitem{Barreira:2021ukk}
A.~Barreira, T.~Lazeyras, and F.~Schmidt, {\it {Galaxy bias from forward
  models: linear and second-order bias of IllustrisTNG galaxies}},  {\em JCAP}
  {\bf 08} (2021) 029, [\href{http://arxiv.org/abs/2105.02876}{{\tt
  arXiv:2105.02876}}].

\bibitem{Matthewson:2020rdt}
W.~L. Matthewson and R.~Durrer, {\it {The Flat Sky Approximation to Galaxy
  Number Counts}},  {\em JCAP} {\bf 02} (2021) 027,
  [\href{http://arxiv.org/abs/2006.13525}{{\tt arXiv:2006.13525}}].

\bibitem{Matthewson:2021rmb}
W.~L. Matthewson and R.~Durrer, {\it {Small scale effects in the observable
  power spectrum at large angular scales}},  {\em JCAP} {\bf 03} (2022), no.~03
  035, [\href{http://arxiv.org/abs/2107.00467}{{\tt arXiv:2107.00467}}].

\bibitem{Foreman-Mackey:2012any}
D.~Foreman-Mackey, D.~W. Hogg, D.~Lang, and J.~Goodman, {\it {emcee: The MCMC
  Hammer}},  {\em Publ. Astron. Soc. Pac.} {\bf 125} (2013) 306--312,
  [\href{http://arxiv.org/abs/1202.3665}{{\tt arXiv:1202.3665}}].

\bibitem{Gebhardt:2020imr}
H.~S. Grasshorn~Gebhardt and D.~Jeong, {\it {Nonlinear redshift-space
  distortions in the harmonic-space galaxy power spectrum}},  {\em Phys. Rev.
  D} {\bf 102} (2020), no.~8 083521,
  [\href{http://arxiv.org/abs/2008.08706}{{\tt arXiv:2008.08706}}].

\bibitem{Montanari:2015rga}
F.~Montanari and R.~Durrer, {\it {Measuring the lensing potential with
  tomographic galaxy number counts}},  {\em JCAP} {\bf 1510} (2015), no.~10
  070, [\href{http://arxiv.org/abs/1506.01369}{{\tt arXiv:1506.01369}}].

\bibitem{Jalilvand:2019brk}
M.~Jalilvand, B.~Ghosh, E.~Majerotto, B.~Bose, R.~Durrer, and M.~Kunz, {\it
  {Nonlinear contributions to angular power spectra}},  {\em Phys. Rev. D} {\bf
  101} (2020), no.~4 043530, [\href{http://arxiv.org/abs/1907.13109}{{\tt
  arXiv:1907.13109}}].

\bibitem{ipythonref}
F.~Perez and B.~E. Granger, {\it Ipython: A system for interactive scientific
  computing},  {\em Computing in Science \& Engineering} {\bf 9} (2007), no.~3
  21--29.

\bibitem{numpyref}
S.~van~der Walt, S.~C. Colbert, and G.~Varoquaux, {\it The numpy array: A
  structure for efficient numerical computation},  {\em Computing in Science \&
  Engineering} {\bf 13} (2011), no.~2 22--30.

\bibitem{numpyref2}
C.~R. {Harris}, K.~J. {Millman}, S.~J. {van der Walt}, R.~{Gommers},
  P.~{Virtanen}, D.~{Cournapeau}, E.~{Wieser}, J.~{Taylor}, S.~{Berg}, N.~J.
  {Smith}, R.~{Kern}, M.~{Picus}, S.~{Hoyer}, M.~H. {van Kerkwijk}, M.~{Brett},
  A.~{Haldane}, J.~F. {del R{\'\i}o}, M.~{Wiebe}, P.~{Peterson},
  P.~{G{\'e}rard-Marchant}, K.~{Sheppard}, T.~{Reddy}, W.~{Weckesser},
  H.~{Abbasi}, C.~{Gohlke}, and T.~E. {Oliphant}, {\it {Array programming with
  NumPy}},  {\em Nature} {\bf 585} (Sept., 2020) 357--362,
  [\href{http://arxiv.org/abs/2006.10256}{{\tt arXiv:2006.10256}}].

\bibitem{scipyref}
P.~{Virtanen}, R.~{Gommers}, T.~E. {Oliphant}, M.~{Haberland}, T.~{Reddy},
  D.~{Cournapeau}, E.~{Burovski}, P.~{Peterson}, W.~{Weckesser}, J.~{Bright},
  S.~J. {van der Walt}, M.~{Brett}, J.~{Wilson}, K.~J. {Millman}, N.~{Mayorov},
  A.~R.~J. {Nelson}, E.~{Jones}, R.~{Kern}, E.~{Larson}, C.~J. {Carey},
  {\.I}.~{Polat}, Y.~{Feng}, E.~W. {Moore}, J.~{VanderPlas}, D.~{Laxalde},
  J.~{Perktold}, R.~{Cimrman}, I.~{Henriksen}, E.~A. {Quintero}, C.~R.
  {Harris}, A.~M. {Archibald}, A.~H. {Ribeiro}, F.~{Pedregosa}, P.~{van
  Mulbregt}, and {SciPy 1. 0 Contributors}, {\it {SciPy 1.0: fundamental
  algorithms for scientific computing in Python}},  {\em Nature Methods} {\bf
  17} (Feb., 2020) 261--272, [\href{http://arxiv.org/abs/1907.10121}{{\tt
  arXiv:1907.10121}}].

\bibitem{matplotlibref}
J.~D. Hunter, {\it Matplotlib: A 2d graphics environment},  {\em Computing in
  Science \& Engineering} {\bf 9} (2007), no.~3 90--95.

\end{thebibliography}\endgroup

\end{document}